\documentclass[12pt,twoside, a4paper]{article}
\pdfoutput=1
\def\pd{\partial}
\def\mc{\mathcal}
\def\ul{\underline}

\usepackage[dvips]{graphicx}
\usepackage{amssymb}
\usepackage{amssymb,amsmath}
\usepackage{graphicx}
\usepackage{dsfont}
\usepackage{caption}
\usepackage{subcaption}
\usepackage{mathtools}
\usepackage{verbatim}
\usepackage{graphicx}
\usepackage{multirow}
\usepackage[outline]{contour}
\usepackage{xcolor,colortbl}
\usepackage{pdflscape}
\input{epsf.sty}  
\contourlength{.1pt}
\graphicspath{ {./Figures/} }
\numberwithin{equation}{section}

\begin{document}
\begin{center}
\Large{\textbf{D4-branes wrapped on topological disks from matter-coupled $F(4)$ gauged supergravity}}
\end{center}
\vspace{1 cm}
\begin{center}
\large{\textbf{Patharadanai Nuchino}$^a$ and \textbf{Parinya Karndumri}$^b$}
\end{center}
\begin{center}
$^a$ Quantum and Gravity Theory Research Group, Department of Physics, Faculty of Science, Ramkhamhaeng University, 282 Ramkhamhaeng Road, Bang Kapi, Bangkok 10240, Thailand\\
E-mail: danai.nuchino@hotmail.com \\
$^b$ String Theory and Supergravity Group, Department
of Physics, Faculty of Science, Chulalongkorn University, 254 Phayathai Road, Pathumwan, Bangkok 10330, Thailand\\
E-mail: parinya.ka@hotmail.com \vspace{1 cm} \\
\end{center}
\begin{abstract}
We study a number of supersymmetric $AdS_4\times \Sigma$ solutions with $\Sigma$ being a topological disk with non-trivial $U(1)$ holonomy on the boundary or a ``half-spindle'' from matter-coupled $F(4)$ gauged supergravity. The gauged supergravity is coupled to three vector multiplets with $SO(3)\times SO(3)$ gauge group. The resulting solutions preserve eight supercharges and $SO(2)\times SO(2)$ or $SO(2)_{\text{diag}}$ symmetries and are expected to be dual to $N=2$ SCFTs in three dimensions arising from compactifications of five-dimensional $N=2$ SCFT on a half-spindle. All of the solutions lie within the $U(1)\times U(1)$ subsector of the matter-coupled $F(4)$ gauged supergravity that can be embedded in massive type IIA theory. After uplifted to ten dimensions, the solutions can be interpreted as a system of D4-D8-branes wrapped on $\Sigma$. Some of the solutions are asymptotic to a locally $AdS_6$ geometry and can be identified as codimension-2 conformal defects within the $N=2$ SCFT in five dimensions. For these solutions, the circle inside $\Sigma$ decompactifies in the $AdS_6$ limit rendering the holographic free energy infinite. There also exist solutions with finite free energy in the dual three-dimensional SCFTs. In addition, we show that the results both extend the previously known solutions and provide a novel class of solutions.     
\end{abstract}
\newpage
\tableofcontents
\section{Introduction}
One of the most interesting aspects of the AdS/CFT correspondence \cite{maldacena,Gubser_AdS_CFT,Witten_AdS_CFT} is to provide a holographic description of strongly coupled superconformal field theories (SCFTs) compactified on a compact manifold. In many cases, this leads to a number of new SCFTs in lower dimensions. Holographic solutions realizing the compactification on a Riemann surface of $N=4$ SYM and $N=(2,0)$ SCFT in six dimensions have been studied long ago in \cite{Maldacena_Twist}. The results have been extended to more general solutions interpolating between $AdS_{m+n}$ and $AdS_m\times M^n$ solutions within gauged supergravity in $m+n$ dimensions by a large number of previous works, see \cite{flow_across_Gauntlett1}-\cite{7D_Max_twist} for an incomplete list. All of these solutions rely on the topological twist as a mechanism for preserving some part of the original supersymmetry.
\\
\indent It turns out that there are other mechanisms in which supersymmetry can be  preserved for solutions of the form $AdS_n\times \Sigma$ with $\Sigma$ being a two-dimensional compact manifold with one or two orbifold singularities. These new mechanisms have been found in \cite{Ferro_D3} and \cite{Bah_AD}, see also \cite{higher_D_orbifold1}-\cite{M5_WCP2} for extensions to higher-dimensional orbifolds. For the case of two singularities, $\Sigma$ is topologically a two-sphere with orbifold singularities at the poles \cite{Hosseini_D3, Boido_D3,Ferrero_M2, Cassani_M2,Ferrero_M5,D4_spindle,Gaunlett_super_spindle} while, in the case of one singularity, $\Sigma$ is a topological disk with non-trivial $U(1)$ holonomy on the boundary and an orbifold singularity at the origin \cite{Bah_M5,Suh_D3, Suh_D4, Suh_M2,D3_disk,7D_N2_disk}. The latter is sometimes called a ``half-spindle", see \cite{Bah_AD} for an explicit illustration. All of these solutions lead to a holographic description of higher-dimensional SCFTs compactified on a spindle or a half-spindle resulting in a large number of new SCFTs in lower dimensions. In many cases, the solutions from gauged supergravity can be uplifted to ten or eleven dimensions leading to various brane configurations realizing these new SCFTs in the context of string/M-theory. 
\\
\indent In this work, we are interested in supersymmetric $AdS_4\times \Sigma$ solutions from matter-coupled $N=(1,1)$ or $F(4)$ gauged supergravity in six dimensions constructed in \cite{Auria_F4_SUGRA} and \cite{Andrianopoli_F4_SUGRA}. Since the solutions with $\Sigma$ being a spindle have already been found in \cite{D4_spindle}, we will focus on the case of $\Sigma$ being a topological disk. A similar solution in pure $F(4)$ gauged supergravity \cite{Pure_F4} has also appeared in \cite{Suh_D4}, see also \cite{Wrapped_D4_Kim} for solutions describing D4-branes wrapped on various supersymmetric cycles. In this paper, we will extend this result to the matter-coupled $F(4)$ gauged supergravity coupled to three vector multiplets with $SO(3)\times SO(3)$ gauge group. A number of holographic solutions from this gauged supergravity have been extensively studied in \cite{PK_F4_RG_flows,PK_F4_SYM,PK_SO3xSO3_Twist, PK_SO3xSO3_BH,6D_Janus_RG,F4_defect}. In the present work, we will add to this collection supersymmetric $AdS_4\times \Sigma$ solutions for $\Sigma$ being a half-spindle. A number of possible solutions of this type have been classified in \cite{D4_4D_Orbifolds}. However, apart from the solution found in \cite{Suh_D4}, most of these solutions have not been constructed to date. In addition to providing explicit solutions within this classification, the present results lead to new solutions that do not seem to lie in the space of solutions given in \cite{D4_4D_Orbifolds}. Moreover, it turns out that all the solutions found here lie within the $U(1)\times U(1)$ truncation of the matter-coupled $F(4)$ gauged supergravity that can in turn be embedded in massive type IIA theory. Accordingly, the solutions can be uplifted to massive type IIA theory in which the interpretation in terms of D4-D8-brane systems can be given.     
\\
\indent The paper is organized as follows. In section \ref{6D_F(4)_review}, we give a brief review of six-dimensional $N=(1,1)$ gauged supergravity coupled to vector multiplets. Supersymmetric $AdS_4\times\Sigma$ solutions with $SO(2)\times SO(2)$ and $SO(2)_{\textrm{diag}}$ symmetries are considered in sections \ref{SO(2)xSO(2)_Sol} and \ref{SO(2)d_SO(3)xSO(3)gg_Sec}, respectively. Some conclusions and comments are given in section \ref{Conclus}. We also include more details on various parts of the analysis in appendices. In appendix \ref{App_field_eqs}, all bosonic field equations for the matter-coupled $F(4)$ gauged supergravity under consideration here are given. In appendix \ref{Spindle_map}, we give some relations between the solutions given here and those involving spindles studied in \cite{D4_spindle}. We also show that some of the solutions lie in the solution space given in \cite{D4_4D_Orbifolds} but some do not indicating new solutions. Finally, in appendix \ref{uplift}, we collect relevant formulae to uplift six-dimensional solutions to massive type IIA theory and needed formula for computing holographic free energy in the three-dimensional SCFTs dual to the $AdS_4\times \Sigma$ solutions. 

\section{Matter-coupled $F(4)$ gauged supergravity}\label{6D_F(4)_review}
We first review the general structure of matter-coupled $F(4)$ gauged supergravity in six dimensions. All the conventions are mostly the same as \cite{Auria_F4_SUGRA} and \cite{Andrianopoli_F4_SUGRA}, to which we refer for more details. In this paper, we will mainly present relevant formulae for obtaining supersymmetric $AdS_4\times \Sigma$ solutions. The field content of the half-maximal supergravity coupled to $n$ vector multiplets is collectively given by
\begin{equation}
(e^{\hat{\mu}}_\mu, \psi^A_\mu, A^\Lambda_\mu, B_{\mu\nu}, \chi^A,\lambda^I_A, \sigma,\phi^{\alpha I}).
\end{equation}
The metric signature is $(- + + + + +)$ with spacetime and tangent space indices denoted respectively by $\mu,\nu=0,\ldots,5$ and $\hat{\mu},\hat{\nu}=0,\ldots,5$. The bosonic fields are given by the graviton $e^{\hat{\mu}}_\mu$, a two-form field $B_{\mu\nu}$, $(4+n)$ vector fields $A^\Lambda_\mu=(A^\alpha_\mu, A^I_\mu)$, the dilaton $\sigma$, and $4n$ scalars $\phi^{\alpha I}$. The fermionic fields are given by two gravitinos $\psi^A_\mu$, two spin-1/2 fields $\chi^A$ and $2n$ gauginos $\lambda^I_A$. Indices $\alpha = (0,i)=0,1,2,3$ are split into $SU(2)_R\sim USp(2)_R\sim SO(3)_R$ R-symmetry singlet $(0)$ and adjoint index $(i=1,2,3)$ while $I=1,2,\ldots,n$ denotes the $n$ vector multiplets. The $SU(2)_R$ fundamental indices $A,B,\ldots=1,2$ are raised and lower by $\epsilon^{AB}$ and $\epsilon_{AB}$ with the convention $T^A=\epsilon^{AB}T_B$ and $T_A=T^B\epsilon_{BA}$.
\\
\indent Including the dilaton, there are $4n+1$ scalar fields described by $\mathbb{R}^+\times SO(4,n)/\left(SO(4)\times SO(n)\right)$ coset with the $\mathbb{R}^+$ factor corresponding to the dilaton. The $4n$ vector multiplet scalars can be parametrized by a coset representative ${L^\Lambda}_{\underline{\Sigma}}$ transforming under the global $SO(4,n)$ and local $SO(4)\times SO(n)$ symmetries respectively by left and right multiplications with indices $\Lambda,\underline{\Sigma}=0,...,n+3$. We can also split the index $\underline{\Sigma}$ transforming under the local $SO(4)\times SO(n)$ as $\underline{\Sigma}=(\alpha,I)=(0,i,I)$. Accordingly, the coset representative can be written as
\begin{equation}
{L^\Lambda}_{\underline{\Sigma}}=({L^\Lambda}_{\alpha},{L^\Lambda}_{I}).
\end{equation}
The inverse of ${L^\Lambda}_{\underline{\Sigma}}$ will be denoted by ${(L^{-1})^{\underline{\Sigma}}}_\Lambda=\left({(L^{-1})^{\alpha}}_\Lambda, {(L^{-1})^{I}}_\Lambda\right)$. $SO(4,n)$ indices $\Lambda,\Sigma,\ldots$ will be raised and lowered by the invariant tensor
\begin{equation}
\eta_{\Lambda\Sigma}=\eta^{\Lambda\Sigma}=(\delta_{\alpha\beta},-\delta_{IJ}).
\end{equation}
\indent We are interested in gauging a compact subgroup of $SO(4,n)$ of the form $G=SO(3)\times G_c\subset SO(4)\times SO(n)$. The $SO(3)$ factor is identified with the R-symmetry $SO(3)_R\sim SU(2)_R$ and gauged by three vector fields $A^i_\mu$ within the gravity multiplet. On the other hand, $G_c$ is a compact subgroup of $SO(n)$ gauged by the vector fields in vector multiplets with $\text{dim}(G_c)\leq n$. In addition to the gauging of $G\subset SO(4,n)$, there is a massive deformation of the two-form field relevant for the existence of supersymmetric $AdS_6$ vacua. With both of these deformations taken into account, the Lagrangian for $N=(1,1)$ gauged supergravity can be written as
\begin{eqnarray}\label{Bos_Lag}
e^{-1}\mathcal{L}&=&\frac{1}{4}R-\partial_\mu\sigma\partial^\mu\sigma-\frac{1}{4}{P_\mu}^{I\alpha} {P^\mu}_{I\alpha}-\frac{1}{8}e^{-2\sigma}\mathcal{N}_{\Lambda\Sigma}\widehat{F}^\Lambda_{\mu\nu}\widehat{F}^{\Sigma\mu\nu}-\frac{3}{64}e^{4\sigma}H_{\mu\nu\rho}H^{\mu\nu\rho}\nonumber\\
&&-V-\frac{1}{64}e^{-1}\epsilon^{\mu\nu\rho\sigma\lambda\tau}B_{\mu\nu}\left(\eta_{\Lambda\Sigma}\widehat{F}^\Lambda_{\rho\sigma}\widehat{F}^\Sigma_{\lambda\tau}
+mB_{\rho\sigma}\widehat{F}^\Lambda_{\lambda\tau}\delta_{\Lambda 0}+\frac{1}{3}m^2B_{\rho\sigma}B_{\lambda\tau}\right)\nonumber\\
\end{eqnarray}
with $e=\sqrt{-g}$. The various field strength tensors appearing in the above Lagrangian are defined by
\begin{eqnarray}
\widehat{F}^\Lambda_{\mu\nu}&=&F^\Lambda_{\mu\nu}-mB_{\mu\nu}\delta^{\Lambda}_0,\\
F^\Lambda_{\mu\nu}&=&\frac{1}{2}\left(\partial_{\mu}A^\Lambda_{\nu}-\partial_{\nu}A^\Lambda_{\mu}-f_{\Sigma\phantom{\Lambda}\Gamma}^{\phantom{\Sigma}\Lambda} A^\Sigma_{\mu}A^\Gamma_{\nu}\right),\\ 
H_{\mu\nu\rho}&=&\frac{1}{3}\left(\partial_{\mu}B_{\nu\rho}+\partial_{\nu}B_{\rho\mu}+\partial_{\rho}B_{\mu\nu}\right).
\end{eqnarray}
The constant $m$ corresponds to the aforementioned massive deformation while $f_{\Sigma\phantom{\Lambda}\Gamma}^{\phantom{\Sigma}\Lambda}$ are related to the structure constants of the gauge group with the corresponding gauge algebra given by
\begin{equation}
[T_\Lambda,T_\Sigma]={f^\Gamma}_{\Lambda\Sigma}T_\Gamma\, .
\end{equation}
In this equation, we have denoted the gauge generators in the fundamental representation of $SO(4,n)$ by $T_\Lambda$. Consistency of the gaugings with supersymmetry also requires
\begin{equation}
f_{\Lambda\Sigma\Gamma}=f_{[\Lambda\Sigma\Gamma]}\, .
\end{equation}
For $SO(3)\times G_c$ gauge group, $f_{\Lambda \Sigma \Gamma}$ takes the form of
\begin{equation}
f_{\Lambda \Sigma \Gamma}=(g_1\epsilon_{ijk},g_2C_{IJK})
\end{equation}
with $g_1$ and $g_2$ being gauge coupling constants. $C_{IJK}$ are structure constants of $G_c$.
\\
\indent The vielbein on $SO(4,n)/\left(SO(4)\times SO(n)\right)$ appearing in the scalar kinetic term can be obtained from the left-invariant one-form
\begin{equation}
\Omega_{\mu\phantom{\underline{\Lambda}}\underline{\Sigma}}^{\phantom{\mu}\underline{\Lambda}}={(L^{-1})^{\underline{\Lambda}}}_{\Pi}D_\mu{L^\Pi}_{\underline{\Sigma}}
\qquad\textrm{with}\qquad D_\mu{L^\Lambda}_{\underline{\Sigma}}=\partial_\mu{L^\Lambda}_{\underline{\Sigma}}-f_{\Gamma\phantom{\Lambda}\Pi}^{\phantom{\Gamma}\Lambda} A^\Gamma_\mu {L^\Pi}_{\underline{\Sigma}}
\end{equation}
via
\begin{equation}
P_{\mu\phantom{I}\alpha}^{\phantom{\mu}I}=(P_{\mu\phantom{I}0}^{\phantom{\mu}I},P_{\mu\phantom{I}i}^{\phantom{\mu}I})
=(\Omega_{\mu\phantom{I}0}^{\phantom{\mu}I},\Omega_{\mu\phantom{I}i}^{\phantom{\mu}I}).
\end{equation}
The other components of $\Omega_{\mu\phantom{\ul{\Lambda}}\underline{\Sigma}}^{\phantom{\mu}\underline{\Lambda}}$ are identified as the $SO(4)\times SO(n)$ composite connections (${\Omega_\mu}^{ij},{\Omega_\mu}^{i0},{\Omega_\mu}^{IJ}$).
\\
\indent The symmetric scalar matrix $\mathcal{N}_{\Lambda\Sigma}$, appearing in the vector kinetic term, is defined by
\begin{equation}
\mathcal{N}_{\Lambda\Sigma}=L_{\Lambda\alpha}{(L^{-1})^{\alpha}}_{\Sigma}-L_{\Lambda I}{(L^{-1})^{I}}_{\Sigma}=(\eta LL^T\eta)_{\Lambda\Sigma}\, .
\end{equation}
The scalar potential is given by
\begin{eqnarray}
V&=&-e^{2\sigma}\left[\frac{1}{36}A^2+\frac{1}{4}B^iB^i+\frac{1}{4}\left({C^I}_iC_{Ii}+4{D^I}_iD_{Ii}\right)\right]+m^2e^{-6\sigma}\mathcal{N}_{00}\nonumber\\&&-me^{-2\sigma}\left[\frac{2}{3}AL_{00}-2B^iL_{0i}\right]
\end{eqnarray}
with various components of fermion-shift matrices defined by
\begin{eqnarray}
A&=&\varepsilon^{ijk}K_{ijk},\qquad B^i=\varepsilon^{ijk}K_{jk0},\nonumber\\
{C_I}^i&=&\varepsilon^{ijk}K_{jIk},\qquad D_{Ii}=K_{0Ii}\, .
\end{eqnarray}
The ``boosted" structure constants are in turn given by
\begin{eqnarray}
K_{ij\alpha}&=&f_{\Lambda\phantom{\Gamma}\Sigma}^{\phantom{\Lambda}\Gamma}{L^\Lambda}_i(L^{-1})_{j\Gamma}{L^\Sigma}_\alpha,\\
K_{\alpha Ii}&=&f_{\Lambda\phantom{\Gamma}\Sigma}^{\phantom{\Lambda}\Gamma}{L^\Lambda}_\alpha(L^{-1})_{I\Gamma}{L^\Sigma}_i\, .
\end{eqnarray}
\indent Supersymmetry transformations of fermionic fields are given by
\begin{eqnarray}
\delta\psi_{\mu A}&=&D_\mu\epsilon_A-\frac{1}{24}\left[Ae^{\sigma}+6me^{-3\sigma}(L^{-1})_{00}\right]\epsilon_{AB}\Gamma_\mu\epsilon^B\nonumber\\
&&-\frac{1}{8}\left[B_ie^\sigma-2me^{-3\sigma}(L^{-1})_{i0}\right]\Gamma_7\Gamma_\mu\sigma^i_{AB}\epsilon^B\nonumber\\
&&+\frac{i}{16}e^{-\sigma}\left[(L^{-1})_{0\Lambda}\widehat{F}^{\Lambda}_{\nu\lambda}\Gamma_7\epsilon_{AB}+(L^{-1})_{i\Lambda}F^{\Lambda}_{\nu\lambda}\sigma^i_{AB}\right]({\Gamma_\mu}^{\nu\lambda}-6\delta^\nu_\mu\Gamma^\lambda)\epsilon^B\nonumber\\
&&+\frac{i}{32}e^{2\sigma}H_{\nu\lambda\rho}\Gamma_7({\Gamma_\mu}^{\nu\lambda\rho}-3\delta_\mu^\nu\Gamma^{\lambda\rho})\epsilon_A,\\
\delta\chi_A&=&\frac{1}{2}\Gamma^\mu\partial_\mu\sigma\epsilon_{AB}\epsilon^B+\frac{1}{24}\left[Ae^{-\sigma}-18me^{-3\sigma}(L^{-1})_{00}\right]\epsilon_{AB}\epsilon^B\nonumber\\
&&-\frac{1}{8}\left[B_ie^\sigma+6me^{-3\sigma}(L^{-1})_{i0}\right]\Gamma_7\sigma^i_{AB}\epsilon^B\nonumber\\
&&+\frac{i}{16}e^{-\sigma}\left[(L^{-1})_{0\Lambda}\widehat{F}^\Lambda_{\mu\nu}\Gamma_7\epsilon_{AB}-(L^{-1})_{i\Lambda}F^\Lambda_{\mu\nu}\sigma^i_{AB}\right]
\Gamma^{\mu\nu}\epsilon^B\nonumber\\
&&-\frac{i}{32}e^{2\sigma}H_{\nu\lambda\rho}\Gamma_7\Gamma^{\nu\lambda\rho}\epsilon_A,\\
\delta\lambda^I_A&=&P_{\mu\phantom{I}i}^{\phantom{\mu}I}\Gamma^\mu\sigma^i_{AB}\epsilon^B+P_{\mu\phantom{I}0}^{\phantom{\mu}I}\Gamma_7\Gamma^\mu\epsilon_{AB}\epsilon^B-e^\sigma\left(2i\Gamma_7{D^I}_i+{C^I}_i\right)\sigma^i_{AB}\epsilon^B\nonumber\\
&&+2me^{-3\sigma}{(L^{-1})^I}_0\Gamma_7\epsilon_{AB}\epsilon^B-\frac{i}{2}e^{-\sigma}{(L^{-1})^I}_\Lambda F^\Lambda_{\mu\nu}\Gamma^{\mu\nu}\epsilon_A\, .
\end{eqnarray}
In these equations, ${\sigma^{iA}}_B$ are the usual Pauli matrices and $\Gamma_\mu=e_\mu^{\hat{\mu}}\Gamma_{\hat{\mu}}$ in which $\Gamma_{\hat{\mu}}$ are six-dimensional spacetime gamma matrices satisfying the $SO(1,5)$ Clifford algebra
\begin{equation}
\{\Gamma_{\hat{\mu}},\Gamma_{\hat{\nu}}\}=2\eta_{\hat{\mu}\hat{\nu}}\mathds{1}_8\qquad\textrm{with}\qquad \eta_{\hat{\mu}\hat{\nu}}=\text{diag}(-1,1,1,1,1,1).
\end{equation}
The chirality matrix is defined by $\Gamma_7=i\Gamma^0\Gamma^1\Gamma^2\Gamma^3\Gamma^4\Gamma^5$ with $\Gamma_7^2=-\mathds{1}_8$ and $\mathds{1}_n$ being an $n\times n$ identity matrix. The supersymmetry parameter $\epsilon_A$ is an eight-component symplectic-Majorana spinor satisfying
\begin{equation}\label{ep_SM_con}
\epsilon^{AB}\epsilon^\ast_B=\mathcal{B}_6\epsilon_A\, .
\end{equation}
The matrix $\mathcal{B}_6$ is defined by
\begin{equation}
\mc{B}_6=i\mc{C}_6\Gamma^0
\end{equation}
with $\mc{C}_6$ being the six-dimensional charge conjugation matrix. The covariant derivative of $\epsilon_A$ takes the form
\begin{equation}
D_\mu\epsilon_A=\partial_\mu\epsilon_A+\frac{1}{4}{\omega_\mu}^{\nu\rho}\Gamma_{\nu\rho}\epsilon_A+Q_{\mu AB}\epsilon^B
\end{equation}
in which the composite connection is defined by
\begin{equation}
Q_{\mu AB}=\frac{i}{2}\sigma^{i}_{AB}\left(\frac{1}{2}\varepsilon_{ijk}{\Omega_\mu}^{jk}-i\Gamma_7\Omega_{\mu i0}\right).
\end{equation}
\indent To give an explicit parametrization of the $SO(4,n)/\left(SO(4)\times SO(n)\right)$ scalar manifold, we can choose the coset representative of the form
\begin{equation} 
L=e^{\phi^{\alpha I}Y_{\alpha I}}\, .
\end{equation}
The $SO(4,n)$ non-compact generators can be written in terms of $GL(n+4,\mathbb{R})$ matrices
\begin{equation}
(e^{\Lambda\Sigma})_{\Gamma\Pi}=\delta^\Lambda_\Gamma\delta^\Sigma_\Pi,\qquad \Lambda,\Sigma,\ldots=0,1,\ldots,n+3
\end{equation}
as
\begin{equation}
Y_{\alpha I}=e^{\alpha,I+3}+e^{I+3, \alpha}\, .
\end{equation}
\indent In this paper, we are interested in $F(4)$ gauged supergravity with the $SO(3)\times SO(3)\sim SU(2)\times SU(2)$ gauge group obtained by coupling the gravity multiplet to $n=3$ vector multiplets. As previously mentioned, the first $SO(3)$ is the R-symmetry gauged by $A^i_\mu$, and the second one is gauged by $A^I_\mu$, $I = 1,2,3$, from the three vector multiplets. Choosing $C_{IJK}=\varepsilon_{IJK}$, we have the following components of the structure constants 
\begin{equation}
f_{\Lambda\Sigma\Gamma}=(g_1\varepsilon_{ijk},g_2\varepsilon_{IJK}).
\end{equation}
Explicitly, $SO(3)_R$ and $SO(3)$ are generated respectively by 
\begin{eqnarray}
SO(3)_R:\qquad J^{ij}&=&e^{ji}-e^{ij},\,\,  \ \ \qquad\qquad\qquad i,j=1,2,3,\label{SO(3)R_J}\\
SO(3):\qquad \tilde{J}^{IJ}&=&e^{J+3,I+3}-e^{I+3,J+3},\qquad I,J=1,2,3\, .\label{SO(3)_J}
\end{eqnarray}
\indent Various types of holographic solutions from this gauged supergravity have been studied in a number of previous works \cite{PK_F4_RG_flows,PK_F4_SYM,PK_SO3xSO3_Twist, PK_SO3xSO3_BH,6D_Janus_RG,F4_defect}. In particular, there are two supersymmetric $AdS_6$ vacua preserving full supersymmetry with $SO(3)\times SO(3)$ and $SO(3)_{\textrm{diag}}$ symmetries. These vacua can be found by considering the $SO(3)_{\textrm{diag}}$ singlet scalars consisting of the dilaton $\sigma$ and another scalar from $SO(4,3)/(SO(4)\times SO(3))$. The latter is given by the following coset representative  
\begin{equation}
L=e^{\phi (Y_{11}+Y_{22}+Y_{33})}\, .
\end{equation}
The $AdS_6$ vacuum with $SO(3)\times SO(3)$ symmetry is given by
\begin{equation}
\phi=0,\qquad \sigma=\frac{1}{4}\ln
\left[\frac{3m}{g_1}\right],\qquad
V_0=-20m^2\left(\frac{g_1}{3m}\right)^{\frac{3}{2}}\, .\label{SO4_AdS6}
\end{equation}
We can also choose $g_1=3m$ or equivalently shifting the value of $\sigma$ such that $\sigma=0$ at the vacuum. $V_0$ denotes the cosmological constant at the vacuum. The other $AdS_6$ vacuum with $SO(3)_{\textrm{diag}}$ symmetry reads 
\begin{eqnarray}
\phi&=&\frac{1}{2}\ln \left[\frac{g_1+g_2}{g_2-g_1}\right],\qquad
\sigma=\frac{1}{4}\ln
\left[\frac{3m\sqrt{g_2^2-g_1^2}}{g_1g_2}\right],\nonumber \\
V_0&=&-20m^2\left[\frac{g_1g_2}{3m\sqrt{g_2^2-g_1^2}}\right]^{\frac{3}{2}}\,
.\label{SO3_AdS6}
\end{eqnarray}
According to the AdS/CFT correspondence, these vacua would correspond to $N=2$ SCFTs in five dimensions. In this paper, we are interested in supersymmetric solutions of the form $AdS_4\times \Sigma$ for $\Sigma$ being a topological disk with a non-trivial $U(1)$ holonomy on the boundary. The solutions are holographically dual to three-dimensional SCFTs arising from compactifications of these $N=2$ SCFTs in five dimensions on $\Sigma$.  

\section{$AdS_4\times\Sigma$ solutions with $SO(2)\times SO(2)$ symmetry}\label{SO(2)xSO(2)_Sol}
We begin with $AdS_4\times \Sigma$ solutions preserving $SO(2)\times SO(2)$ symmetry generated by $J^{12}$ and $\tilde{J}^{12}$. In this case, as in $AdS_4\times \Sigma_g$ solutions for $\Sigma_g$ being a Riemann surface considered in \cite{PK_SO3xSO3_Twist}, the resulting solutions lie within the $U(1)\times U(1)$ truncation of matter-coupled $F(4)$ gauged supergravity that can in turn be embedded in massive type IIA theory. Accordingly, we will first find the solutions from the six-dimensional gauged supergravity and then uplift the solutions to ten dimensions via a consistent truncation given in \cite{D4_4D_Orbifolds}.
 
\subsection{Analysis of the BPS equations} 
We now look for supersymmetric $AdS_4\times \Sigma$ solutions by solving the relevant BPS equations obtained from setting the supersymmetry transformations of fermionic fields to zero. Apart from the dilaton $\sigma$, there are two singlet scalars from the $SO(4,3)/\left(SO(4)\times SO(3)\right)$ with coset representative given by
\begin{equation}\label{SO(2)xSO(2)_singlet_in_SO(3)xSO(3)gg_main}
L=e^{\phi_1 Y_{03}}e^{\phi_2 Y_{33}}\, .
\end{equation}
Following \cite{Bah_M5}, we take the ansatz for the six-dimensional metric to be
\begin{equation}\label{6Dmetrix1}
ds^2_6=f(r)ds^2_{AdS_4}+h_1(r)dr^2+h_2(r)d\theta^2
\end{equation}
where the metric on $AdS_4$ with unit radius takes the form
\begin{equation}
ds^2_{AdS_4}=\frac{1}{\rho^2}(dx^2_{1,2}+d\rho^2)
\end{equation}
with $dx^2_{1,2}=\eta_{mn}dx^m dx^n$, $m,n= 0,1,2$ being the flat metric on the three-dimensional Minkowski space. $r$ and $\theta$ are respectively radial and angular coordinates on the topological disk $\Sigma$. The six-dimensional curved and flat spacetime indices will be split into $\mu=(m,\rho,r,\theta)$ and $\hat{\mu}=(\hat{m},\hat{\rho},\hat{r},\hat{\theta})$, respectively. 
\\
\indent With the following vielbein one-forms
\begin{eqnarray}
& &e^{\hat{m}}_{(1)}=\frac{\sqrt{f(r)}}{\rho}dx^{m},\qquad e^{\hat{\rho}}_{(1)}=\frac{\sqrt{f(r)}}{\rho}d\rho,\nonumber \\
& & e^{\hat{r}}_{(1)}=\sqrt{h_1(r)}dr,\qquad\ \ e^{\hat{\theta}}_{(1)}=\sqrt{h_2(r)}d\theta, \label{6Dvielbein}
\end{eqnarray}
we can straightforwardly compute all non-vanishing components of the spin connections
\begin{eqnarray}
& &\omega_{(1)}^{\hat{m}\hat{\rho}}=-\frac{1}{\sqrt{f}}e^{\hat{m}},\qquad\omega_{(1)}^{\hat{m}\hat{r}}=\frac{f'}{2f\sqrt{h_1}}e^{\hat{m}},\nonumber \\
& &\omega_{(1)}^{\hat{\rho}\hat{r}}=\frac{f'}{2f\sqrt{h_1}}e^{\hat{\rho}},\qquad\omega_{(1)}^{\hat{\theta}\hat{r}}=\frac{h'_2}{2h_2\sqrt{h_1}}e^{\hat{\theta}}\, .\label{spinCon}
\end{eqnarray}
From now on, we will use primes to denote $r$-derivatives and mostly suppress arguments of the $r$-dependent functions for convenience.
\\
\indent In addition, there are two $SO(2)\times SO(2)$ gauge fields $A^3$ and $A^6$ with the following ansatz 
\begin{equation}
A^3=A_1(r)d\theta\qquad \textrm{and}\qquad A^6=A_2(r)d\theta\, .
\end{equation}
The corresponding two-form field strengths are then given by
\begin{equation}\label{SO(2)xSO(2)_GaugeFS}
F^{3}=\frac{A'_1}{2}dr\wedge d\theta\qquad\text{ and }\qquad  F^{6}=\frac{A'_2}{2}dr\wedge d\theta\, .
\end{equation}
\indent  To find supersymmetric $AdS_4\times \Sigma$ solutions, we decompose the six-dimensional gamma matrices as
\begin{equation}\label{6D_gamma_Dec}
\Gamma_{\hat{a}}=\gamma_{\hat{a}}\otimes\mathds{1}_2,\qquad\quad \Gamma_{\hat{r}}=\gamma_\ast\otimes\sigma_1, \qquad\quad \Gamma_{\hat{\theta}}=\gamma_\ast\otimes\sigma_2
\end{equation}
in which $\sigma_1$, $\sigma_2$, and $\sigma_3$ are Pauli matrices, while $\gamma_{\hat{a}}$, for $\hat{a}=(\hat{m},\hat{\rho})=0,1,2,3$, are four-dimensional $4\times4$ gamma matrices satisfying the Clifford algebra 
\begin{equation}
\{\gamma_{\hat{a}},\gamma_{\hat{b}}\}=2\eta_{\hat{a}\hat{b}}\mathds{1}_4 \qquad \textrm{for}\qquad \eta_{\hat{a}\hat{b}}=\text{diag}(-1,1,1,1).
\end{equation}
In this basis, the six-dimensional chirality matrix takes the form
\begin{equation}
\Gamma_7=i\gamma_\ast\otimes\sigma_3
\end{equation}
where the four-dimensional chirality matrix is defined by $\gamma_\ast=i\gamma^0\gamma^1\gamma^2\gamma^3$ with $\gamma_\ast^2=\mathds{1}_4$. The six-dimensional matrix $\mc{B}_6$ is given by $\mathcal{B}_6=\mathcal{B}_4\otimes\sigma_2$. The ansatz for the symplectic-Majorana Killing spinors $\epsilon_A$ is taken to be
\begin{equation}\label{6DKilling}
\epsilon_A=\left(\vartheta_+\otimes\eta_{+,A}\right)+\left(\vartheta_-\otimes\eta_{-,A}\right)
\end{equation}
where $\vartheta=\vartheta(x^{\hat{a}})$ is a Majorana Killing spinor on $AdS_4$ with $\vartheta_\pm$ being its chiral components, i.e., $\vartheta=\vartheta_++\vartheta_-$. Thus, we have $\gamma_\ast\vartheta_\pm=\pm\vartheta_\pm$, $\vartheta^\ast_\pm=\mathcal{B}_4\vartheta_\mp$, and
\begin{equation}\label{4DKilling}
\nabla_{\hat{a}}^{AdS_4}\vartheta_\pm=\frac{k}{2}\gamma_{\hat{a}}\vartheta_\mp.
\end{equation}
The real parameter $k=\pm1$ is an arbitrary sign choice. The spinors $\eta_{\pm,A}=\eta_{\pm,A}(r,\theta)$ are two-component Dirac spinors defined on the topological disk. Note that $\eta_{\pm,A}$ are not chiral spinors. The sign $\pm$ is just a label. Given these definitions, the symplectic-Majorana condition \eqref{ep_SM_con} reduces to
\begin{equation}\label{eta_SM_con}
\epsilon^{AB}\eta^\ast_{\pm,B}=\sigma_2\eta_{\mp,A}\, .
\end{equation}
\indent Setting the two-form field $B_{\mu\nu}$ to zero and using the aforementioned ansatz for the gauge fields, we find that the field equation of the two-form field \eqref{H_field_eqs} reduces to
\begin{eqnarray}
0=me^{-2\sigma}\mathcal{N}_{0\Sigma}F^{\Sigma\mu\nu}
\end{eqnarray}
which gives
\begin{eqnarray}
0&=&\mathcal{N}_{03}F^{3r\theta}+\mathcal{N}_{06}F^{6r\theta}\nonumber\\
&=&A'_1\sinh\phi_1\sinh2\phi_2-A'_2\cosh^2\phi_2\sinh2\phi_1\, .
\end{eqnarray}
This implies either $\phi_1=0$ or $\phi_2=A_2=0$. The latter is a subsector of $SO(2)_R$ invariant sector which leads to the same solution as in pure $F(4)$ gauged supergravity studied in \cite{Suh_D4}. In particular, consistency of the field equations for the $SO(2)_R$ gauge field requires that all scalars from vector multiplets must vanish. Accordingly, we will set $\phi_1=0$ in order for the two-form field to be consistently truncated out. 
\\
\indent With $\phi_1=0$ and $H_{\mu\nu\rho}=0$ as $B_{\mu\nu}=0$, the gauge field equation \eqref{F_field_eqs} leads to
\begin{eqnarray}
A''_1&=&2\phi'_2A'_2-\left(\frac{4f'}{f}-\frac{h'_1}{h_1}-\frac{h'_2}{h_2}-4\sigma'\right)\frac{A'_1}{2},\label{SO(2)xSO(2)_Diff_eq1}\\
A''_2&=&2\phi'_2A'_1-\left(\frac{4f'}{f}-\frac{h'_1}{h_1}-\frac{h'_2}{h_2}-4\sigma'\right)\frac{A'_2}{2}\, .\label{SO(2)xSO(2)_Diff_eq2}
\end{eqnarray}
The most general solution to these equations is given by 
\begin{eqnarray}
& &A'_1=(a_1e^{4\phi_2}+a_2)e^{2\sigma-2\phi_2}\frac{\sqrt{h_1h_2}}{2f^2},\label{Gen_SO(2)xSO(2)_Ansatz1} \\
\textrm{and} \qquad & & A'_2=(a_1e^{4\phi_2}-a_2)e^{2\sigma-2\phi_2}\frac{\sqrt{h_1h_2}}{2f^2}\label{Gen_SO(2)xSO(2)_Ansatz2}
\end{eqnarray}
for two integration constants $a_1$ and $a_2$.
\\
\indent In order to find supersymmetric solutions, we consider first-order BPS equations derived from the supersymmetry transformations of fermionic fields. For later convenience, we first introduce some notations for the fermion-shift matrices as follows
\begin{eqnarray}
S_{[AB]}&=&\frac{1}{24}\left[Ae^{\sigma}+6me^{-3\sigma}(L^{-1})_{00}\right]\epsilon_{AB},\\
S_{(AB)}&=&\frac{1}{8}\left[B_ie^\sigma-2me^{-3\sigma}(L^{-1})_{i0}\right]\sigma^i_{AB},\\
N_{[AB]}&=&\frac{1}{24}\left[Ae^{-\sigma}-18me^{-3\sigma}(L^{-1})_{00}\right]\epsilon_{AB},\\
N_{(AB)}&=&\frac{1}{8}\left[B_ie^\sigma+6me^{-3\sigma}(L^{-1})_{i0}\right]\sigma^i_{AB}
\end{eqnarray}
together with the dressed field strength tensors
\begin{equation}
T_{[AB]}=\epsilon_{AB}(L^{-1})_{0\Lambda}F^{\Lambda}_{\hat{r}\hat{\theta}},\qquad T_{(AB)}=\sigma^i_{AB}(L^{-1})_{i\Lambda}F^{\Lambda}_{\hat{r}\hat{\theta}},\qquad T^I={(L^{-1})^I}_\Lambda F^\Lambda_{\hat{r}\hat{\theta}}.
\end{equation}
\indent With the coset representative given in \eqref{SO(2)xSO(2)_singlet_in_SO(3)xSO(3)gg_main} together with the two-form field strengths \eqref{SO(2)xSO(2)_GaugeFS}, the explicit form of these quantities reads 
\begin{eqnarray}
S_{[AB]}&=&\mathbb{S}_{\text{a}}\epsilon_{AB}=\frac{1}{4}(me^{-3\sigma}\cosh\phi_1+g_1e^{\sigma}\cosh\phi_2)\epsilon_{AB},\\
S_{(AB)}&=&\mathbb{S}_{\text{s}}\sigma^3_{AB}=-\frac{m}{4}e^{-3\sigma}\sinh\phi_1\sinh\phi_2\sigma^3_{AB},\\
N_{[AB]}&=&\mathbb{N}_{\text{a}}\epsilon_{AB}=-\frac{1}{4}(3me^{-3\sigma}\cosh\phi_1-g_1e^{\sigma}\cosh\phi_2)\epsilon_{AB},\\
N_{(AB)}&=&\mathbb{N}_{\text{s}}\sigma^3_{AB}=\frac{3m}{4}e^{-3\sigma}\sinh\phi_1\sinh\phi_2\sigma^3_{AB},\\
T_{[AB]}&=&\mathbb{F}_{\text{a}}\epsilon_{AB}=-\frac{1}{2\sqrt{h_1h_2}}A'_2\sinh\phi_1\epsilon_{AB},\\
T_{(AB)}&=&\mathbb{F}_{1}\sigma^3_{AB}=\frac{1}{2\sqrt{h_1h_2}}(A'_1\cosh\phi_2-A'_2\cosh\phi_1\sinh\phi_2)\sigma^3_{AB},\ \qquad\\
T^I&=&\mathbb{F}_{2}\delta^I_3=\frac{1}{2\sqrt{h_1h_2}}(A'_2\cosh\phi_1\cosh\phi_2-A'_1\sinh\phi_2)\delta^I_3
\end{eqnarray}
with $\mathbb{S}_{\text{a}}$, $\mathbb{S}_{\text{s}}$, $\mathbb{N}_{\text{a}}$, $\mathbb{N}_{\text{s}}$, $\mathbb{F}_{\text{a}}$, $\mathbb{F}_{1}$, and $\mathbb{F}_{2}$ being their scalar components.
\\
\indent We begin with the gravitino variation $\delta\psi_{\mu A}$ along $m$, $\rho$, $r$, and $\theta$ directions. These are respectively given by
\begin{eqnarray}
0&=&\partial_m\epsilon_A-\frac{1}{2\sqrt{f}}\Gamma_m\Gamma_{\hat{\rho}}\epsilon_A+\frac{f'}{4f\sqrt{h_1}}\Gamma_m\Gamma_{\hat{r}}\epsilon_A-\left(S_{[AB]}+S_{(AB)}\Gamma_7\right)\Gamma_m\epsilon^B\nonumber\\
&&+\frac{i}{8}e^{-\sigma}\left(T_{[AB]}\Gamma_7+T_{(AB)}\right)\Gamma_m\Gamma^{\hat{r}\hat{\theta}}\epsilon^B,\label{BPS_m}\\
0&=&\partial_\rho\epsilon_A+\frac{f'}{4f\sqrt{h_1}}\Gamma_\rho\Gamma_{\hat{r}}\epsilon_A-\left(S_{[AB]}+S_{(AB)}\Gamma_7\right)\Gamma_\rho\epsilon^B\nonumber\\
&&+\frac{i}{8}e^{-\sigma}\left(T_{[AB]}\Gamma_7+T_{(AB)}\right)\Gamma_\rho\Gamma^{\hat{r}\hat{\theta}}\epsilon^B,\label{BPS_rho}\\
0&=&\partial_r\epsilon_A+Q_{r AB}\epsilon^B-\left(S_{[AB]}+S_{(AB)}\Gamma_7\right)\Gamma_r\epsilon^B\nonumber\\
&&-\frac{3i}{8}\sqrt{h_1}e^{-\sigma}\left(T_{[AB]}\Gamma_7+T_{(AB)}\right)\Gamma^{\hat{\theta}}\epsilon^B,\label{BPS_r}\\
0&=&\partial_\theta\epsilon_A+Q_{\theta AB}\epsilon^B+\frac{h'_2}{4h_2\sqrt{h_1}}\Gamma_\theta\Gamma_{\hat{r}}\epsilon_A-\left(S_{[AB]}+S_{(AB)}\Gamma_7\right)\Gamma_\theta\epsilon^B\nonumber\\
&&+\frac{3i}{8}\sqrt{h_2}e^{-\sigma}\left(T_{[AB]}\Gamma_7+T_{(AB)}\right)\Gamma^{\hat{r}}\epsilon^B.\label{BPS_theta}
\end{eqnarray}
The next condition from $\delta\chi_A$ yields
\begin{equation}
0\ =\ \frac{\sigma'}{2}\Gamma^{r}\epsilon_A+\left(N_{[AB]}-N_{(AB)}\Gamma_7\right)\epsilon^B+\frac{i}{8}e^{-\sigma}\left(T_{[AB]}\Gamma_7-T_{(AB)}\right)\Gamma^{\hat{r}\hat{\theta}}\epsilon^B\label{BPS_sigma}
\end{equation}
while the supersymmetry transformation of the gauginos $\delta\lambda^I_A$ gives rise to
\begin{eqnarray}
0&=&{{P_r}^{I}}_i\Gamma^{r}\sigma^i_{AB}\epsilon^B-{{P_r}^{I}}_0\Gamma_7\Gamma^{r}\epsilon_A-e^{\sigma}\left(2i\Gamma_7{D^I}_i+{C^I}_i\right)\sigma^i_{AB}\epsilon^B\nonumber\\
&&-2me^{-3\sigma}{(L^{-1})^I}_0\Gamma_7\epsilon_{A}-ie^{-\sigma}T^I\Gamma^{\hat{r}\hat{\theta}}\epsilon_A\, .\label{BPS_phi}
\end{eqnarray}
\indent With the six-dimensional gamma matrices \eqref{6D_gamma_Dec} and the supersymmetry parameter \eqref{6DKilling}, the first two BPS equations given in \eqref{BPS_m} and \eqref{BPS_rho} reduce to a single equation 
\begin{eqnarray}
0&=&\left(\gamma_{\hat{a}}\vartheta_+\right)\otimes\left\{\frac{k}{2\sqrt{f}}\eta_{-,A}+\frac{f'}{4f\sqrt{h_1}}\sigma_1\eta_{+,A}-S_{[AB]}\eta^B_+-\frac{e^{-\sigma}}{8}T_{(AB)}\sigma_3\eta^B_+\right.\nonumber\\&&\left.\quad\qquad\qquad+iS_{(AB)}\eta^B_+-\frac{ie^{-\sigma}}{8}T_{[AB]}\eta^B_+\right\}\nonumber\\
&&+\left(\gamma_{\hat{a}}\vartheta_-\right)\otimes\left\{\frac{k}{2\sqrt{f}}\eta_{+,A}-\frac{f'}{4f\sqrt{h_1}}\sigma_1\eta_{-,A}-S_{[AB]}\eta^B_--\frac{e^{-\sigma}}{8}T_{(AB)}\sigma_3\eta^B_-\right.\nonumber\\&&\left.\quad\qquad\qquad-iS_{(AB)}\eta^B_-+\frac{ie^{-\sigma}}{8}T_{[AB]}\eta^B_-\right\}
\end{eqnarray}
in which we have used the Killing spinor equation \eqref{4DKilling}. The two parts are linearly independent and have to vanish separately giving rise to the following four BPS equations on the two-component spinors $\eta^A_{\pm}$
\begin{eqnarray}
0&=&\frac{k}{2\sqrt{f}}\eta^1_-+\frac{f'}{4f\sqrt{h_1}}\sigma_1\eta^1_++\mathbb{S}_{\text{a}}\eta^1_+-\frac{e^{-\sigma}}{8}\mathbb{F}_{1}\sigma_3\eta^1_++i\mathbb{S}_{\text{s}}\eta^1_++\frac{ie^{-\sigma}}{8}\mathbb{F}_{\text{a}}\eta^1_+,\qquad\label{BPS_m_eta1p}\\
0&=&\frac{k}{2\sqrt{f}}\eta^2_-+\frac{f'}{4f\sqrt{h_1}}\sigma_1\eta^2_++\mathbb{S}_{\text{a}}\eta^2_++\frac{e^{-\sigma}}{8}\mathbb{F}_{1}\sigma_3\eta^2_+-i\mathbb{S}_{\text{s}}\eta^2_++\frac{ie^{-\sigma}}{8}\mathbb{F}_{\text{a}}\eta^2_+,\qquad\label{BPS_m_eta2p}\\
0&=&\frac{k}{2\sqrt{f}}\eta^1_+-\frac{f'}{4f\sqrt{h_1}}\sigma_1\eta^1_-+\mathbb{S}_{\text{a}}\eta^1_--\frac{e^{-\sigma}}{8}\mathbb{F}_{1}\sigma_3\eta^1_--i\mathbb{S}_{\text{s}}\eta^1_--\frac{ie^{-\sigma}}{8}\mathbb{F}_{\text{a}}\eta^1_-,\qquad\label{BPS_m_eta1m}\\
0&=&\frac{k}{2\sqrt{f}}\eta^2_+-\frac{f'}{4f\sqrt{h_1}}\sigma_1\eta^2_-+\mathbb{S}_{\text{a}}\eta^2_-+\frac{e^{-\sigma}}{8}\mathbb{F}_{1}\sigma_3\eta^2_-+i\mathbb{S}_{\text{s}}\eta^2_--\frac{ie^{-\sigma}}{8}\mathbb{F}_{\text{a}}\eta^2_-.\label{BPS_m_eta2m}
\end{eqnarray}
However, $\eta^A_{\pm}$ are related to $(\epsilon_{AB}\eta^B_{\mp})^\ast$ via the symplectic-Majorana condition \eqref{eta_SM_con}. Consistency between \eqref{BPS_m_eta1p} and \eqref{BPS_m_eta2m} implies that $\mathbb{S}_{\text{s}}=\mathbb{F}_{\text{a}}=0$. These two equations then reduce to the complex conjugate of each other. Similarly, \eqref{BPS_m_eta1m} becomes the complex conjugate of \eqref{BPS_m_eta2p}. We see again that $\phi_1$ must vanish in consistent with the previous result obtained from truncating out the two-form field. In addition, with $\phi_1=0$, the quantities $P_{r\phantom{I}0}^{\phantom{r}I}$, ${D^I}_i$, and ${(L^{-1})^I}_0$ also vanish. 
\\
\indent With $\phi_1=0$, the explicit form of non-vanishing components of the fermion-shift matrices and the dressed field strength tensors are given by 
\begin{eqnarray}
\mathbb{S}_{\text{a}}&=&\frac{1}{4}e^{-3\sigma}(m+g_1e^{4\sigma}\cosh\phi_2),\label{SO(2)xSO(2)_Sa}\\
\mathbb{N}_{\text{a}}&=&-\frac{1}{4}e^{-3\sigma}(3m-g_1e^{4\sigma}\cosh\phi_2),\label{SO(2)xSO(2)_Na}\\
\mathbb{F}_{1}&=&\frac{1}{2\sqrt{h_1h_2}}(A'_1\cosh\phi_2-A'_2\sinh\phi_2),\label{SO(2)xSO(2)_Ansatz1}\\
\mathbb{F}_{2}&=&\frac{1}{2\sqrt{h_1h_2}}(A'_2\cosh\phi_2-A'_1\sinh\phi_2)\label{SO(2)xSO(2)_Ansatz2}
\end{eqnarray}
together with
\begin{equation}\label{SO(2)xSO(2)_PandC}
{{P_r}^{I}}_i=\phi'_2\delta^I_3\delta_{i3}\qquad\text{ and }\qquad {C^I}_i=2g_1\sinh\phi_2\delta^I_3\delta_{i3}.
\end{equation}
The composite connections $Q_{rAB}=0$ and $Q_{\theta AB}$ take the form of
\begin{equation}
Q_{\theta AB}=-\frac{i}{2}g_1A^3_\theta\sigma^3_{AB}=-\frac{i}{2}g_1A_1(r)\sigma^3_{AB}\, .
\end{equation}
We also note that these involve only the $SO(2)_R\subset SO(3)_R$ gauge field $A^3$ since the supersymmetry parameters are charged exclusively under $SO(3)_R$. 
\\
\indent We now consider \eqref{BPS_m_eta1p} to \eqref{BPS_m_eta2m} with $\mathbb{S}_s=\mathbb{F}_a=0$ due to $\phi_1=0$. By further imposing the following projector
\begin{equation}\label{add_proj}
\eta^2_-=\sigma_3\eta^2_+,
\end{equation}
we can find only one independent BPS equation
\begin{equation}
0=\frac{k}{2\sqrt{f}}\sigma_3\eta^2_++\frac{f'}{4f\sqrt{h_1}}\sigma_1\eta^2_++\mathbb{S}_{\text{a}}\eta^2_++\frac{e^{-\sigma}}{8}\mathbb{F}_{1}\sigma_3\eta^2_+,\label{AdS4_con}
\end{equation}
while the four two-component spinors $\eta^A_{\pm}$ can be expressed in terms of a single independent spinor, which we choose to be $\eta^2_+$, by means of the relations \eqref{add_proj} together with
\begin{equation}
\eta^1_-=-\sigma_2\left(\eta^2_+\right)^\ast\qquad\text{ and }\qquad \eta^1_+=-i\sigma_1\left(\eta^2_+\right)^\ast.
\end{equation}
Repeating the same procedure for the remaining conditions, we find the following BPS equations on the two-component spinor $\eta^2_+$
\begin{eqnarray}
0&=&\frac{1}{\sqrt{h_1}}\partial_r\eta^2_++\mathbb{S}_{\text{a}}\sigma_1\eta^2_++\frac{3}{8}e^{-\sigma}\mathbb{F}_{1}(i\sigma_2)\eta^2_+,\label{partial_r_con}\quad\\
0&=&\frac{i}{\sqrt{h_2}}\left(\partial_\theta\eta^2_++\frac{ig_1A_1}{2}\eta^2_+\right)+\frac{h'_2}{4h_2\sqrt{h_1}}\sigma_3\eta^2_++\mathbb{S}_{\text{a}}(i\sigma_2)\eta^2_+\nonumber\\&&+\frac{3}{8}e^{-\sigma}\mathbb{F}_{1}(\sigma_1)\eta^2_+,\label{partial_theta_con}\\
0&=&\frac{\sigma'}{2\sqrt{h_1}}\sigma_1\eta^2_+-\mathbb{N}_{\text{a}}\eta^2_+-\frac{e^{-\sigma}}{8}\mathbb{F}_{1}\sigma_3\eta^2_+,\label{sigma_con}\\
0&=&\frac{\phi'_2}{\sqrt{h_1}}\sigma_1\eta^2_+-2g_1e^{\sigma}\sinh\phi_2\eta^2_+-e^{-\sigma}\mathbb{F}_{2}\sigma_3\eta^2_+\label{key_BPS_4}
\end{eqnarray}
\indent We now perform the analysis of the resulting BPS equations. Following \cite{Bah_M5}, we assume a definite charge of the two-component spinor $\eta^2_+(r,\theta)$ under the $U(1)_\theta$ isometry. This leads to the ansatz for $\eta^2_+$ of the form
\begin{equation}\label{eta_ansatz}
\eta^2_+(r,\theta)=e^{iq\theta}\widehat{\eta}(r)
\end{equation}
with a constant $q$. We find that the combination $\left(\partial_\theta\eta^2_++\frac{ig_1A_1}{2}\eta^2_+\right)$ in \eqref{partial_theta_con} is invariant under the transformations
\begin{equation}
A_1\rightarrow A_1-\frac{2\alpha_0}{g_1}\qquad\text{ and }\qquad \eta^2_+\rightarrow e^{i\alpha_0\theta}\eta^2_+
\end{equation}
where $\alpha_0$ is an arbitrary constant. It is then convenient to define
\begin{equation}\label{AandA}
g_1\widehat{A}_1=2q+g_1A_1
\end{equation}
with $\widehat{A}'_1=A'_1$.
\\
\indent With all these and suitable left-multiplications by Pauli matrices, we can combine \eqref{AdS4_con} and \eqref{partial_theta_con} with \eqref{sigma_con} to obtain the following non-trivial set of the BPS equations 
\begin{eqnarray}
0&=&\frac{k}{\sqrt{f}}\eta^2_++\frac{1}{\sqrt{h_1}}\left(\frac{f'}{2f}+\sigma'\right)(i\sigma_2)\eta^2_++2me^{-3\sigma}\sigma_3\eta^2_+,\label{key_BPS_1}\\
0&=&\frac{1}{\sqrt{h_1}}\partial_r\eta^2_++\frac{1}{4}e^{-3\sigma}(m+g_1e^{4\sigma}\cosh\phi_2)\sigma_1\eta^2_++\frac{3}{8}e^{-\sigma}\mathbb{F}_1(i\sigma_2)\eta^2_+,\label{key_r_con}\\
0&=&\frac{g_1\widehat{A}_1}{\sqrt{h_2}}\sigma_1\eta^2_++\frac{1}{\sqrt{h_1}}\left(\frac{h'_2}{2h_2}+\sigma'\right)(i\sigma_2)\eta^2_++2me^{-3\sigma}\sigma_3\eta^2_+-e^{-\sigma}\mathbb{F}_1\eta^2_+,\qquad \label{key_BPS_2}\\
0&=&\frac{\sigma'}{\sqrt{h_1}}\sigma_3\eta^2_++\frac{1}{2}e^{-3\sigma}(3m-g_1e^{4\sigma}\cosh\phi_2)(i\sigma_2)\eta^2_++\frac{1}{4}e^{-\sigma}\mathbb{F}_1\sigma_1\eta^2_+,\label{key_BPS_3}\\
0&=&\frac{\phi'_2}{\sqrt{h_1}}\sigma_3\eta^2_+-2g_1e^{\sigma}\sinh\phi_2(i\sigma_2)\eta^2_++e^{-\sigma}\mathbb{F}_2\sigma_1\eta^2_+\, .\label{key_BPS_4}
\end{eqnarray}
Equation \eqref{key_r_con} determines the explicit form of $\eta^2_+(r,\theta)$ while the other four equations are of the form $M^{(x)}\eta^2_+=0$ where the index $x=1,2,3,4$ respectively labels the BPS equations in \eqref{key_BPS_1}, \eqref{key_BPS_2}, \eqref{key_BPS_3}, and \eqref{key_BPS_4}. The four $2\times2$ matrices $M^{(x)}$ can be parametrized by
\begin{equation}
M^{(x)}=X^{(x)}_0\mathds{1}_2+X^{(x)}_1\sigma_1+X^{(x)}_2(i\sigma_2)+X^{(x)}_3\sigma_3\, .
\end{equation}
Following \cite{Bah_M5}, for each matrix $M^{(x)}$, we define the two-component vectors
\begin{equation}
v^{(x)}=\begin{pmatrix} X^{(x)}_1+X^{(x)}_2 \\ -X^{(x)}_0-X^{(x)}_3 \end{pmatrix}\qquad \textrm{and}\qquad w^{(x)}=\begin{pmatrix} X^{(x)}_0-X^{(x)}_3 \\ -X^{(x)}_1+X^{(x)}_2 \end{pmatrix}
\end{equation}
together with
\begin{equation}
\mathcal{A}^{xy}=\text{det}(v^{(x)}|w^{(y)}),\qquad\mathcal{B}^{xy}=\text{det}(v^{(x)}|v^{(y)}),\qquad\mathcal{C}^{xy}=\text{det}(w^{(x)}|w^{(y)}).
\end{equation}
The notation $(v|w)$ denotes a $2\times2$ matrix obtained from a juxtaposition of the two-column vectors $v$ and $w$. As pointed out in \cite{Bah_M5}, the vanishing of $\mathcal{A}^{xy}$, $\mathcal{B}^{xy}$, and $\mathcal{C}^{xy}$ gives a number of necessary conditions for the existence of a non-trivial solution for $\eta^2_+$ as well as BPS equations for all the other fields. We will solve the conditions from the supergravity and vector multiplets separately by splitting the index $x$ as $(\bar{x},4)$ with $\bar{x}=1,2,3$. 
\\
\indent Starting from the first matrix $\mathcal{A}^{xy}$ that is asymmetric in $xy$ indices, the vanishing of the diagonal components $\mathcal{A}^{\bar{x}\bar{x}}$ gives the following conditions
\begin{eqnarray}
0&=&\frac{1}{f}+\frac{1}{4h_1}\left(\frac{f'}{f}+2\sigma'\right)^2-4 m^2 e^{-6 \sigma},\label{DABPS1}\\
0&=&(\mathbb{F}_1)^2 e^{-2 \sigma}+\frac{1}{4 h_1}\left(\frac{h'_2}{h_2}+2 \sigma '\right)^2-4 m^2 e^{-6 \sigma}-\frac{(g_1 \widehat{A}_1)^2}{h_2},\label{DABPS2}\\
0&=&\frac{16(\sigma')^2}{ h_1}-4 e^{-6 \sigma} \left(g_1 e^{4 \sigma} \cosh{\phi_2}-3m\right)^2+(\mathbb{F}_1)^2 e^{-2 \sigma}.\label{DABPS3}
\end{eqnarray}
The vanishing of the off-diagonal symmetric components $\mathcal{A}^{\bar{x}\bar{y}}+\mathcal{A}^{\bar{y}\bar{x}}$ for $\bar{x}\neq \bar{y}$ yields 
\begin{eqnarray}
0&=&\frac{1}{2h_1}\left(\frac{f'}{f}+2\sigma'\right) \left(\frac{h'_2}{h_2}+2\sigma'\right)-\frac{2k}{\sqrt{f}}\mathbb{F}_1e^{-\sigma}-8 m^2 e^{-6 \sigma},\label{SABPS1}\\
0&=&8m\sigma'+\left(\frac{f'}{f}+2\sigma'\right)\left(g_1e^{4 \sigma}\cosh \phi_2-3m\right),\label{SABPS2}\\
0&=&\frac{8 m\sigma'}{\sqrt{h_1}}+\frac{1}{\sqrt{h_1}}\left(\frac{h'_2}{h_2}+2\sigma'\right)\left(g_1e^{4 \sigma} \cosh\phi_2-3 m\right)+\frac{g_1\widehat{A}_1}{\sqrt{h_2}} \mathbb{F}_1e^{2 \sigma}\ \label{SABPS3}
\end{eqnarray}
while the off-diagonal antisymmetric components $\mathcal{A}^{\bar{x}\bar{y}}-\mathcal{A}^{\bar{y}\bar{x}}=0$ give
\begin{eqnarray}
0&=&\frac{4k m}{\sqrt{f}}e^{-3 \sigma}+\frac{g_1\widehat{A}_1}{\sqrt{h_1h_2}}\left(\frac{f'}{f}+2\sigma'\right)+4m\mathbb{F}_1 e^{-4 \sigma},\label{AABPS1}\\
0&=&\frac{8 k \sigma'}{\sqrt{f}}+\left(\frac{f'}{f}+2\sigma'\right)\mathbb{F}_1e^{-\sigma} ,\label{AABPS2}\\
0&=&\frac{4g_1\widehat{A}_1}{\sqrt{h_2}}\left(g_1e^{4 \sigma} \cosh\phi_2-3m\right)+\frac{1}{\sqrt{h_1}}\left(\frac{h'_2}{h_2}-6\sigma '\right)\mathbb{F}_1e^{2\sigma}.\label{AABPS3}
\end{eqnarray}
\indent On the other hand, matrices $\mathcal{B}^{xy}$ and $\mathcal{C}^{xy}$ are both antisymmetric in $xy$ indices. It turns out that the BPS conditions arising from the combinations $\mathcal{B}^{xy}\pm\mathcal{C}^{xy}=0$ take a simpler form. The combinations $\mathcal{B}^{\bar{x}\bar{y}}+\mathcal{C}^{\bar{x}\bar{y}}=0$ give the following conditions
\begin{eqnarray}
0&=&\frac{e^{-\sigma}}{\sqrt{h_1}}\left(\frac{f'}{f}+2 \sigma'\right)\mathbb{F}_1+\frac{k}{\sqrt{fh_1}}\left(\frac{h'_2}{h_2}+2\sigma'\right)+\frac{4m}{\sqrt{h_2}}g_1\widehat{A}_1e^{-3 \sigma},\label{ABCBPS1}\\
0&=&m\mathbb{F}_1e^{-\sigma}-\frac{k}{\sqrt{f}}\left(g_1e^{4 \sigma} \cosh\phi_2-3 m\right),\label{ABCBPS2}\\
0&=&\frac{2g_1\widehat{A}_1\sigma'}{\sqrt{h_1h_2}}-\mathbb{F}_1\left(g_1\cosh\phi_2-2me^{-4 \sigma}\right)\label{ABCBPS3}
\end{eqnarray}
while the combinations $\mathcal{B}^{\bar{x}\bar{y}}-\mathcal{C}^{\bar{x}\bar{y}}=0$ lead to
\begin{eqnarray}
0&=&\frac{m e^{-3 \sigma}}{\sqrt{h_1}}\left(\frac{f'}{f}-\frac{h'_2}{h_2}\right)-\frac{kg_1\widehat{A}_1}{\sqrt{fh_2}},\label{SBCBPS1}\\
0&=&\frac{ke^{-\sigma}}{\sqrt{f}}\mathbb{F}_1-\frac{2\sigma'}{h_1}\left(\frac{f'}{f}+2\sigma'\right)-4m e^{-6 \sigma} \left(g_1e^{4 \sigma} \cosh\phi_2-3 m\right),\label{SBCBPS2}\\
0&=&e^{-2 \sigma}(\mathbb{F}_1)^2+\frac{2\sigma'}{h_1}\left(\frac{h'_2}{h_2}+2\sigma'\right)+4me^{-6 \sigma}(g_1e^{4 \sigma}\cosh\phi_2-3m).\label{SBCBPS3}
\end{eqnarray}
Accordingly, there are in total fifteen algebraic conditions obtained from the BPS equations of the supergravity multiplet \eqref{key_BPS_1}, \eqref{key_BPS_2}, and \eqref{key_BPS_3}.  
\\
\indent Extending this procedure to the BPS equation \eqref{key_BPS_4} from vector multiplets gives additional BPS conditions derived from the vanishing of $\mathcal{A}^{x4}$, $\mathcal{B}^{\bar{x}4}$, and $\mathcal{C}^{\bar{x}4}$. The first condition obtained from $\mathcal{A}^{44}=0$ takes the form
\begin{equation}
0=\frac{(\phi'_2)^2}{h_1}-4e^{2 \sigma} \left(g_1\sinh\phi_2\right)^2+(\mathbb{F}_2)^2 e^{-2 \sigma}.\label{A44_BPS_con}
\end{equation}
The symmetric part $\mathcal{A}^{\bar{x}4}+\mathcal{A}^{4\bar{x}}=0$ gives 
\begin{eqnarray}
0&=&2m\phi'_2+g_1e^{4 \sigma} \sinh\phi_2 \left(\frac{f'}{f}+2\sigma'\right),\label{SABPS4}\\
0&=&\frac{2m\phi'_2}{\sqrt{h_1}}+\frac{g_1}{\sqrt{h_1}}e^{4\sigma}\sinh\phi_2\left(\frac{h'_2}{h_2}+2\sigma'\right)+\frac{g_1\widehat{A}_1}{\sqrt{h_2}}\mathbb{F}_2e^{2\sigma},\label{SABPS5}\\
0&=&\frac{4 \sigma '\phi'_2}{h_1}e^{2 \sigma}-4g_1\sinh\phi_2\left(g_1e^{4 \sigma}\cosh\phi_2-3m\right)+\mathbb{F}_1\mathbb{F}_2\label{SABPS6}
\end{eqnarray}
while the antisymmetric part $\mathcal{A}^{\bar{x}4}-\mathcal{A}^{4\bar{x}}=0$ results in
\begin{eqnarray}
0&=&\frac{2 k \phi'_2}{\sqrt{f}}+\left(\frac{f'}{f}+2\sigma'\right)\mathbb{F}_2e^{-\sigma},\label{AABPS4}\\
0&=&\frac{4 g_1^2\widehat{A}_1}{\sqrt{h_2}}e^{2 \sigma} \sinh\phi_2+\frac{1}{\sqrt{h_1}}\left(\frac{h'_2}{h_2}+2\sigma'\right)\mathbb{F}_2-\frac{2\phi'_2}{\sqrt{h_1}}\mathbb{F}_1,\label{AABPS5}\\
0&=&g_1\mathbb{F}_1e^{4\sigma}\sinh\phi_2-(g_1e^{4\sigma}\cosh\phi_2-3m)\mathbb{F}_2\, .\label{AABPS6}
\end{eqnarray}
\indent Moreover, the combinations $\mathcal{B}^{\bar{x}4}+\mathcal{C}^{\bar{x}4}=0$ and $\mathcal{B}^{\bar{x}4}-\mathcal{C}^{\bar{x}4}=0$ respectively lead to the following sets of extra conditions 
\begin{eqnarray}
0&=&\frac{kg_1}{\sqrt{f}}e^{5 \sigma}\sinh\phi_2-m\mathbb{F}_2,\label{ABCBPS4}\\
0&=&2g_1\mathbb{F}_1\sinh\phi_2+2m\mathbb{F}_2e^{-4 \sigma}-\frac{g_1\widehat{A}_1\phi'_2}{\sqrt{h_1h_2}},\label{ABCBPS5}\\
0&=&4\mathbb{F}_2\sigma'-\mathbb{F}_1\phi'_2\label{ABCBPS6}
\end{eqnarray}
and
\begin{eqnarray}
0&=&\frac{\phi'_2}{h_1}\left(\frac{f'}{f}+2\sigma'\right)+8mg_1e^{-2 \sigma} \sinh\phi_2-\frac{2 k e^{-\sigma}}{\sqrt{f}}\mathbb{F}_2,\label{SBCBPS4}\\
0&=&\frac{\phi'_2}{h_1}\left(\frac{h'_2}{h_2}+2\sigma'\right)+8mg_1e^{-2\sigma}\sinh\phi_2+2e^{-2\sigma}\mathbb{F}_1\mathbb{F}_2,\label{SBCBPS5}\\
0&=&\phi'_2\left(g_1e^{4 \sigma} \cosh\phi_2-3m\right)-4g_1 e^{4 \sigma} \sigma '\sinh\phi_2.\label{SBCBPS6}
\end{eqnarray}

\subsection{Supersymmetric $AdS_4\times \Sigma$ solutions} 
We now solve all these conditions in order to find supersymmetric $AdS_4\times \Sigma$ solutions. With $\mathbb{F}_1$ and $\mathbb{F}_2$ given in \eqref{SO(2)xSO(2)_Ansatz1} and \eqref{SO(2)xSO(2)_Ansatz2} together with the explicit forms of $A'_1$ and $A'_2$ given in \eqref{Gen_SO(2)xSO(2)_Ansatz1} and \eqref{Gen_SO(2)xSO(2)_Ansatz2}, we can solve equation \eqref{AABPS6} and find the solution for $\phi_2$ of the form
\begin{equation}
\phi_2=\ln \left[\frac{(a_1-a_2) g_1e^{4 \sigma}+\sqrt{(a_1-a_2)^2 g_1^2 e^{8 \sigma}+36 a_1 a_2 m^2}}{6 a_1 m}\right].\label{phi2_gen_sol}
\end{equation}
Similarly, equations \eqref{ABCBPS2} and \eqref{DABPS3} can be solved for $f$ and $h_1$ as functions of $\sigma$ and $\phi_2$ as follows
\begin{eqnarray}
f&=&\frac{m^{\frac{2}{3}}e^{\frac{2}{3}(\sigma-\phi_2)}\left(a_1 e^{2\phi_2}+a_2\right)^{\frac{2}{3}}}{\left(4k\left(g_1e^{4 \sigma} \cosh\phi_2-3 m\right)\right)^{\frac{2}{3}}},\label{f_gen_sol}\\
h_1&=&\frac{16m^{\frac{8}{3}}\left(a_1 e^{2\phi_2}+a_2\right)^{\frac{2}{3}}\left(g_1e^{4 \sigma}\cosh\phi_2-3 m\right)^{-2}e^{6\sigma}(\sigma')^2}{4m^{\frac{8}{3}}\left(a_1 e^{2\phi_2}+a_2\right)^{\frac{2}{3}}-e^{\frac{2}{3} (8 \sigma+\phi_2)} \left(4k\left(g_1e^{4 \sigma} \cosh\phi_2-3 m\right)\right)^{\frac{2}{3}}}\, .\qquad\quad\ \label{h1_gen_sol}
\end{eqnarray}
\indent We now determine the function $\widehat{A}_1$ defined in \eqref{AandA}. From \eqref{ABCBPS3}, we find the following expression for $\sqrt{h_1h_2}$ of the form
\begin{equation}
\sqrt{h_1h_2}=\frac{8 g_1 f^2 \widehat{A}_1 e^{2 \sigma+\phi_2} \sigma'}{\left(a_1 e^{2\phi_2}+a_2\right) \left(g_1 e^{4 \sigma} \cosh \phi_2-2m\right)}.\label{sqrth1h1_1}
\end{equation}
With $A'_1=\widehat{A}'_1$, we also find another expression for $\sqrt{h_1h_2}$ from \eqref{Gen_SO(2)xSO(2)_Ansatz1}
\begin{equation}
\sqrt{h_1h_2}=\frac{2 f^2 \widehat{A}'_1 e^{2\phi_2-2 \sigma}}{a_1 e^{4 \phi_2}+a_2}.\label{sqrth1h1_2}
\end{equation}
These two equations together with all the previous results lead to an ordinary differential equation for $\widehat{A}_1$ of the form
\begin{equation}
\widehat{A}_1'=\frac{72e^{4 \sigma} a_1g_1 m^2 \left(a_1 \mathbb{K}^4+a_2\right) \left[(a_1-a_2)g_1e^{4 \sigma} \mathbb{K} +6 a_2 m\right]^{-1} \sigma'\widehat{A}_1}{(a_1-a_2) g_1^2 e^{8 \sigma}\mathbb{K} +3 (a_1+a_2) g_1 m e^{4 \sigma}-12 a_1 m^2 \mathbb{K}}
\end{equation}
where
\begin{equation}
\mathbb{K}=e^{\phi_2}=\frac{(a_1-a_2) g_1e^{4 \sigma}+\sqrt{(a_1-a_2)^2 g_1^2 e^{8 \sigma}+36 a_1 a_2 m^2}}{6 a_1 m}\, .
\end{equation}
Solving this equation leads to the following solution 
\begin{eqnarray}
\widehat{A}_1&=&\frac{\mathbb{C}}{6} e^{8 \sigma}\sqrt{\frac{4 w^4-(5 a_1+a_2)(a_1+5 a_2)g_1^2w^2+g_1^4(\Delta a^2)^2}{a_1 a_2}}\nonumber\\
&&\times\left\{\frac{\left[u^2 \left(w^2-g_1^2\Delta a^2\right)+8 a_1 a_2 \left(g_1^2\Delta a^2 -4 w^2\right)\right]/(6 a_1 a_2 y)}{2 \sqrt{2}g_1 w u \Delta a^2+g_1^2 \Delta a^2 \left(u^2-8 a_1 a_2\right)+w^2(u^2-32 a_1 a_2)}\right\}^{\frac{\left(y+a_1+a_2\right)}{2 \sqrt{2} u}}\nonumber\\
&&\times\left\{\frac{\left[v^2 \left(w^2-g_1^2\Delta a^2\right)+8 a_1 a_2 \left(g_1^2\Delta a^2-4 w^2\right)\right]/(48 a_1 a_2 )}{2 \sqrt{2} g_1 w v\Delta a^2-g_1^2\Delta a^2(v^2-8 a_1 a_2)-w^2(v^2-32a_1 a_2)}\right\}^{\frac{\left(y-a_1-a_2\right)}{2 \sqrt{2} v}}\nonumber\\
&&\times\left\{\frac{e^{8 \sigma} g_1 v \Delta a^2\left(y-a_1-a_2\right)  \left(w^2-\Delta a^2 g_1^2\right)}{48 \sqrt{2} a_1 a_2}\right\}^{\frac{\left(y-a_1-a_2\right)}{2 \sqrt{2} v}}\label{Gen_A1_soln}
\end{eqnarray}
with an integration constant $\mathbb{C}$. In this equation, we have defined the following quantities 
\begin{equation}
\Delta a^2=(a_1-a_2)^2 
\end{equation}
and
\begin{eqnarray}
u&=&\sqrt{a_1^2+18 a_1 a_2+a_2^2+(c_1+c_2) y},\\
v&=&\sqrt{a_1^2+18 a_1 a_2+a_2^2-(c_1+c_2) y},\\
w&=&\sqrt{(a_1-a_2)^2 g_1^2 e^{8 \sigma}+36 a_1 a_2 m^2},\\
y&=&\sqrt{a_1^2+34 a_1 a_2+a_2^2}\, .
\end{eqnarray}
\indent Substituting these solutions into \eqref{sqrth1h1_1}, we can also find the solution for $h_2$ in terms of the other ones. However, the explicit form of this solution is far more complicated, so we refrain from giving it here. Moreover, finding the solution for $A_2$ is much more difficult since all the previously obtained results lead to a highly complicated differential equation for $A_2'$. The situation is similar to that of \cite{7D_N2_disk} in which the $AdS_5\times \Sigma$ solution with $SO(2)\times SO(2)$ symmetry has been studied in $N=2$ seven-dimensional gauged supergravity. In any case, it should also be noted that $A_2$ does not appear in any BPS equations in which only $A_1'$ and $A_2'$ appear. In particular, we have verified that all the BPS conditions are satisfied by the above solution without the explicit solution for $A_2$. 
\\
\indent To make subsequent analysis of the solutions more tractable, we will further simplify the solution by setting 
\begin{equation}
a_1=-a_2\, .
\end{equation}
It should be noted that a more obvious choice with $a_1=a_2$ leads to a trivial solution for equation \eqref{AABPS6} namely $\phi_2=0$. With $\phi_2=0$ and $a_1=a_2$, we find that $A_2'=0$, see \eqref{Gen_SO(2)xSO(2)_Ansatz2}, giving rise to constant $A_2$ that can be set to zero. Therefore, all the fields from vector multiplets are truncated out recovering the BPS equations studied in \cite{Suh_D4} for pure $F(4)$ gauged supergravity up to notational and conventional differences.
\\
\indent In the case of $a_1=-a_2=b$, we find non-vanishing components of the dressed field strengths given by
\begin{equation}
\mathbb{F}_1=\frac{b e^{2 \sigma}}{2 f^2}\sinh\phi_2\qquad\text{ and }\qquad \mathbb{F}_2=\frac{b e^{2 \sigma}}{2 f^2}\cosh\phi_2\, .
\end{equation}
The previously obtained solution now takes a much simpler form
\begin{eqnarray}
\phi_2&=&\ln \left[\frac{g_1 e^{4 \sigma}+\sqrt{g_1^2 e^{8 \sigma}-9 m^2}}{3 m}\right],\label{SO(3)xSO(3)_SO(2)xSO(2)_ph2_Soln}\\
f&=&\frac{b^{\frac{2}{3}} m^{\frac{2}{3}}e^{\frac{2 \sigma}{3}}}{k^{\frac{2}{3}}2^{\frac{2}{3}} \left(g_1^2 e^{8 \sigma}-9 m^2\right)^{\frac{1}{3}}},\label{SO(3)xSO(3)_SO(2)xSO(2)_f_Soln}\\
h_1&=&\frac{9\times 2^{\frac{10}{3}} m^2 e^{6 \sigma} \left(\frac{m^4b}{k}\right)^{\frac{2}{3}} (\sigma')^2}{\left(g_1^2 e^{8 \sigma}-9 m^2\right)^2 \left[2^{\frac{4}{3}} \left(\frac{m^4 b}{k}\right)^{\frac{2}{3}}-e^{\frac{16 \sigma}{3}} \left(g_1^2 e^{8 \sigma}-9 m^2\right)^{\frac{1}{3}}\right]},\qquad\\
h_2&=&\frac{9C^2 g_1^2 m^2 e^{\frac{2 \sigma}{3}}\left[2^{\frac{4}{3}} \left(\frac{m^4 b}{k}\right)^{\frac{2}{3}}-e^{\frac{16 \sigma}{3}} \left(g_1^2 e^{8 \sigma}-9 m^2\right)^{\frac{1}{3}}\right]}{4\left(g_1^2 e^{8 \sigma}-9 m^2\right)^{\frac{1}{3}}},\\
\widehat{A}_1&=&C \left(g_1^2 e^{8 \sigma}-6 m^2\right)
\end{eqnarray}
where $C$ is an integration constant. It can also be verified that all of the BPS equations and the field equations are satisfied for 
\begin{equation}\label{alter_SO(2)xSO(2)_sign_Con}
g_1^2 e^{8 \sigma}-9 m^2>0\, .
\end{equation}
\indent In addition, equation \eqref{Gen_SO(2)xSO(2)_Ansatz2} leads to the following differential equation
\begin{equation}
A'_2=\frac{4 g_1 C e^{4 \sigma} \left(2 g_1^2 e^{8 \sigma}-9 m^2\right)\sigma'}{\sqrt{g_1^2 e^{8 \sigma}-9 m^2}}
\end{equation}
which can be readily solved by
\begin{equation}
A_2=g_1C e^{4 \sigma} \sqrt{g_1^2 e^{8 \sigma}-9 m^2}+c_2
\end{equation}
for another integration constant $c_2$. We also note an explicit form of $A_1$ which reads 
\begin{equation}
A_1=C \left(g_1^2 e^{8 \sigma}-6 m^2\right)-\frac{2q}{g_1}\, .
\end{equation}
\indent Finally, the BPS equations \eqref{key_BPS_1}, \eqref{key_BPS_2}, \eqref{key_BPS_3}, and \eqref{key_BPS_4} are satisfied by the following form of the two-component spinor 
\begin{equation}
\eta^2_+=e^{iq\theta}Y\begin{pmatrix} \sqrt{2^{\frac{2}{3}} \left(\frac{m^4b}{k}\right)^{\frac{1}{3}}-k e^{\frac{8 \sigma}{3}} \left(g_1^2 e^{8 \sigma}-9 m^2\right)^{\frac{1}{6}}} \\ \sqrt{2^{\frac{2}{3}} \left(\frac{m^4b}{k}\right)^{\frac{1}{3}}+k e^{\frac{8 \sigma}{3}} \left(g_1^2 e^{8 \sigma}-9 m^2\right)^{\frac{1}{6}}}\end{pmatrix}
\end{equation}
with the function $Y$ being the solution of an ordinary differential equation given in \eqref{key_r_con}. The explicit form of the solution for $Y$ can be written as
\begin{equation}
Y=\frac{\zeta e^{\frac{\sigma}{6}}}{(g_1^2 e^{8 \sigma}-9 m^2)^{\frac{1}{12}}}
\end{equation}
in which $\zeta$ is an integration constant. 
\\
\indent It should be noted that the dilaton $\sigma(r)$ is not determined by the BPS equations as in \cite{Bah_M5, Suh_D3, Suh_D4, Suh_M2,7D_N2_disk} due to general covariance. We will use this freedom to fix the form of the solution for $\sigma$ to be
\begin{equation}\label{sigma_soln}
\sigma=\frac{1}{8}\ln r\, .
\end{equation}
 
\subsection{Regular solutions}
\begin{figure}[h!]
  \centering
  \begin{subfigure}[b]{0.326\linewidth}
    \includegraphics[width=\linewidth]{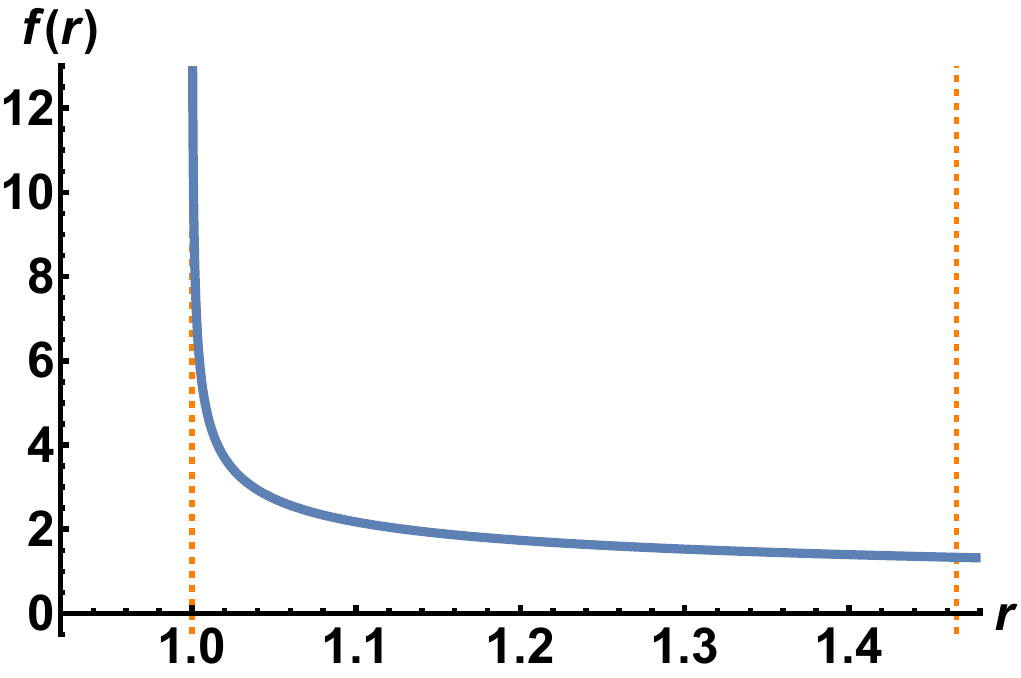}
\caption{$f$ solution}
  \end{subfigure}
  \begin{subfigure}[b]{0.326\linewidth}
    \includegraphics[width=\linewidth]{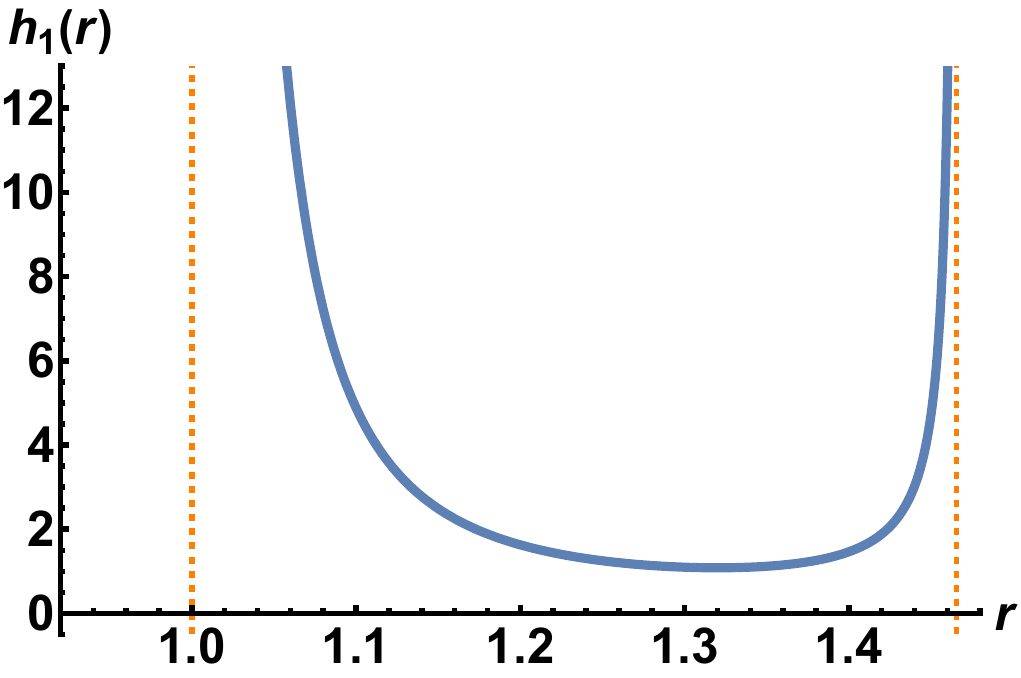}
\caption{$h_1$ solution}
  \end{subfigure}
  \begin{subfigure}[b]{0.326\linewidth}
    \includegraphics[width=\linewidth]{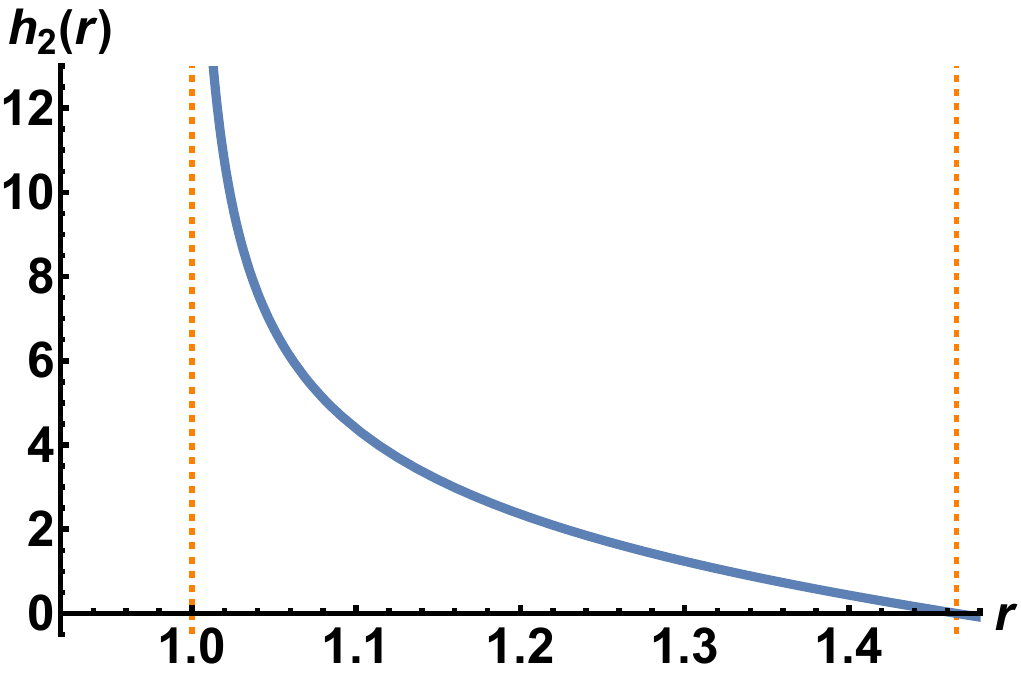}
\caption{$h_2$ solution}
  \end{subfigure}
  \caption{Numerical plots of the warp factors for $SO(2)\times SO(2)$ symmetric solution with $B=G=\mathcal{C}=1$ and $m=\frac{1}{2}$. The solution is regular in the range $\frac{1}{G^2}=1<r<r_1=1.46$ with the two vertical dashed lines representing the two boundaries.}
  \label{Fig1}
\end{figure}

We now consider the range of $r$ for obtaining regular solutions with all the metric functions positive definite and all scalar fields real. For convenience, we will work with the simplified solution obtained in the previous section. By introducing the following parameters
\begin{equation}\label{SO(2)xSO(2)_para_redefine}
B=\frac{2^{\frac{4}{3}} b^{\frac{2}{3}} m^2}{3^{\frac{2}{3}} k^{\frac{2}{3}}},\qquad G=\frac{g_1}{3m},\qquad \mathcal{C}=\frac{3}{2}g_1 m^2C, 
\end{equation}
we can rewrite the six-dimensional metric as
\begin{eqnarray}
ds^2_6&=&\frac{Br^{\frac{1}{12}}}{4m^2(G^2r-1)^{\frac{1}{3}}}\left[ds^2_{AdS_4}+ds^2_{\Sigma}\right]\label{SO(2)xSO(2)_6D_Met}\\
\textrm{with}\qquad ds^2_{\Sigma}&=&\frac{r^{-\frac{4}{3}}}{36W(G^2r-1)^{\frac{5}{3}}}dr^2+\frac{4\mathcal{C}^2W}{B}d\theta^2\, .\label{SO(2)xSO(2)_Sigma_Met}
\end{eqnarray}
The function $W(r)$ is defined by
\begin{equation}\label{SO(2)xSO(2)_Def_W}
W=B-r^{\frac{2}{3}}(G^2r-1)^{\frac{1}{3}}\, .
\end{equation}
\\
\indent The solutions for the two scalar fields $\sigma$ and $\phi_2$ are given by
\begin{equation}
\sigma= \frac{1}{8}\ln r\qquad\text{ and }\qquad \phi_2=\ln\left(\sqrt{G^2 r-1}+G \sqrt{r}\right)\label{scalar_sol11}
\end{equation}
while the non-vanishing components of the two $SO(2)\times SO(2)$ gauge fields take the form
\begin{equation}
A^{3}_\theta=\frac{2}{3 G m}\left[\mathcal{C}(3 G^2 r-2)-q\right]\qquad\text{ and }\qquad  A^{6}_\theta=\frac{2 \mathcal{C}}{m}\sqrt{r(G^2 r-1)}+c\, .
\end{equation}
It is also convenient to note the corresponding gauge field strength tensors  
\begin{equation}
\widehat{F}^\Lambda_{r\theta}=F^\Lambda_{r\theta}=\frac{G\mathcal{C}}{m}\delta^\Lambda_3+\frac{\mathcal{C}(2G^2r-1)}{2m\sqrt{r(G^2 r-1)}}\delta^\Lambda_6\, .
\end{equation}
At this point, we observe that the angular coordinate $\theta$ always enters the six-dimensional metric and the field strength $F^\Lambda_{r\theta}$ in the combination $\mathcal{C} d\theta$. 
\\
\indent We find that regular solutions are possible only in the following range
\begin{equation}\label{SO(3)xSO(3)_SO(2)xSO(2)_Pos_Range}
\frac{1}{G^2}<r<r_1
\end{equation}
for $B>0$, $G>0$, and $m>0$ where $r_1$ is the real root of $W(r_1)=0$ given by 
\begin{eqnarray}
r_1&=&\frac{1}{3G^2}(1+X+X^{-1})\label{r_Roots}
\end{eqnarray}
for
\begin{equation}
X=\left[x+\sqrt{x (x+2)}+1\right]^{\frac{1}{3}}\label{X_SO(2)xSO(2)_def}
\end{equation}
with
\begin{equation} 
x=\frac{27}{2}B^3G^4\, .\label{x_SO(2)xSO(2)_def}
\end{equation}
We note that there are also two complex roots of the equation $W(r)=0$. In terms of $X$ defined in \eqref{X_SO(2)xSO(2)_def}, these are given by 
\begin{equation}
r_{c\pm}=\frac{(X-1)}{6 G^2 X}\left[(1-X)\pm i \sqrt{3} (X+1)\right].
\end{equation} 
Since $X$ is a non-negative real number, both of these roots, corresponding to the upper and lower signs, are complex and are not relevant to the present analysis.
\\
\indent For an illustrative purpose, we plot representative solutions of $f(r)$, $h_1(r)$, and $h_2(r)$ with $B=G=\mathcal{C}=1$ and $m=\frac{1}{2}$ in figure \ref{Fig1}. As can be seen from the figure, all the warp factors in the metric \eqref{6Dmetrix1} are positive in the range $\frac{1}{G^2}<r<r_1$. 

\begin{figure}[h!]
  \centering
    \includegraphics[width=0.6\linewidth]{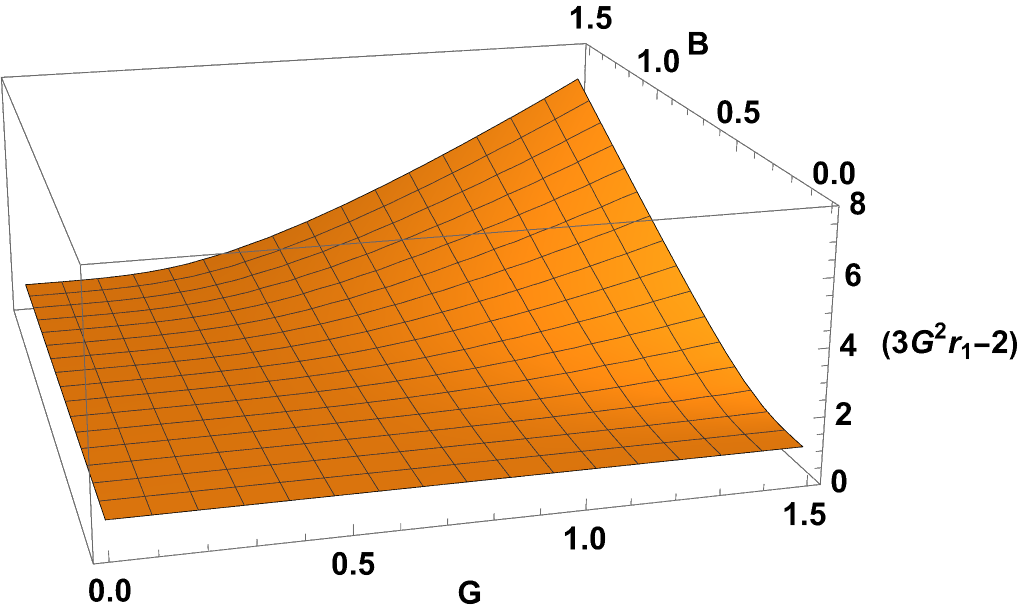}
  \caption{A numerical plot of the function $\left(3 G^2 r_1-2\right)$ in condition \eqref{SO(3)xSO(3)_SO(2)xSO(2)_smooth_con}. Note that $\left(3 G^2 r_1-2\right)\geq 1$ for all $B>0$ and $G>0$.}
  \label{Fig2}
\end{figure}

This result is very similar to the $SO(2)\times SO(2)$ symmetric solution found in seven-dimensional $N=2$ gauged supergravity in \cite{7D_N2_disk}. As $r\rightarrow r_1$, the six-dimensional metric is approximately given by
\begin{equation}\label{SO(4)_SO(2)xSO(2)_spindle_metric}
ds_6^2\approx\frac{B r_1^{\frac{1}{12}}}{4 m^2 (G^2 r_1-1)^{\frac{1}{3}}}\left[ds^2_{AdS_4}+\frac{d\rho^2+4 \mathcal{C}^2(3 G^2 r_1-2)^2 \rho ^2d\theta^2}{-9W'(r_1) r_1^{\frac{4}{3}} (G^2 r_1-1)^{\frac{5}{3}}}\right]
\end{equation}
in which $\rho$ is the new radial coordinate defined by $\rho=\sqrt{r_1-r}$. The $\theta$-circle shrinks smoothly giving rise to an $\mathbb{R}^2/\mathbb{Z}_l$ orbifold at $r=r_1$ if we impose
\begin{equation}\label{SO(3)xSO(3)_SO(2)xSO(2)_smooth_con}
\mathcal{C}=\frac{1}{2 l \left(3 G^2 r_1-2\right)},\qquad l=1,2,3,\ldots\, .
\end{equation}
In this case, the function $(3G^2 r_1-2)$ depends on the two constants $G$ and $B$. However, its explicit form obtained from equation \eqref{r_Roots} is rather complicated, so we will only show that $(3G^2 r_1-2)$ is strictly positive in the range of the $r$ coordinate under consideration by giving a numerical plot of $(3G^2 r_1-2)$ in figure \ref{Fig2}. 

On the other hand, as $r\rightarrow \frac{1}{G^2}$, the six-dimensional metric becomes conformal to a product of $AdS_4$ and a cylinder. With the new radial coordinate $R$ given by $r=\frac{1}{G^2}+64\mathcal{C}^6G^2B^3R^6$, the metric near $R=0$ is given by  
\begin{equation}\label{SO(3)xSO(3)_SO(2)xSO(2)_AdS4xdisk_metric}
ds_6^2\approx\frac{1}{16 \mathcal{C}^2 G^{\frac{3}{2}} m^2 R^2}\left[ds^2_{AdS_4}+4\mathcal{C}^2(dR^2+d\theta^2)\right].
\end{equation}
It should be noted that in this limit the $S^1$ parametrized by $\theta$ decompactifies as also pointed out in \cite{Bah_M5}. In addition, by changing to another coordinate given by 
\begin{equation}
\rho=\frac{R^2}{4\mc{C}^2},
\end{equation}  
we find, up to an overall constant and some rescaling in $ds^2_{AdS_4}$,
\begin{equation}
ds^2_6\sim \frac{1}{\rho}\left(ds^2_{AdS_4}+d\theta^2\right)+\frac{d\rho^2}{4\rho^2}\, .
\end{equation}
This is an $AdS_6$ space in Fefferman-Graham coordinate with the holographic boundary given by $AdS_4\times S^1$. This is similar to the seven-dimensional solution studied in \cite{Gutperle_coD2_defect} that leads to a holographic dual of a codimension-2 defect within $N=(1,0)$ SCFT in six dimensions. Therefore, the present solution could also be interpreted holographically as a codimension-2 defect within the five-dimensional $N=2$ SCFT dual to the $AdS_6$ geometry with $SO(3)\times SO(3)$ symmetry. We also note that in the limit $r\rightarrow \frac{1}{G^2}=\frac{9m^2}{g_1^2}$, the scalar fields become
\begin{equation}
\sigma=\frac{1}{4}\ln\left[\frac{3m}{g_1}\right]\qquad \textrm{and}\qquad \phi_2=0
\end{equation}
which are precisely the values at the $AdS_6$ vacuum.
\\
\indent Topologically, the space $\Sigma$ with metric \eqref{SO(2)xSO(2)_Sigma_Met} forms a disk with an origin at $r = r_1$ and a boundary at $r = \frac{1}{G^2}$. On this boundary, the $\theta$-circle does not collapse while there is a $\mathbb{Z}_l$ orbifold singularity at the center of the disk. By using the Gauss-Bonnet theorem and the fact that the boundary is a geodesic, we can calculate the Euler characteristic of $\Sigma$ from the metric $ds^2_{\Sigma}$ given in \eqref{SO(2)xSO(2)_Sigma_Met}
\begin{eqnarray}
\chi(\Sigma)&=&\frac{1}{4\pi}\int_{\Sigma}R_{\Sigma}\text{vol}_{\Sigma}\ =\ \frac{2 \mathcal{C}(G^2r_1-1)^{\frac{1}{6}}r_1^{\frac{1}{3}} (3G^2 r_1-2)}{\sqrt{B}}\nonumber\\&=&2 \mathcal{C} \left(3 G^2 r_1-2\right)\ =\ \frac{1}{l}\label{SO(2)xSO(2)_chi}
\end{eqnarray}
which corresponds to a disk with an $\mathbb{R}^2/\mathbb{Z}_l$ orbifold singularity at $r= r_1$. Note that the integration has been performed on the interval $\frac{1}{G^2}<r<r_1$ and $0<\theta<2\pi$, and we have used $B=r_1^{\frac{2}{3}} (G^2r_1-1)^{\frac{1}{3}}$ as obtained from $W(r_1)=0$.
\\
\indent In order for the $SO(2)\times SO(2)$ gauge fields to vanish at $r=r_1$, we fix the constants $q$ and $c_2$ to be
\begin{eqnarray}
q=\mathcal{C} \left(3 G^2 r_1-2\right)\qquad \textrm{and}\qquad c_2=-\frac{2 \mathcal{C}}{m}\sqrt{r_1(G^2r_1-1)}
\end{eqnarray}
leading to
\begin{eqnarray}
& &A^{3}_\theta=\frac{2 \mathcal{C}G}{m}(r-r_1)\nonumber \\
\textrm{and}\qquad & & A^{6}_\theta=\frac{2 \mathcal{C}}{m}\left(\sqrt{r(G^2 r-1)}-\sqrt{r_1(G^2 r_1-1)}\right).
\end{eqnarray}
We also note that $A^3_\theta$ is well-defined for all values of $r$ while $A^6_\theta$ is complex for $r<\frac{1}{G^2}$. 
\\
\indent Finally, we give the explicit form of the Killing spinors corresponding to the unbroken supersymmetry. For the coordinate $r$ in the range $\frac{1}{G^2}<r<r_1$, the two-component Killing spinor $\eta^2_+$ takes the form
\begin{equation}
\eta^2_+(r,\theta)=\frac{\zeta\, e^{iq\theta}r^{\frac{1}{48}}}{(G^2 r-1)^{\frac{1}{12}}}\begin{pmatrix} \sqrt{\sqrt{B}- k\left[r^{2} \left(G^2 r-1\right)\right]^{\frac{1}{6}}} \\ \sqrt{\sqrt{B}+k\left[r^{2} \left(G^2 r-1\right)\right]^{\frac{1}{6}}}\end{pmatrix}
\end{equation}
where $\zeta$ is a complex constant. We can now determine the number of supersymmetries preserved by the $AdS_4\times\Sigma$ solution. Because $\vartheta$ is a Majorana spinor, it carries four real degrees of freedom, whereas the spinors $\eta^2_+$ are completely fixed by the complex constant $\zeta$. This yields a total of eight real independent supercharges, so the solution is $\frac{1}{2}$-BPS. Furthermore, the two-component Killing spinor $\eta^2_+$ never vanishes as can be seen from its norm
\begin{equation}\label{SO(2)xSO(2)_eta_norm}
\left(\eta^2_+\right)^\dagger\eta^2_+=\frac{2  \left|\zeta\right|^2 r^{\frac{1}{24}}\sqrt{B}}{\left(G^2 r-1\right)^{\frac{1}{6}}}.
\end{equation}
In the limit $r\rightarrow r_1$, the Killing spinor exhibits a definite chirality for $k=\pm1$, given by
\begin{equation}
\eta^2_+(r_1,\theta)=\zeta\, e^{iq\theta}r_1^{3/16}\begin{pmatrix} \sqrt{1- k} \\ \sqrt{1+k}\end{pmatrix}.
\end{equation}
As noted previously, the condition $B=\left[r_1^{2} \left(G^2 r_1-1\right)\right]^{\frac{1}{3}}$ as $r\rightarrow r_1$ ensures that the norm of $\eta^2_+$ still satisfies \eqref{SO(2)xSO(2)_eta_norm}. It is evident that $\eta^2_+$ remains regular near $r=r_1$ which confirms that $\eta^2_+$ is a globally defined spinor on the disk $\Sigma$.

\subsection{Uplift to massive type IIA theory}
As previously mentioned, the resulting solution lies within the $U(1)\times U(1)$ sector of the matter-coupled $F(4)$ gauged supergravity which can be embedded in massive type IIA theory by a consistent truncation on a half four-sphere. The corresponding truncation ansatz is reviewed in appendix \ref{uplift}. Using the formulae in appendix \ref{uplift}, we find the ten-dimensional metric in string frame of the form
\begin{eqnarray}
ds^2_{10}&=&\lambda^2\cos^{-\frac{1}{3}}\xi e^{-\frac{\sigma}{2}}\Delta^{\frac{1}{2}}\left\{\frac{Br^{\frac{1}{12}}}{4m^2(r-1)^{\frac{1}{3}}}ds^2_{AdS_4}+\frac{Br^{-\frac{5}{4}}}{36W(r-1)^{\frac{5}{3}}}dr^2+\frac{\mc{C}Wr^{\frac{1}{12}}}{m^2(r-1)^{\frac{1}{3}}} \right.\nonumber \\
& &+\frac{\Delta^{-1}e^{-\sigma}}{9m^2}\left[\sin^2\xi(e^{-\phi_2}\cos^2\eta+e^{\phi_2}\sin^2\eta)d\eta^2-\sin2\xi \sin2\eta\sinh\phi_2 d\xi d\eta \right. \nonumber \\
& &+ [e^{4\sigma}\sin^2\xi+\cos^2\xi (e^{-\phi_2}\sin^2\eta+e^{\phi_2}\cos^2\eta)]d\xi^2\nonumber \\
& &\left. \phantom{\frac{1^{\frac{1}{1}}}{1^{\frac{1}{1}}}}\left. +\sin^2\xi [\cos^2\eta e^{\phi_2}(d\varphi_2-g_1\mc{A}_2)^2+\sin^2\eta e^{-\phi_2}(d\varphi_1-g_1 \mc{A}_1)^2] \right] \right\}\quad
\end{eqnarray}
in which for convenience, we have set $g_1=3m$ leading to $G=1$. The gauge fields $\mc{A}_1$ and $\mc{A}_2$ are given by
\begin{eqnarray}
\mc{A}_1&=&\frac{\mc{C}}{m}\left\{r-r_1+\sqrt{r(r-1)}-\sqrt{r_1(r_1-1)}\right\}d\theta\nonumber \\
\textrm{and}\qquad \mc{A}_2&=&\frac{\mc{C}}{m}\left\{r-r_1-\sqrt{r(r-1)}+\sqrt{r_1(r_1-1)}\right\}d\theta\, .\label{A12_11}
\end{eqnarray}
The ten-dimensional dilaton takes the form
\begin{equation}
e^{\Phi}=\lambda^2\cos^{-\frac{5}{6}}\xi \Delta^{\frac{1}{4}}e^{-\frac{5\sigma}{4}}\, .\label{10D_dilaton1}
\end{equation}
\indent We can also use the formulae given in appendix \ref{uplift} to compute the holographic free energy of the dual SCFT in three dimensions. However, the solution leads to an infinite free energy due to the decompactified $\theta$-circle at $r=1$. The infinity arises near the $AdS_6$ asymptotic geometry as pointed out in \cite{D4_4D_Orbifolds}. Following \cite{Gutperle_coD2_defect}, the solution could be holographically interpreted as the presence of a three-dimensional conformal defect within the $N=2$ SCFT in five dimensions. It would be interesting to further study the solution along this line and perform some holographic computation as in \cite{Gutperle_coD2_defect} and \cite{Conti}.  
\\
\indent To complete the analysis of the uplifted ten-dimensional solution, we consider flux quantization of the four-form field strength along the internal directions. In the present case, the uplifted four-form field is given by
\begin{eqnarray}
F_{(4)}&=&\frac{\lambda}{2g_1^3\Delta^2}\cos^{\frac{1}{3}}\xi \sin^3\xi\sin2\eta \left\{Ud\eta\wedge d\xi\wedge D\varphi_1\wedge D\varphi_2\right.\nonumber \\
& &\phi_2'e^{-6\sigma}\left[e^{-2\phi_2}\sin2\xi\left(e^{4\phi_2}\sin^2\eta-\cos^2\eta\right)dr\wedge d\eta \right.\nonumber \\
& &\left.\left.-\sin2\eta\left(2\cos^2\xi\cosh2\phi_2+e^{8\sigma}\sin^2\xi\right)dr\wedge d\xi\right]\right\}\wedge D\varphi_1\wedge D\varphi_2\nonumber \\
& &+\frac{\lambda e^{-3\sigma}}{2g_1^2\Delta}\cos^{\frac{1}{3}}\xi\sin\xi \left\{\sin2\eta\sin\xi\cos\xi(\mc{F}_1\wedge d\varphi_2-\mc{F}_2\wedge d\varphi_1)\wedge d\eta\right.\nonumber\\
& &+\cos^2\eta\left(\cos^2\xi+e^{4\sigma-\phi_2}\sin^2\xi\right)\mc{F}_1\wedge d\xi\wedge d\varphi_1\nonumber \\
& &\left.+2\sin^2\eta\left(\cos^2\xi+e^{4\sigma+\phi_2}\sin^2\xi\right)\mc{F}_2\wedge d\xi\wedge d\varphi_2 \right\}
\end{eqnarray}
with $D\varphi_i=d\varphi_i-g_1\mc{A}_i$, $i=1,2$. We also recall that a real parameter $\lambda$ is introduced in order to properly quantize all the fluxes along the internal space, see more detail in \cite{D4_spindle}. The function $U$ is given by
\begin{eqnarray}
U&=&\frac{2}{3}e^{-6\sigma}\cos^2\xi-\frac{2}{3}e^{-2\sigma}\left[2\cosh\phi_2(2+\cos2\xi)\right.\nonumber \\
& &\left. +\sin^2\xi(3e^{4\sigma}-2\cos2\eta\sinh\phi_2)\right].
\end{eqnarray}
The three-form potential reads
\begin{eqnarray}
C_{(3)}&=&\frac{\lambda}{2g_1^3}\cos^{-\frac{2}{3}}\left\{\left(\frac{1}{2}\sin2\xi\cos2\eta-\frac{e^\sigma}{\Delta}\cos\xi\sin^3\xi\sin^22\eta\sinh\phi_2\right)d\xi\right. \nonumber \\
& &\left.-\left[\sin^2\xi+\frac{e^\sigma}{\Delta}\sin^4\xi(\cos2\eta\sinh\phi_2-\cosh\phi_2)\right]\sin2\eta d\eta\right\}\wedge\nonumber \\
& &(d\varphi_1\wedge d\varphi_2-g_1\mc{A}_{1\theta}d\theta\wedge d\varphi_2-g_1\mc{A}_{2\theta}d\varphi_1\wedge \theta)\nonumber \\
& &-\frac{3\lambda}{8g_1^2}\cos^{\frac{4}{3}}\xi \left(\mc{A}'_{1\theta}dr\wedge d\theta\wedge d\varphi_2+\mc{A}'_{2\theta}dr\wedge d\theta\wedge d\varphi_1\right).
\end{eqnarray}
We also recall that the solutions for $\sigma$ and $\phi_2$ are give in \eqref{scalar_sol11}. The remaining ten-dimensional field is given by the Romans mass
\begin{equation}
F_{(0)}=\frac{m}{\lambda^3}=\frac{g_1}{3\lambda^3}\, .
\end{equation}
\indent To determine the parameter $\lambda$, we follow \cite{D4_spindle} and consider the flux quantization of $F_{(0)}$ and $F_{(4)}$ of the form
\begin{eqnarray}
2\pi \ell_s F_{(0)}=n_0=8-N_f\qquad \textrm{and}\qquad \frac{1}{(2\pi \ell_s)^3}\int F_{(4)}=N
\end{eqnarray}
for intergers $n_0$, $N$ and $N_f$. We also recall that $N$ and $N_f$ correspond to the numbers of D4-branes and D8-branes in the configuration, respectively. The first condition immediately gives
\begin{equation}
n_0=\frac{2\pi \ell_sg_1}{3\lambda^3}\, .\label{quantize1}
\end{equation}
By considering the integral along the half $S^4$ directions, we find 
\begin{eqnarray}
\frac{1}{(2\pi \ell_s)^3}\int F_{\eta\xi \varphi_1\varphi_2}d\eta\wedge d\xi \wedge d\varphi_1\wedge d\varphi_2&=& \frac{\lambda(2\pi)^2}{(2\pi \ell_s)^3}\int_0^{\frac{\pi}{2}}\int_0^{\frac{\pi}{2}}\frac{U\cos^{\frac{1}{3}}\xi \sin^3\xi \sin2\eta}{2g_1^3\Delta^2}d\eta d\xi\nonumber \\
&=&-\frac{3\lambda}{8\pi g_1^3\ell_s^3}\label{S4_flux}
\end{eqnarray}
which gives
\begin{equation}
N=\frac{3\lambda}{8\pi g_1^3\ell_s^3}\, .\label{quantize2}
\end{equation}  
For consistency, we have identified the flux of $F_{(4)}$ as minus the number of D4-branes $N$ as in \cite{D4_spindle}. We also note that the result in \eqref{S4_flux} is valid for any value of $r$.
\\
\indent The two conditions in \eqref{quantize1} and \eqref{quantize2} then give rise to 
\begin{equation}
\lambda^8=\frac{\pi^2}{9n_0N}\qquad \textrm{and}\qquad g_1^8=\frac{9\pi^6}{(2\pi \ell_s)^8N^3n_0}\, .\label{lambda_g1}
\end{equation}
To express the parameter $r_1$ in terms of integers corresponding to fluxes of $F_{(4)}$, we consider another four-cycle along the direction $(r\xi\theta\varphi_2)$ with $\eta=0$. In this case, we have
\begin{equation}
F_{r\xi\theta\varphi_2}=\frac{\pd C_{\xi\theta\varphi_2}}{\pd r}-\frac{\pd C_{r\theta\varphi_2}}{\pd \xi}\, .
\end{equation} 
Using $\mc{A}_{1\theta}$ and $\mc{A}_{2\theta}$ from \eqref{A12_11}, we find 
\begin{equation}
\int F_{r\xi\theta\varphi_2}dr\wedge d\xi \wedge d\theta\wedge d\varphi_2=\frac{3\lambda\mc{C}}{4g_1^2m}[1-r_1-\sqrt{r_1(r_1-1)}].
\end{equation}
Setting this equal to $-M$ for $M\in \mathbb{Z}$ and using \eqref{quantize2} as well as \eqref{SO(3)xSO(3)_SO(2)xSO(2)_smooth_con}, we eventually find
\begin{equation}
\frac{r_1-1+\sqrt{r_1(r_1-1)}}{3r_1-2}=\frac{8Ml}{3N}
\end{equation}
which gives
\begin{equation}
r_1=\frac{3N^2-80NMl+256M^2l^2\pm N\sqrt{9N^2+96NMl-512M^2l^2}}{96Ml(4Ml-N)}\, .
\end{equation}
\section{$AdS_4\times \Sigma$ solutions with $SO(2)_{\text{diag}}$ symmetry}\label{SO(2)d_SO(3)xSO(3)gg_Sec}
In this section, we repeat the same analysis within a truncation to $SO(2)_{\text{diag}}$ invariant sector. There are four singlet scalars from $SO(4,3)/(SO(4)\times SO(3))$ coset under $SO(2)_{\text{diag}}$ symmetry generated by $J^{12}+\tilde{J}^{12}$. These singlets can be described by the coset representative
\begin{equation}\label{SO(2)diag_singlet_in_SO(3)xSO(3)gg_main}
L=e^{\phi_0 Y_{03}}e^{\phi_1(Y_{11}+Y_{22})}e^{\phi_2 Y_{33}}e^{\phi_3(Y_{12}-Y_{21})}\, .
\end{equation}
We also note that, in general, this sector is larger than the $U(1)\times U(1)$ truncation considered previously. Therefore, solutions with non-vanishing $\phi_0$, $\phi_1$, and $\phi_3$ currently cannot be embedded in massive type IIA theory.  
\\
\indent To implement the $SO(2)_{\textrm{diag}}$ symmetry, we impose the following constraint on the two $SO(2)\times SO(2)$ gauge fields
\begin{equation}
 g_1A^3=-g_2A^6\, .
\end{equation} 
We now repeat the same procedure as in the previous section to find supersymmetric $AdS_4\times \Sigma$ solutions. In particular, we can still consistently set $B_{\mu\nu}=0$ as in the previous section.

\subsection{BPS equations}
In this case, the field equation for the two-form field \eqref{H_field_eqs} becomes
\begin{equation}
0=\frac{m e^{-2 \sigma}}{2 g_2} A'_1\left(g_1 \sinh 2\phi _0 \cosh ^2\phi_2+g_2 \sinh \phi_0 \sinh2\phi_2\right)
\end{equation}
which implies $\phi_0$ has to vanish. For the vector field equation \eqref{F_field_eqs}, the radial component ($\mu=r$) gives rise to the following non-trivial relation between the two scalar fields
\begin{equation}\label{phi'1_and_phi'3}
\phi'_3=\frac{\phi'_1}{2} \coth 2\phi_1 \sinh 4\phi_3
\end{equation}
while the azimutal component ($\mu=\theta$), with $A_2=-\frac{g_1}{g_2}A_1$ and $\phi_0=0$, gives rise to
\begin{eqnarray}
A''_1&=&-\left(\frac{4 f'}{f}-\frac{h'_1}{h_1}-\frac{h'_2}{h_2}-4 \sigma'+\frac{4 g_1 \phi'_2}{g_2}\right)\frac{A'_1}{2},\\
A''_1&=&-\left(\frac{4 f'}{f}-\frac{h'_1}{h_1}-\frac{h'_2}{h_2}-4 \sigma'+\frac{4g_2\phi'_2}{g_1} \right)\frac{A'_1}{2}.\qquad
\end{eqnarray}
Consistency between the two equations for $\phi'_2\neq0$ requires
\begin{equation}\label{SO(2)diag_Con}
g_2=\pm g_1
\end{equation}
leading to a differential equation for the $A_1(r)$ function
\begin{equation}
A''_1=-\left(\frac{4 f'}{f}-\frac{h'_1}{h_1}-\frac{h'_2}{h_2}-4 \sigma'\pm4\phi'_2\right)\frac{A'_1}{2}\, .\label{A1pp2}
\end{equation}
The plus/minus sign correponds to that in the relation \eqref{SO(2)diag_Con}. From now on, we will choose $g_2=-g_1$ for definiteness. This choice leads to $A_1=A_2$, and equation \eqref{A1pp2} can be solved by
\begin{equation}
A'_1=A'_2=be^{2(\sigma+\phi_2)}\frac{\sqrt{h_1h_2}}{f^{2}}
\end{equation}
where $b$ is an integration constant.
\\
\indent Repeating the same analysis as in the previous case, we find that the BPS equations in terms of fermion-shift matrices take the same form as in \eqref{AdS4_con} and \eqref{partial_r_con} to \eqref{sigma_con}. In particular, consistency still requires $\mathbb{S}_{\text{s}}$ and $\mathbb{F}_{\text{a}}$ to vanish in order to obtain non-trivial solutions. With the coset representative given in \eqref{SO(2)diag_singlet_in_SO(3)xSO(3)gg_main}, these components take the following form
\begin{eqnarray}
\mathbb{S}_{\text{s}}&=&-\frac{\sinh\phi_0 }{4e^{3 \sigma}}\left[m \sinh\phi_2+\frac{g_1}{2} e^{4\sigma} (\cosh2\phi_1 \cosh 2\phi_3-1)\right],\,\\
\mathbb{F}_{\text{a}}&=&-\frac{A'_1 }{2} \sinh \phi_0\, .
\end{eqnarray}
We again see that $\phi_0$ must vanish as previously required by the two-form field equation. Moreover, the quantities ${{P_r}^{I}}_0$, ${D^I}_i$, and ${(L^{-1})^I}_0$ also vanish when $\phi_0=0$. As a result, the BPS equations take the same form as those given in \eqref{AdS4_con} and \eqref{partial_r_con} to \eqref{sigma_con} with the only difference being in the explicit form of the remaining components of the fermion-shift matrices. With $\phi_0=0$, $g_2=-g_1$, and $A'_1=A'_2$, these components are given by
\begin{eqnarray}
\mathbb{S}_{\text{a}}&=&\frac{1}{8}e^{-3 \sigma}\left[g_1 e^{4\sigma} \left(e^{-\phi_2} \cosh 2\phi_1 \cosh 2 \phi_3+e^{\phi_2}\right)+2m\right],\label{SO(2)d_Sa}\\
\mathbb{N}_{\text{a}}&=&\frac{1}{8}e^{-3 \sigma}\left[g_1 e^{4\sigma} \left(e^{-\phi_2} \cosh 2\phi_1 \cosh 2 \phi_3+e^{\phi_2}\right)-6m\right],\label{SO(2)d_Na}\\
\mathbb{F}_1&=&\mathbb{F}_2=\frac{A'_1e^{-\phi_2}}{2\sqrt{h_1h_2}}\label{SO(2)d_F}
\end{eqnarray}
together with
\begin{eqnarray}
{{P_r}^{I}}_i&=&\phi'_1 \cosh 2\phi_3(\delta^I_1\delta_{i1}+\delta^I_2\delta_{i2})-\phi'_3(\delta^I_1\delta_{i2}-\delta^I_2\delta_{i1})+\phi'_2\delta^I_3\delta_{i3},\label{SO(2)d_P}\\
{C^I}_i&=&g_1e^{-\phi_2}\left[\sinh 2\phi_1(\delta^I_1\delta_{i1}+\delta^I_2\delta_{i2})+\cosh 2\phi_1 \sinh 2\phi_3(\delta^I_1\delta_{i2}-\delta^I_2\delta_{i1})\right.\nonumber\\&&\left.+ \left(e^{2\phi_2}-\cosh 2\phi_1 \cosh 2\phi_3\right)\right]\delta^I_3\delta_{i3}.\label{SO(2)d_C}
\end{eqnarray}
With all these, we finally obtained the explicit form of the BPS equations on $\eta^2_+$ as follows
\begin{eqnarray}
0&=&\frac{k}{\sqrt{f}}\eta^2_++\frac{1}{\sqrt{h_1}}\left(\frac{f'}{2f}+\sigma'\right)(i\sigma_2)\eta^2_++2me^{-3\sigma}\sigma_3\eta^2_+,\label{dkey_BPS_1}\\
0&=&\frac{1}{\sqrt{h_1}}\partial_r\eta^2_++\frac{1}{8}e^{-3 \sigma}\left[g_1 e^{4\sigma} \left(e^{-\phi_2} \cosh 2\phi_1 \cosh 2 \phi_3+e^{\phi_2}\right)+2m\right]\sigma_1\eta^2_+\nonumber\\
&&+\frac{3}{8}e^{-\sigma}\mathbb{F}_1(i\sigma_2)\eta^2_+,\label{dkey_r_con}\quad\\
0&=&\frac{g_1\widehat{A}_1}{\sqrt{h_2}}\sigma_1\eta^2_++\frac{1}{\sqrt{h_1}}\left(\frac{h'_2}{2h_2}+\sigma'\right)(i\sigma_2)\eta^2_++2me^{-3\sigma}\sigma_3\eta^2_+-e^{-\sigma}\mathbb{F}_1\eta^2_+,\quad \label{dkey_BPS_2}\\
0&=&\frac{\sigma'}{\sqrt{h_1}}\sigma_3\eta^2_+-\frac{1}{4}e^{-3 \sigma}\left[g_1 e^{4\sigma} \left(e^{-\phi_2} \cosh 2\phi_1 \cosh 2 \phi_3+e^{\phi_2}\right)-6m\right](i\sigma_2)\eta^2_+\quad\nonumber\\
&&+\frac{1}{4}e^{-\sigma}\mathbb{F}_1\sigma_1\eta^2_+,\label{dkey_BPS_3}\\
0&=&\frac{\phi'_2}{\sqrt{h_1}}\sigma_3\eta^2_+-g_1e^{\sigma-\phi_2}\left(e^{2\phi_2}-\cosh 2\phi_1 \cosh 2\phi_3\right)(i\sigma_2)\eta^2_++e^{-\sigma}\mathbb{F}_2\sigma_1\eta^2_+,\ \qquad\label{dkey_BPS_4}\\
0&=&\frac{\phi'_1}{\sqrt{h_1}}\sigma_3\eta^2_+-g_1e^{\sigma-\phi_2}\sinh2\phi_1\,\text{sech}\,2\phi_3(i\sigma_2)\eta^2_+,\label{dkey_BPS_5}\\
0&=&\frac{\phi'_3}{\sqrt{h_1}}\sigma_3\eta^2_+-g_1e^{\sigma-\phi_2}\cosh2\phi_1\sinh2\phi_3(i\sigma_2)\eta^2_+\, .\label{dkey_BPS_6}
\end{eqnarray}
Note that \eqref{dkey_BPS_5} and \eqref{dkey_BPS_6} can be mapped to each other through the relation \eqref{phi'1_and_phi'3}. Therefore, in this case, there are only five independent equations of the form $M^{(x)}\eta^2_+=0$ where $x=1,2,3,4,5$ respectively labels the BPS equations in \eqref{dkey_BPS_1}, \eqref{dkey_BPS_2}, \eqref{dkey_BPS_3}, \eqref{dkey_BPS_4}, and \eqref{dkey_BPS_5}. On the other hand, equation \eqref{dkey_r_con} determines the explicit form of $\eta^2_+(r,\theta)$ as in the previous case.
\\
\indent After defining $\mathcal{A}^{xy}$, $\mathcal{B}^{xy}$, and $\mathcal{C}^{xy}$ matrices as in the previous case, we find that the vanishing of $\mathcal{A}^{55}$ gives the following condition
\begin{equation}
0=\frac{(\phi'_1)^2}{h_1}-g_1^2 e^{2 (\sigma-\phi_2)} \sinh ^22\phi_1\, \text{sech}^22\phi_3.\label{add_BPS_dcon1}
\end{equation}
The symmetric part $\mathcal{A}^{\tilde{x}5}+\mathcal{A}^{5\tilde{x}}=0$ with $\tilde{x}=1,2,3,4$ gives 
\begin{eqnarray}
0&=&\frac{e^{-\sigma-\phi_2}}{\sqrt{h_1}}\left[g_1e^{4 \sigma} \sinh 2\phi_1\, \text{sech}\,2\phi_3 \left(\frac{f'}{f}+2\sigma'\right)+4 m e^{\phi_2} \phi'_1\right],\\
0&=&\frac{e^{-\sigma-\phi_2}}{\sqrt{h_1}}\left[g_1 e^{4 \sigma} \sinh 2\phi_1\, \text{sech}\,2\phi_3 \left(\frac{h'_2}{h_2}+2 \sigma'\right)+4 m e^{\phi_2} \phi'_1\right],\\
0&=&\frac{\sigma' \phi'_1}{h_1}-\frac{g_1\left[2 e^{\phi_2} \sinh 2\phi_1\, \text{sech}\,2\phi_3 \left(g_1 e^{4 \sigma+\phi_2}-6 m\right)+g_1e^{4 \sigma} \sinh 4\phi_1\right]}{8e^{2 (\sigma+\phi_2)}},\qquad\ \, \\
0&=&\frac{2\phi'_1 \phi'_2}{h_1}-g_1^2 e^{2 (\sigma-\phi_2)} \left(2 e^{2\phi_2} \sinh 2\phi_1\, \text{sech}\,2\phi_3-\sinh 4\phi_1\right).
\end{eqnarray}
The antisymmetric part $\mathcal{A}^{\tilde{x}5}-\mathcal{A}^{5\tilde{x}}=0$ yields
\begin{eqnarray}
0&=&\frac{k \phi'_1}{\sqrt{fh_1}},\label{add_BPS_dcon6}\\
0&=&\frac{\phi'_1}{\sqrt{h_1}}e^{-\sigma}\mathbb{F}_1-\frac{g_1^2 \widehat{A}_1 }{\sqrt{h_2}}e^{\sigma-\phi_2} \sinh 2\phi_1\, \text{sech}\,2 \phi_3,\\
0&=&g_1 \mathbb{F}_1 e^{-\phi_2} \sinh 2 \phi_1\, \text{sech}\,2 \phi_3,\label{add_BPS_dcon8}\\
0&=&g_1 \mathbb{F}_2 e^{-\phi_2} \sinh 2\phi_1\, \text{sech}\,2 \phi_3\, .\label{add_BPS_dcon9}
\end{eqnarray}
From $\mathcal{B}^{\tilde{x}5}+\mathcal{C}^{\tilde{x}5}=0$, we find
\begin{eqnarray}
0&=&\frac{kg_1}{\sqrt{f}}e^{\sigma-\phi_2} \sinh 2\phi_1\, \text{sech}\,2\phi_3,\label{add_BPS_dcon10}\\
0&=&\frac{g_1 \widehat{A}_1 \phi'_1}{\sqrt{h_1h_2}}-g_1 \mathbb{F}_1 e^{-\phi_2} \sinh 2\phi_1\, \text{sech}\,2\phi_3,\\
0&=&\frac{\mathbb{F}_1e^{-\sigma} \phi'_1}{\sqrt{h_1}},\label{add_BPS_dcon12}\\
0&=&\frac{\mathbb{F}_2e^{-\sigma} \phi'_1}{\sqrt{h_1}}.\label{add_BPS_dcon13}
\end{eqnarray}
Finally, $\mathcal{B}^{\tilde{x}5}-\mathcal{C}^{\tilde{x}5}=0$ gives 
\begin{eqnarray}
0&=&\frac{\phi'_1}{h_1}\left(\frac{f'}{f}+2 \sigma'\right)+4g_1m e^{-2 \sigma-\phi_2} \sinh 2\phi_1\, \text{sech}\,2\phi_3,\\
0&=&\frac{\phi'_1}{h_1}\left(\frac{h'_2}{h_2}+2 \sigma'\right)+4 g_1 m e^{-2 \sigma-\phi_2} \sinh 2\phi_1\, \text{sech}\,2 \phi_3,\\
0&=&\frac{\phi'_1}{4 \sigma'} \left[\left(\cosh 2\phi_1 \cosh 2\phi_3+e^{2 \phi_2}\right)-\frac{6me^{\phi_2}}{g_1e^{4\sigma}}\right]-\sinh 2\phi_1\, \text{sech}\,2 \phi_3,\\
0&=&\frac{2 g_1e^{\sigma-\phi_2}}{\sqrt{h_1}}\left[\phi'_1 \left(e^{2\phi_2}-\cosh 2\phi_1 \cosh 2\phi_3\right)-\sinh 2\phi_1 \phi'_2\, \text{sech}\,2 \phi_3\right].\quad\label{add_BPS_dcon17}
\end{eqnarray}
\indent From these results, we can readily see that setting $\phi_1=0$ satisfies all of these conditions without any additional constraint. Moreover, from \eqref{phi'1_and_phi'3}, we also find that $\phi'_3=0$ for $\phi_1'=0$ giving rise to constant $\phi_3$ that can be set to zero. With $\phi_1=\phi_3=0$, we find that $\mathbb{S}_{\text{a}}$, $\mathbb{N}_{\text{a}}$, ${{P_r}^{I}}_i$, and ${C^I}_i$ from \eqref{SO(2)d_Sa}, \eqref{SO(2)d_Na}, \eqref{SO(2)d_P}, and \eqref{SO(2)d_C} reduce to \eqref{SO(2)xSO(2)_Sa}, \eqref{SO(2)xSO(2)_Na}, and  \eqref{SO(2)xSO(2)_PandC}, respectively. Accordingly, the non-trivial BPS equations as well as the conditions resulted from the vanishing of $\mathcal{A}^{\tilde{x}\tilde{y}}$, $\mathcal{B}^{\tilde{x}\tilde{y}}$, and $\mathcal{C}^{\tilde{x}\tilde{y}}$ are also the same as in \eqref{key_BPS_1} to \eqref{key_BPS_4} and \eqref{DABPS1} to \eqref{SBCBPS6} obtained in the previous case with $\mathbb{F}_1$ and $\mathbb{F}_2$ given in \eqref{SO(2)d_F}. 
\\
\indent As a result, the $SO(2)_{\text{diag}}$ symmetric solution is then given by
\begin{eqnarray}
f&=&\frac{e^{\frac{10 \sigma}{3}}(b g_1 m)^{\frac{2}{3}}}{\left(k\left(g_1^2 e^{8 \sigma}-9 m^2\right)\right)^{\frac{2}{3}}},\\
h_1&=&\frac{576 m^4 e^{6 \sigma} (b g_1 m)^{\frac{2}{3}} (\sigma')^2}{\left(g_1^2 e^{8 \sigma}-9 m^2\right)^2 \left[4 m^2 (b g_1 m)^{\frac{2}{3}}-e^{\frac{8 \sigma}{3}} \left(k \left(g_1^2 e^{8 \sigma}-9 m^2\right)\right)^{\frac{2}{3}}\right]},\\
h_2&=&\frac{9 C^2 g_1^2 m^2 e^{\frac{10 \sigma}{3}}}{\left(k\left(g_1^2e^{8 \sigma}-9 m^2\right)\right)^{\frac{2}{3}}} \left[4 m^2\left(bg_1 m\right)^{\frac{2}{3}}-e^{\frac{8 \sigma}{3}}\left(k\left(g_1^2e^{8 \sigma}-9 m^2\right)\right)^{\frac{2}{3}}\right],\qquad\ \\
\phi_2&=&4 \sigma+\ln \left[\frac{g_1}{3 m}\right],\\
A_1&=&C \left(g_1^2 e^{8 \sigma}-3 m^2\right)-\frac{2q}{g_1}
\end{eqnarray}
where $C$ is an integration constant. It can also be verified that all of the BPS conditions and the field equations are satisfied for
\begin{equation}\label{alter_SO(2)d_sign_Con}
b g_1 m>0,\qquad s\left(g_1^2 e^{8 \sigma}-9 m^2\right)>0,\qquad \frac{g_1}{3 m}>0\, .
\end{equation}
\indent Finally, solving equation \eqref{dkey_r_con}, we find the Killing spinor given by
\begin{equation}
\eta^2_+=\frac{\zeta\, e^{i \theta  q} e^{\frac{5 \sigma}{6}}}{\left(g_1^2 e^{8 \sigma}-9 m^2\right)^{\frac{1}{6}}}\begin{pmatrix} \sqrt{2 m k \left(bg_1 m\right)^{\frac{1}{3}}-e^{\frac{4 \sigma}{3}} \left(k \left(g_1^2 e^{8 \sigma}-9 m^2\right)\right)^{\frac{1}{3}}} \\ \sqrt{2 m k \left(bg_1 m\right)^{\frac{1}{3}}+e^{\frac{4 \sigma}{3}} \left(k \left(g_1^2 e^{8 \sigma}-9 m^2\right)\right)^{\frac{1}{3}}} \end{pmatrix}.
\end{equation}
As in the previous case, all functions are determined in terms of the dilaton $\sigma(r)$. We will choose the solution for $\sigma$ as in \eqref{sigma_soln}.

\subsection{Regular solutions}
In this case, the six-dimensional metric reads
\begin{eqnarray}
ds^2_6&=&\frac{B r^{\frac{5}{12}}}{4 m^2 \left(k \left(G^2 r-1\right)\right)^{\frac{2}{3}}}\left[ds^2_{AdS_4}+ds^2_{\Sigma}\right]\label{SO(2)d_6D_Met}\\
\textrm{with} \qquad ds^2_{\Sigma}&=&\frac{r^{-\frac{5}{3}}}{9W\left(k \left(G^2 r-1\right)\right)^{\frac{4}{3}}}dr^2+\frac{4\mathcal{C}^2W}{B}d\theta^2\label{SO(2)d_Sigma_Met}
\end{eqnarray}
where the function $W(r)$ is now given by
\begin{equation}\label{SO(2)d_Def_W}
W=B-r^{\frac{1}{3}} \left(k \left(G^2 r-1\right)\right)^{\frac{2}{3}}.
\end{equation}
The three parameters appearing in the above equations are defined by
\begin{equation}\label{SO(2)d_para_redefine}
B=\left(\frac{8 b g_1 m^2}{9}\right)^{\frac{2}{3}},\qquad G=\frac{g_1}{3m},\qquad \mathcal{C}=3g_1 m^2C.
\end{equation}
\\
\indent The solutions for the dilaton $\sigma$ and the only non-vanishing scalar from the vector multiplets $\phi_2$ are given by
\begin{equation}
\sigma= \frac{1}{8}\ln r\qquad\text{ and }\qquad \phi_2=\frac{1}{2}\ln r+\ln G\, .\label{scalar_sol_SO2d}
\end{equation}
At this point, we recall that the analysis performed previously also leads to a relation between the two $SO(3)$ coupling constants namely $g_2=-g_1$. However, when $\phi_1$ and $\phi_3$ vanish, the coupling constant $g_2$ no longer appears in the BPS equations. Accordingly, the resulting solutions still lie within the $U(1)\times U(1)$ sector of the matter-coupled $F(4)$ gauged supergravity and can be uplifted to massive type IIA theory as in the previous section.
\\
\indent However, in this case, the two gauge fields $A^{3}$ and $A^{6}$ are equal with non-vanishing components given by
\begin{equation}
A^{3}_\theta=A^{6}_\theta=\frac{\mathcal{C}(3G^2 r-1)-2q}{3 G m}\, .
\end{equation}
The two-component Killing spinor $\eta^2_+$ takes the form of
\begin{equation}
\eta^2_+(r,\theta)=\frac{\zeta\, e^{iq\theta} r^{\frac{5}{48}}}{\left(k\left(G^2 r-1\right)\right)^{\frac{1}{6}}}\begin{pmatrix} \sqrt{\sqrt{B}-k r^{\frac{1}{6}} \left(k\left(G^2 r-1\right)\right)^{\frac{1}{3}}} \\ \sqrt{\sqrt{B}+ k r^{\frac{1}{6}} \left(k\left(G^2 r-1\right)\right)^{\frac{1}{3}}}\end{pmatrix}
\end{equation}
with $\zeta$ being a complex constant. Consequently, the $SO(2)_{\text{diag}}$ symmetric solutions analyzed in this section also preserve half of the supersymmetries.
\\
\indent We now determine the range of the coordinate $r$ leading to regular solutions. Unlike in the previous section, for suitable values of the parameters $B$ and $G$, there can be up to three roots for $W(r)=0$ equation. For $\left(27 B^3 G^2-4\right)>0$ or $B>\frac{2^{\frac{2}{3}}}{3G^{\frac{2}{3}}}$, there is only one root given by
\begin{equation}
r_0=\frac{1}{3} \left[\frac{2}{G^2}+\frac{1}{\tilde{Z}}+\frac{\tilde{Z}}{G^4}\right]
\end{equation}
with 
\begin{equation}
\tilde{Z}=\left[\frac{27 B^3 G^8}{2}-G^6+\frac{3}{2} \sqrt{3} \sqrt{B^3 G^{14} \left(27 B^3 G^2-4\right)}\right]^{\frac{1}{3}}.
\end{equation}
\indent On the other hand, for $\left(27 B^3 G^2-4\right)<0$ or $B<\frac{2^{\frac{2}{3}}}{3G^{\frac{2}{3}}}$, there are three roots given by
\begin{eqnarray}
r_0&=&\frac{1}{3} \left[\frac{2}{G^2}+\frac{1}{Z}+\frac{1}{Z^*}\right]\label{r0_def}\\
\textrm{and}\qquad r_\pm&=&\frac{1}{6} \left[\frac{4}{G^2}-\frac{1}{Z}-\frac{1}{Z^*}\pm \sqrt{3}i\left(\frac{1}{Z^*}-\frac{1}{Z}\right)\right]\label{rpm_def}
\end{eqnarray}
for
\begin{equation}
Z=\left[\frac{27 B^3 G^8}{2}-G^6+i\frac{3\sqrt{3}}{2}\sqrt{B^3 G^{14} \left(4-27 B^3 G^2\right)}\right]^{\frac{1}{3}}.
\end{equation}
We also note that for $B=\frac{2^{\frac{2}{3}}}{3G^{\frac{2}{3}}}$ the two roots $r_+$ and $r_-$ degenerate. Furthermore, the behaviors of the solution for $k = +1$ and $k = -1$ are different, so we will discuss each case separately. 

\subsection{Solutions with $k=+1$}
\begin{figure}[h!]
  \centering
  \begin{subfigure}[b]{0.326\linewidth}
    \includegraphics[width=\linewidth]{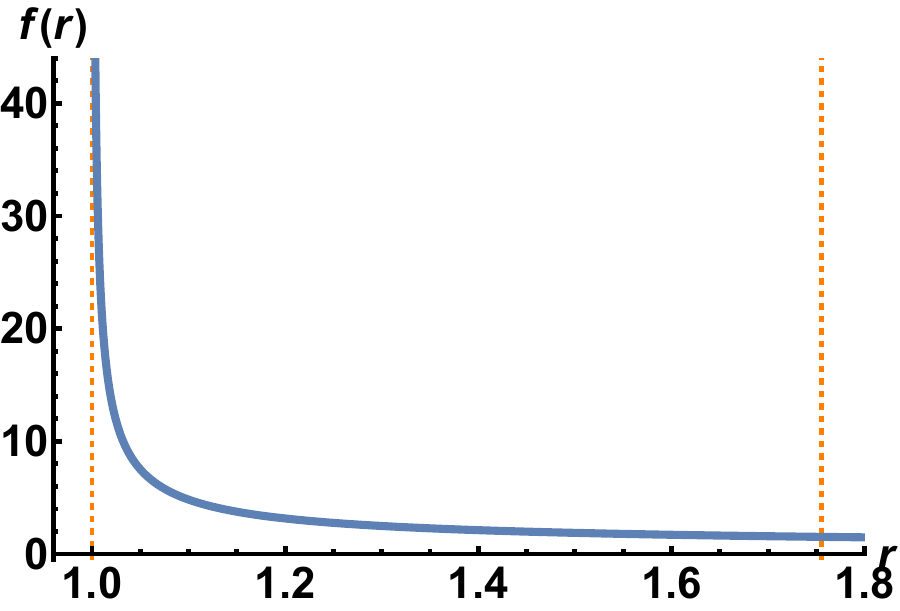}
\caption{$f$ solution}
  \end{subfigure}
  \begin{subfigure}[b]{0.326\linewidth}
    \includegraphics[width=\linewidth]{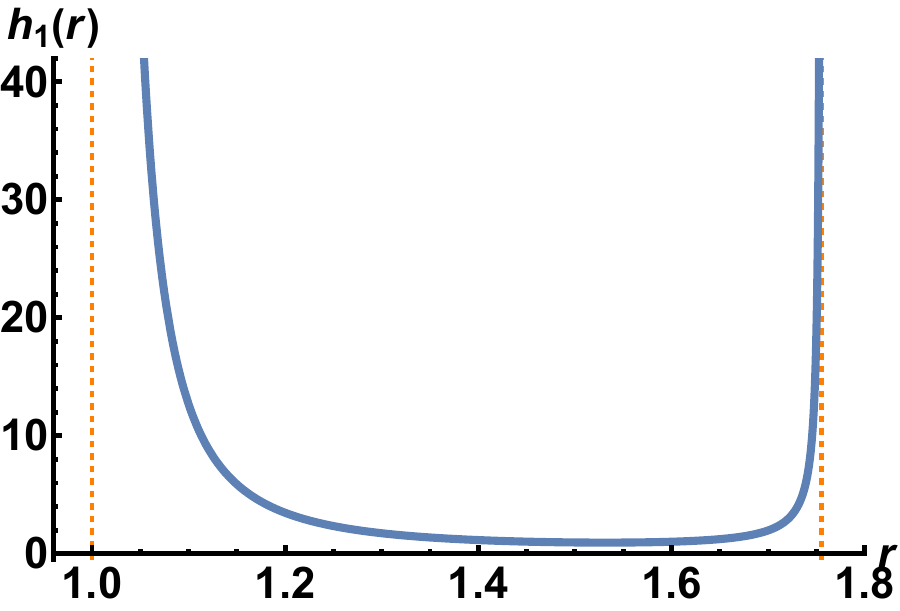}
\caption{$h_1$ solution}
  \end{subfigure}
  \begin{subfigure}[b]{0.326\linewidth}
    \includegraphics[width=\linewidth]{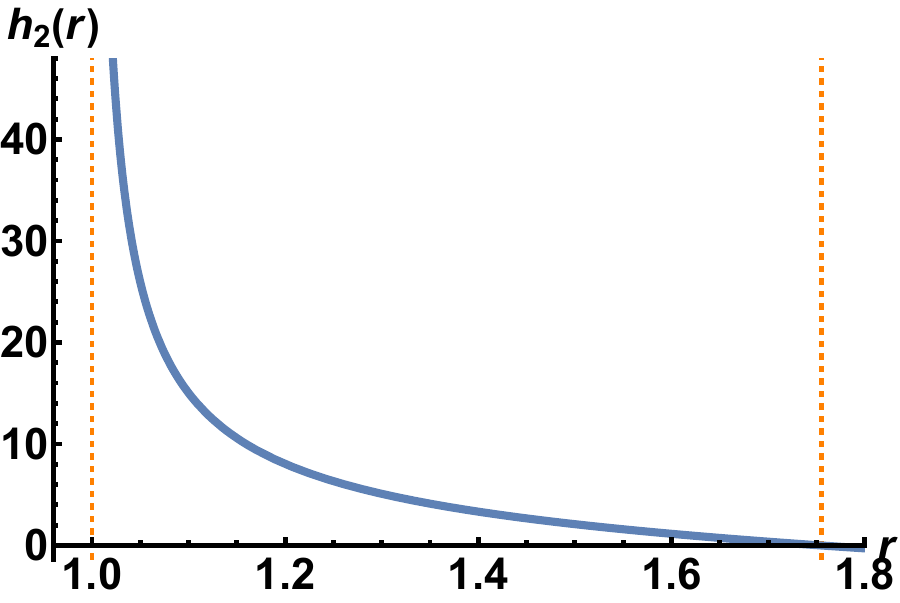}
\caption{$h_2$ solution}
  \end{subfigure}
  \caption{Numerical plots of the warp factors for the $SO(2)_{\text{diag}}$ symmetric solution with $k=1$, $B=G=\mathcal{C}=1$, and $m=\frac{1}{2}$. The solution is regular in the range $\frac{1}{G^2}=1<r<r_0=1.755$ with the two vertical dashed lines representing the two boundaries.}
  \label{Fig3}
\end{figure}

For the case of $k=+1$, we find a regular solution in the range
\begin{equation}\label{SO(3)d_Case+_SO(2)xSO(2)_Pos_Range}
\frac{1}{G^2}<r<r_0
\end{equation}
for $B>0$, $G>0$, and $m>0$. An example of numerical plots for the three warp factors with $k=1$, $B=G=\mathcal{C}=1$, and $m=\frac{1}{2}$ is shown in figure \ref{Fig3}. This is very similar to the solution with $SO(2)\times SO(2)$ symmetry considered in the previous section. As $r\rightarrow\frac{1}{G^2}$, the six-dimensional metric is conformal to a product of $AdS_4$ and a cylinder with the same metric given in \eqref{SO(3)xSO(3)_SO(2)xSO(2)_AdS4xdisk_metric} but with the new radial coordinate $R$ given by $r=\frac{1}{G^2}+\frac{8\mathcal{C}^3B^{\frac{3}{2}}R^3}{G}$. As in the previous section, the $S^1$ with the coordinate $\theta$ decompactifies in this limit leading to a locally asymptotic $AdS_6$ geometry with a holographic boundary given by $AdS_4\times S^1$. In this limit, the scalar fields become
\begin{equation}
\sigma=\frac{1}{4}\ln\left[\frac{3m}{g_1}\right]\qquad \textrm{and}\qquad \phi_2=0
\end{equation}
as expected at the $AdS_6$ vacuum with $SO(3)\times SO(3)$ symmetry.

\begin{figure}[h!]
  \centering
    \includegraphics[width=0.6\linewidth]{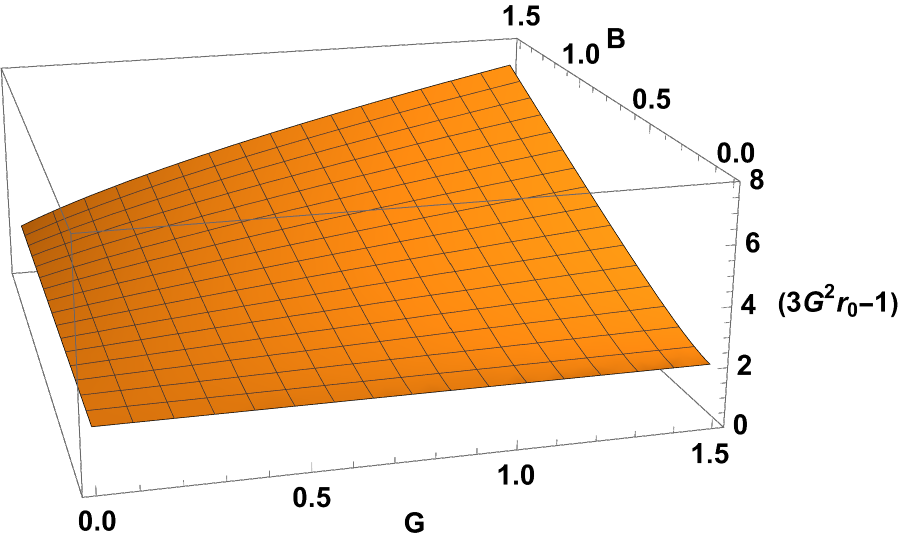}
  \caption{A numerical plot of the function $\left(3 G^2 r_0-1\right)$ under the condition \eqref{Case+_smooth_con}. Note that $\left(3 G^2 r_0-1\right)\geq 2$ for all $B>0$ and $G>0$.}
  \label{Fig4}
\end{figure}

On the other hand, as $r\rightarrow r_0$, we find the six-dimensional metric of the form
\begin{equation}\label{Case+_spindle_metric}
ds_6^2\approx\frac{B r_0^{\frac{5}{12}}}{4 m^2 \left(G^2 r_0-1\right)^{\frac{2}{3}}}\left[ds^2_{AdS_4}+\frac{4\left[d\rho^2+\mathcal{C}^2\left(3 G^2 r_0-1\right)^2\rho^2d\theta^2\right]}{\left[-9W'(r_0)r_0^{\frac{5}{3}}\left(G^2 r_0-1\right)^{\frac{4}{3}}\right]}\right]
\end{equation}
after changing to the new radial coordinate $\rho=\sqrt{r_0-r}$. In this case, the $\theta$-circle shrinks smoothly near the endpoint at $r=r_0$ if we choose
\begin{equation}\label{Case+_smooth_con}
\mathcal{C}=\frac{1}{l\left(3 G^2 r_0-1\right)},\qquad l=1,2,3,\ldots\, .
\end{equation}
Since the explicit form of $\left(3 G^2 r_0-1\right)$ derived from \eqref{r0_def} is complicated, we only numerically verify that this function is always greater than $2$ within the regularity range as shown in figure \ref{Fig4}. 
\\
\indent As in the previous case, we can compute the Euler characteristic of $\Sigma$ from the metric \eqref{SO(2)d_Sigma_Met}. The result is given by
\begin{eqnarray}
\chi(\Sigma)&=&\frac{1}{4\pi}\int_{\Sigma}R_{\Sigma}\text{vol}_{\Sigma}\ =\ \frac{\mathcal{C} r_0^{1/6} (G^2 r_0-1)^{1/3} \left(3 G^2 r_0-1\right)}{\sqrt{B}}\nonumber\\&=&\mathcal{C} \left(3 G^2 r_0-1\right)\ =\ \frac{1}{l}\label{case+_chi}
\end{eqnarray}
which again shows that $\Sigma$ is a disk with an $\mathbb{R}^2/\mathbb{Z}_l$ orbifold singularity at $r= r_0$.

To obtain the $SO(2)_{\text{diag}}$ gauge field that vanishes at $r=r_0$, we choose the constant $q$ to be
\begin{equation}
q=\frac{\mathcal{C}}{2} \left(3 G^2 r_0-1\right)=\frac{1}{2l}
\end{equation}
resulting in 
 \begin{equation}\label{CaseIII_SO(2)R_soln}
A^{3}_\theta=A^{6}_\theta=\frac{\mathcal{C}G}{m}\left(r-r_0\right).
\end{equation}
Finally, the Killing spinor $\eta^2_+$ at the end point $r=r_0$ is given by
\begin{equation}
\eta^2_+(r_0,\theta)=\sqrt{2}\xi\, e^{\frac{i\theta}{2l}} r_0^{\frac{3}{16}}\begin{pmatrix} 0 \\ 1\end{pmatrix}.
\end{equation}

\subsection{Solutions with $k=-1$ and $0<B<\frac{2^{\frac{2}{3}}}{3G^{\frac{2}{3}}}$}
\begin{figure}[h!]
  \centering
  \begin{subfigure}[b]{0.326\linewidth}
    \includegraphics[width=\linewidth]{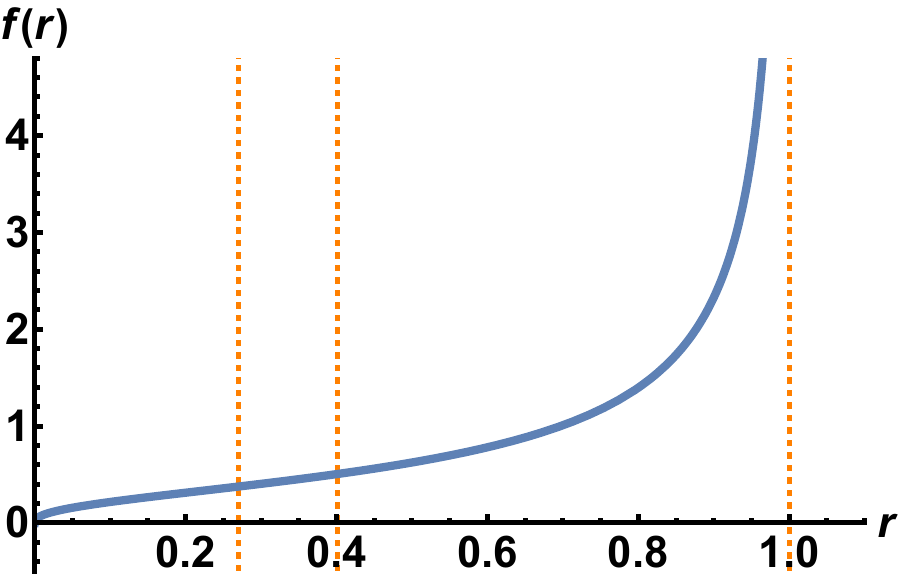}
\caption{$f$ solution}
  \end{subfigure}
  \begin{subfigure}[b]{0.326\linewidth}
    \includegraphics[width=\linewidth]{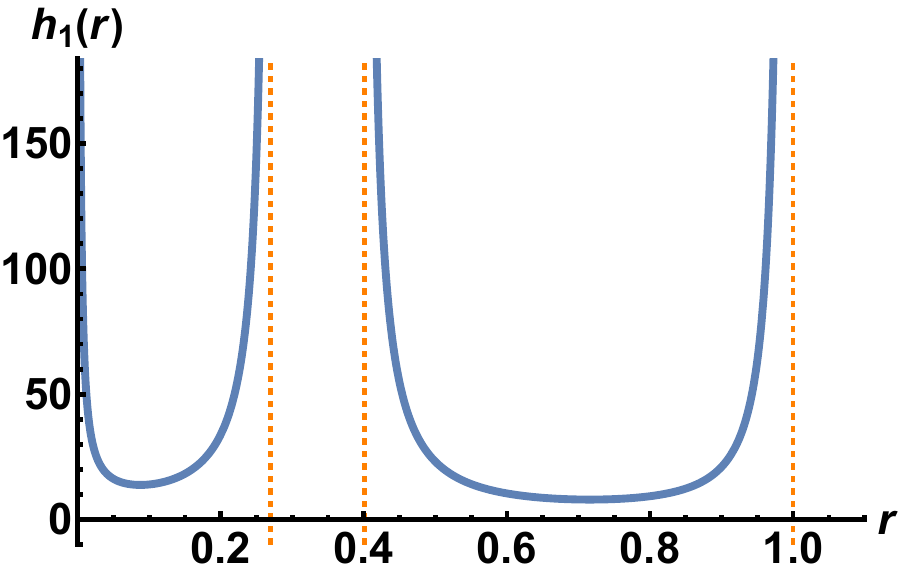}
\caption{$h_1$ solution}
  \end{subfigure}
  \begin{subfigure}[b]{0.326\linewidth}
    \includegraphics[width=\linewidth]{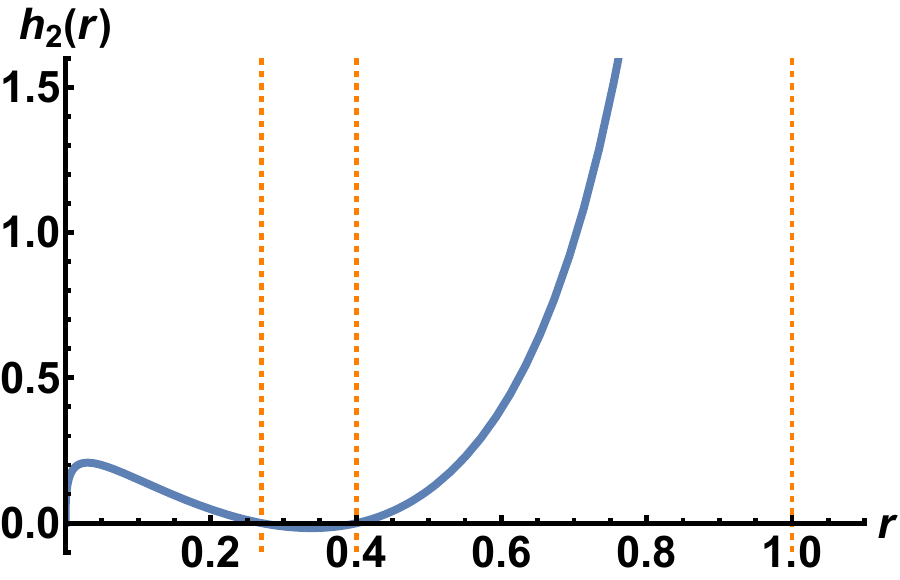}
\caption{$h_2$ solution}
  \end{subfigure}
  \caption{Numerical plots of the warp factors for the $SO(2)_{\text{diag}}$ symmetric solution evaluated at $k=-1$, $B=0.524$, $G=\mathcal{C}=1$, and $m=\frac{1}{2}$. The solution is regular within the intervals $0<r<r_+=0.27$ and $r_-=0.4<r<\frac{1}{G^2}=1$. The three vertical dashed lines mark the boundaries of these regions (excluding $r = 0$).}
  \label{Fig5}
\end{figure}

A more interesting case occurs when $k=-1$ and $0<B<\frac{2^{\frac{2}{3}}}{3G^{\frac{2}{3}}}$. With $G>0$ and $m>0$, we find regular solutions in the following two ranges
\begin{eqnarray}
i)&&0<r<r_+, \label{r_uplift}\\
ii)&& r_-<r<\frac{1}{G^2}\, .
\end{eqnarray}
An example of numerical plots of the three warp factors for $k=-1$, $B=0.524$, $G=\mathcal{C}=1$, and $m=\frac{1}{2}$ is shown in figure \ref{Fig5}. 

\begin{figure}[h!]
  \centering
    \includegraphics[width=0.6\linewidth]{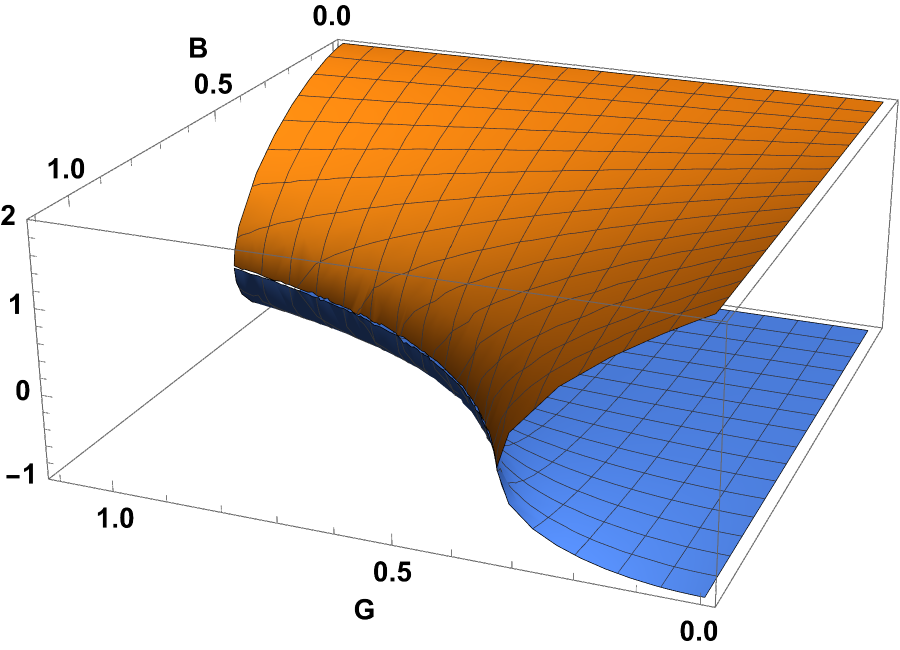}
  \caption{Numerical plot of the functions $\left(1-3 G^2 r_-\right)$ (orange surface) and $\left(1-3 G^2 r_+\right)$ (blue surface) in condition \eqref{Case-_smooth_con}. Both functions approach zero as $B\rightarrow\frac{2^{\frac{2}{3}}}{3G^{\frac{2}{3}}}$.}
  \label{Fig6}
\end{figure}

As $r\rightarrow\frac{1}{G^2}$, we can change to a new radial coordinate $R$ given by $r=\frac{1}{G^2}-\frac{8\mathcal{C}^3B^{\frac{3}{2}}R^3}{G}$ and obtain the six-dimensional metric given in \eqref{SO(3)xSO(3)_SO(2)xSO(2)_AdS4xdisk_metric}. This again leads to the $AdS_6$ asymptotic geometry with an $AdS_4\times S^1$ boundary. On the other hand, as $r\rightarrow0$, the metric near $r=0$ reads  
\begin{equation}\label{SO(3)d_rto0_SO(2)xSO(2)_AdS4xdisk_metric}
ds_6^2\approx\frac{B^{\frac{9}{4}}\mathcal{C}^{\frac{5}{2}}R^{\frac{5}{2}}}{4 m^2}\left[ds^2_{AdS_4}+4\mathcal{C}^2(dR^2+d\theta^2)\right]
\end{equation}
with the new radial coordinate $R$ given by $r=B^3\mathcal{C}^6R^6$. In this case, the six-dimensional metric becomes conformal to a product of $AdS_4$ and a cylinder with the warp factor of $AdS_4$ vanishing. This potentially leads to a curvature singularity. We will see later that the uplifted ten-dimensional solution corresponds to D4-branes smeared over two directions.  
\\
\indent As $r\rightarrow r_\pm$, we can introduce a new radial coordinate $\rho=\sqrt{r_\pm-r}$ and find the six-dimensional metric, as $r\rightarrow r_\pm$ or $\rho\rightarrow0$, of the form  
\begin{equation}\label{Case-1_spindle_metric}
ds_6^2\approx\frac{B r_{\pm }^{\frac{5}{12}}}{4 m^2 \left(1-G^2 r_{\pm }\right)^{\frac{2}{3}}}\left[ds^2_{AdS_4}+\frac{4\left[d\rho^2+\mathcal{C}^2\left(1-3 G^2 r_{\pm }\right)^2\rho ^2d\theta^2\right]}{\left[-9W'(r_{\pm }) r_{\pm }^{\frac{5}{3}} \left(1-G^2 r_{\pm }\right)^{\frac{4}{3}}\right]}\right].
\end{equation}
In this case, the $\theta$-circle shrinks smoothly at $r=r_\pm$ if we impose
\begin{equation}\label{Case-_smooth_con}
\mathcal{C}= \frac{1}{l \left| 1-3 G^2 r_{\pm }\right|},\qquad l=1,2,3,\ldots
\end{equation}
leading to an $\mathbb{R}^2/\mathbb{Z}_l$ orbifold as before. We have used an absolute sign for the function $\left(1-3 G^2 r_{\pm }\right)$ to ensure that the constant $\mathcal{C}$ is positive. In particular, using $r_\pm$ given in \eqref{rpm_def}, we find that $0<\left(1-3 G^2 r_-\right)<2$ and $-1<\left(1-3 G^2 r_+\right)<0$. This can also be seen from a numerical plot shown in figure \ref{Fig6}. 

With the condition \eqref{Case-_smooth_con}, the Euler characteristic of $\Sigma$ can be determined to be the same as in \eqref{SO(2)xSO(2)_chi} and \eqref{case+_chi}. Accordingly, the space $\Sigma$ with the metric \eqref{SO(2)d_Sigma_Met} forms a topological disk for both $r\in (0,r_+)$ and $r \in (r_-,\frac{1}{G^2})$ with a $\mathbb{Z}_l$ orbifold singularity at the center $r=r_\pm$. In order for the $SO(2)_{\text{diag}}$ gauge field to vanish at $r= r_{\pm}$, we fix the constant $q$ such that
\begin{equation}
q=-\frac{1}{2} \mathcal{C} \left(1-3 G^2 r_\pm\right)\qquad \textrm{and}\qquad A^{3}_\theta=A^{6}_\theta=\frac{\mathcal{C}G}{m}\left(r-r_\pm\right)
\end{equation}
which gives rise to the Killing spinor $\eta^2_+$ at $r= r_{\pm}$ of the form
\begin{equation}
\eta^2_+(r_\pm,\theta)=\sqrt{2}\xi\, e^{\pm\frac{i\theta}{2l}}r_\pm^{\frac{3}{16}}\begin{pmatrix} 1 \\ 0\end{pmatrix}.
\end{equation}
In this equation, the positive or negative sign of the exponential $\pm\frac{i\theta}{2l}$ is determined by the sign of $r_\pm$.

\subsection{Solutions with $k=-1$ and $B=\frac{2^{\frac{2}{3}}}{3G^{\frac{2}{3}}}$}
\begin{figure}[h!]
  \centering
  \begin{subfigure}[b]{0.326\linewidth}
    \includegraphics[width=\linewidth]{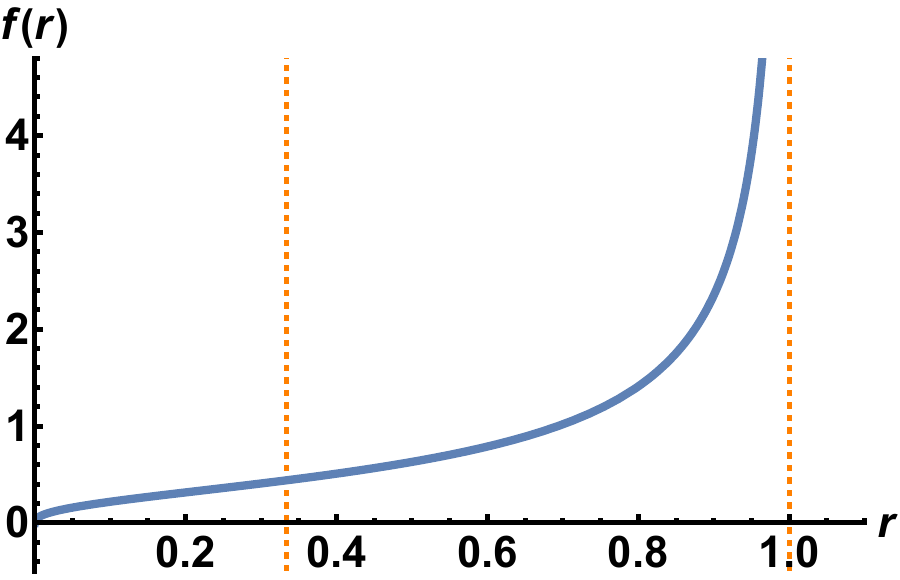}
\caption{$f$ solution}
  \end{subfigure}
  \begin{subfigure}[b]{0.326\linewidth}
    \includegraphics[width=\linewidth]{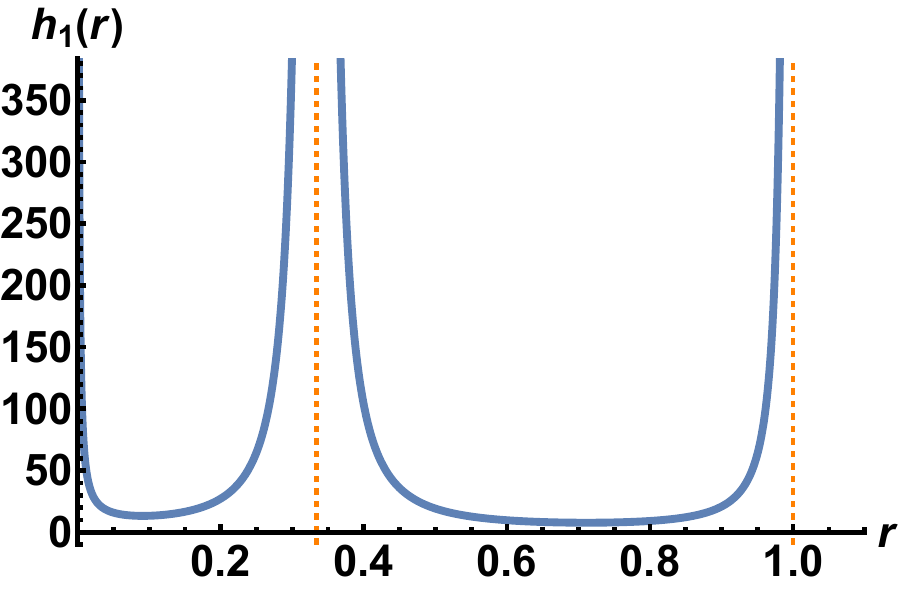}
\caption{$h_1$ solution}
  \end{subfigure}
  \begin{subfigure}[b]{0.326\linewidth}
    \includegraphics[width=\linewidth]{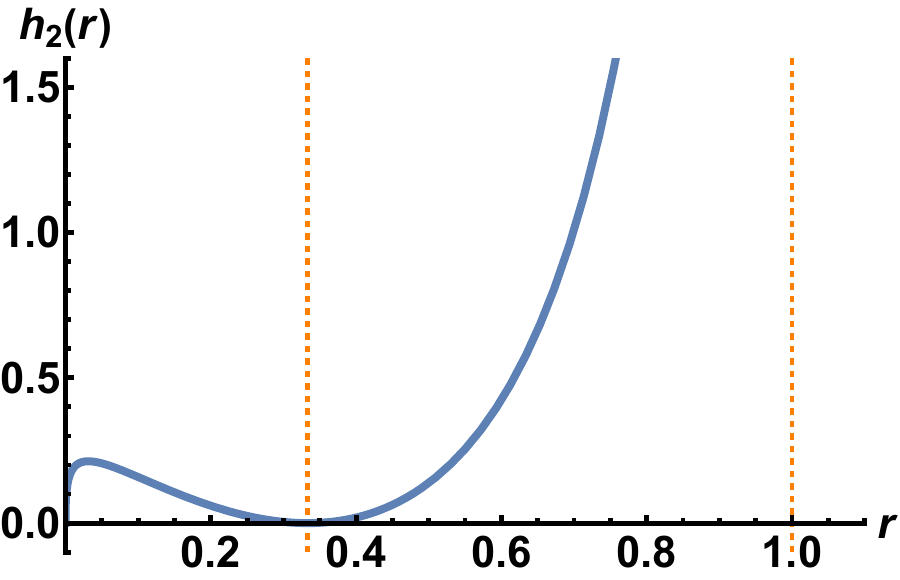}
\caption{$h_2$ solution}
  \end{subfigure}
  \caption{Numerical plots of the warp factors for the $SO(2)_{\text{diag}}$ symmetric solution evaluated at $k=-1$, $B=0.529$, $G=\mathcal{C}=1$, and $m=\frac{1}{2}$. The solution is regular within the intervals $0<r<r_\ast=0.33$ and $r_\ast<r<\frac{1}{G^2}=1$. The two vertical dashed lines mark the boundaries of these regions (excluding $r = 0$).}
  \label{Fig7}
\end{figure}

In this case, setting $k=-1$ and $B=\frac{2^{\frac{2}{3}}}{3G^{\frac{2}{3}}}$, we find that $r_+=r_-=r_\ast$. However, it turns out that the $\theta$-circle cannot shrink smoothly at $r=r_\ast$ since $\left(1-3 G^2 r_{\ast }\right)=0$ when $B=\frac{2^{\frac{2}{3}}}{3G^{\frac{2}{3}}}$. This makes $\mathcal{C}$ diverge according to condition \eqref{Case-_smooth_con}. Instead, at $r=r_\ast$, the six-dimensional metric is asymptotically given by
\begin{equation}\label{Case-2_6D_metric}
ds_6^2\approx\frac{1}{4\times 3^{3/4} G^{3/2} m^2}\left[ds^2_{AdS_4}+\frac{d\rho^2+9 \mathcal{C}^2G^4 d\theta^2}{\rho^2}\right]
\end{equation}
with the new radial coordinate $\rho$ defined by $\rho=\frac{1}{3 G^2}-r$. An example of numerical plots of the three warp factors for $k=-1$, $B=\frac{2^{2/3}}{3G^{2/3}}=0.529$, $G=\mathcal{C}=1$, and $m=\frac{1}{2}$ is given in figure \ref{Fig7}. 

\subsection{Solutions with $k=-1$ and $B>\frac{2^{\frac{2}{3}}}{3G^{\frac{2}{3}}}$}
\begin{figure}[h!]
  \centering
  \begin{subfigure}[b]{0.326\linewidth}
    \includegraphics[width=\linewidth]{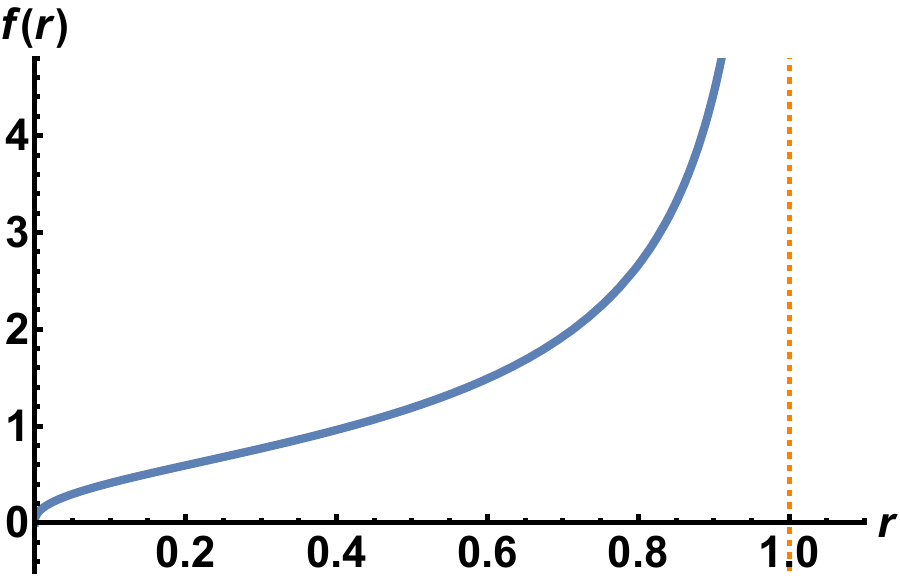}
\caption{$f$ solution}
  \end{subfigure}
  \begin{subfigure}[b]{0.326\linewidth}
    \includegraphics[width=\linewidth]{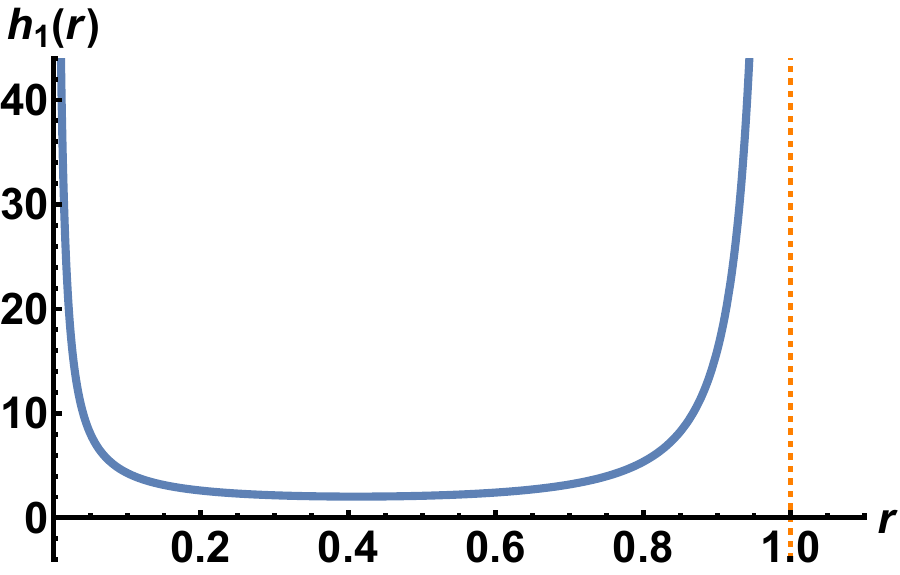}
\caption{$h_1$ solution}
  \end{subfigure}
  \begin{subfigure}[b]{0.326\linewidth}
    \includegraphics[width=\linewidth]{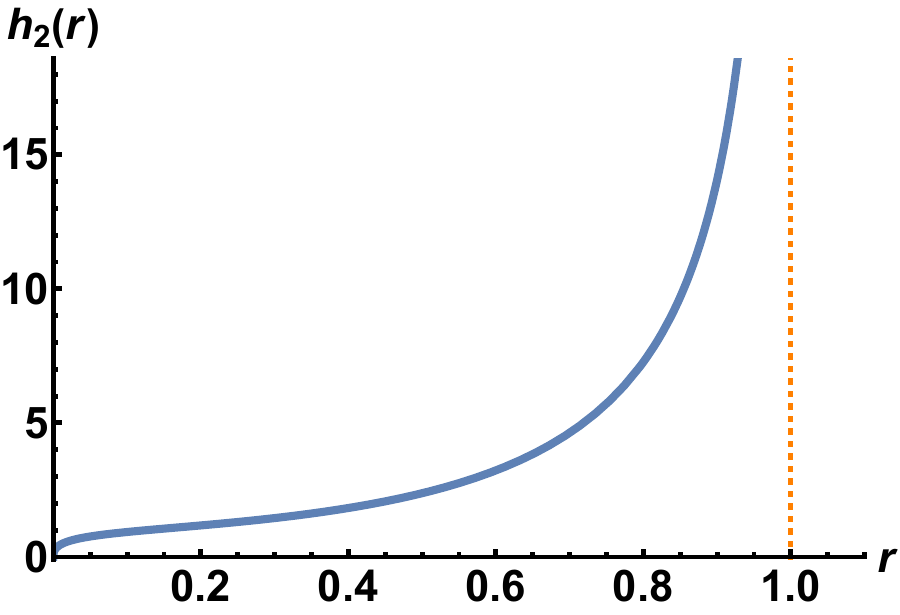}
\caption{$h_2$ solution}
  \end{subfigure}
  \caption{Numerical plots of the warp factors for the $SO(2)_{\text{diag}}$ symmetric solution evaluated at $k=-1$, $B=G=\mathcal{C}=1$, and $m=\frac{1}{2}$. The solution is regular within the interval $0<r<\frac{1}{G^2}=1$. The vertical dashed line marks the boundary of the region at $r = 1$.}
  \label{Fig8}
\end{figure}

Finally, for $k=-1$ and $B>\frac{2^{\frac{2}{3}}}{3G^{\frac{2}{3}}}$, there is only one root to $W(r)=0$ equation, and we find a regular solution only in the range
\begin{equation}\label{SO(3)d_Case-3_SO(2)xSO(2)_Pos_Range}
0<r<\frac{1}{G^2}
\end{equation}
for $G>0$ and $m>0$. As in the previous case, the $\theta$-circle does not shrink, and the solution interpolates between the two end points that are conformal to a product of $AdS_4$ and a cylinder at $r=0$ and $r=\frac{1}{G^2}$ considered above. An example of numerical plots of the warp factors with $k=-1$, $B=G=\mathcal{C}=1$, and $m=\frac{1}{2}$ is shown in figure \ref{Fig8}.

\subsection{Uplift to massive type IIA theory}
We now consider uplifting the six-dimensional solutions to ten-dimensional massive type IIA theory. As in the previous section, at $r=\frac{1}{G^2}$, the $\theta$-circle decompactifies resulting in an $AdS_6$ asymptotic geometry. This again leads to infinite holographic free energy. Accordingly, we will only consider the uplifted solution in the case of $0<r<r_+$ given in \eqref{r_uplift}.    
\\
\indent In the present case, we have 
\begin{equation}
\mc{A}_1=\frac{\mc{C}G}{m}(r-r_+)d\theta\qquad \textrm{and}\qquad \mc{A}_2=0\, . \label{A12_22}
\end{equation}
The ten-dimensional metric is then given by
\begin{eqnarray}
ds^2_{10}&=&\frac{9\lambda^2\cos^{-\frac{1}{3}}\xi e^{-\frac{\sigma}{2}}\Delta^{\frac{1}{2}}Br^{\frac{5}{12}}}{4g_1^2(r-1)^{\frac{2}{3}}}\left\{ds^2_{AdS_4}+\frac{r^{-\frac{5}{3}}}{9W(r-1)^{\frac{4}{3}}}dr^2+\frac{4\mc{C}W}{B}d\theta^2\right.\nonumber \\ 
& & +\frac{4(r-1)^{\frac{2}{3}}e^{-\frac{\sigma}{2}}\Delta^{-1}}{9Br^{\frac{5}{12}}}\left[\sin^2\xi (e^{-\phi_2}\cos^2\eta+e^{\phi_2}\sin^2\eta)d\eta^2-\sin2\xi \sin2\eta\times \right.\nonumber \\
& &\sinh\phi_2 d\eta d\xi+[e^{4\sigma}\sin^2\xi+\cos^2\xi\left(e^{-\phi_2}\sin^2\eta+e^{\phi_2}\cos^2\eta)\right]d\xi^2 \nonumber 
\\
& &\left.\phantom{\frac{1^{\frac{5}{3}}}{1^{\frac{4}{3}}}}\left.+\sin^2\xi(e^{\phi_2}\cos^2\eta d\varphi_2^2+e^{-\phi_2}\sin^2\eta(d\varphi_1-g_1\mc{A}_1)^2)\right]\right\}.
\end{eqnarray} 
In this equation, we have set $m=\frac{g_1}{3}$ such that the parameter $G=1$ for simplicity. The warp factor is given by
\begin{equation}
\Delta=e^{-3\sigma}\left[\cos^2\xi+e^{4\sigma-\phi_2}\sin^2\xi(\cos^2\eta +e^{2\phi_2}\sin^2\eta)\right].
\end{equation}
The ten-dimensional dilaton takes the same form as given in \eqref{10D_dilaton1}
\\
\indent The four-form field strength reads
\begin{eqnarray}
F_{(4)}&=&\frac{\lambda \cos^{-\frac{2}{3}}\xi}{2g_1^3\Delta}\left\{U\Delta^{-1}\sin^3\xi\cos\xi \sin2\eta d\eta \wedge d\xi \wedge D\varphi_1\wedge d\varphi_2\right.\nonumber \\
& &+\frac{\phi_2'e^{-6\sigma}}{\Delta}\sin2\eta \sin^3\xi \cos\xi\left[e^{-2\phi_2}\sin2\xi(e^{4\phi_2}\sin^2\eta-\cos^{2}\eta)dr\wedge d\eta\right. \nonumber \\
& &\left.-\sin2\eta(2\cos^2\xi \cosh2\phi_2+e^{8\sigma}\sin^2\xi)dr\wedge d\xi\right]\wedge D\varphi_1 \wedge d\varphi_2\nonumber \\
& & +\frac{1}{4}g_1e^{-3\sigma}\sin2\xi\mc{F}_1\wedge \left[4e^{-\phi_2}\cos^2\eta(e^{\phi_2}\cos^2\xi+e^{4\sigma}\sin^2\xi)d\xi \right. \nonumber \\
& &\left.\left.-\sin2\eta\sin2\xi d\eta \right]\wedge d\varphi_2 \right\}.
\end{eqnarray}
On the other hand, the three-form potential is given by
\begin{eqnarray}
C_{(3)}&=&\frac{\lambda}{2g_1^3}\cos^{-\frac{2}{3}}\left\{\left(\frac{1}{2}\sin2\xi\cos2\eta-\frac{e^\sigma}{\Delta}\cos\xi\sin^3\xi\sin^22\eta\sinh\phi_2\right)d\xi\right. \nonumber \\
& &\left.-\left[\sin^2\xi+\frac{e^\sigma}{\Delta}\sin^4\xi(\cos2\eta\sinh\phi_2-\cosh\phi_2)\right]\sin2\eta d\eta\right\}\wedge\nonumber \\
& &(d\varphi_1\wedge d\varphi_2-g_1\mc{A}_{1\theta}d\theta\wedge d\varphi_2)-\frac{3\lambda}{8g_1^2}\cos^{\frac{4}{3}}\xi \mc{A}'_{1\theta}dr\wedge d\theta\wedge d\varphi_2\, .
\end{eqnarray}
The flux quantization proceeds along the same direction as in the previous case. In particular, the integral of $F_{(4)}$ along $(\eta\xi\varphi_1\varphi_2)$ and the quantization of $F_{(0)}$ still give the two relations \eqref{lambda_g1}. Following the same procedure as in the previous section but with $\mc{A}_{1\theta}$ and $\mc{A}_{2\theta}$ given in \eqref{A12_22}, we find
\begin{equation}
\int F_{\xi r\theta\varphi_2}dr\wedge d\xi\wedge d\theta\wedge d\varphi_2=\frac{9\lambda \pi^2r_+}{g_1^3(1-3r_+)}=(2\pi \ell_s)^3M
\end{equation}
with $M\in \mathbb{Z}$. Using \eqref{lambda_g1}, we can write $r_+$ as
\begin{equation}
r_+=\frac{Ml}{3(N+Ml)}\, .\label{rp_MN}
\end{equation}
\indent Using the formula in appendix \ref{uplift}, we can compute the holographic free-energy of the dual SCFT in three dimensions. The result is given by
\begin{eqnarray}
\mathbf{F}&=&\frac{9\lambda^4}{80\pi^2g_1^4\ell_s^8}\int_0^{r_+}\frac{B^{\frac{3}{2}}\mc{C}}{24m^4(r-1)^2}dr\nonumber \\
&=&\left(\frac{3}{2}\right)^5\frac{\lambda^4 B^{\frac{3}{2}}\mc{C}r_+}{20\pi^2g_1^8\ell_s^8(1-r_+)}\, .
\end{eqnarray}
Using the relation $B^{\frac{3}{2}}=r_+^{\frac{1}{2}}(r_+-1)$ and $\mc{C}=\frac{1}{(3r_+-1)l}$ as well as \eqref{rp_MN}, we can determine the holographic free energy as
\begin{equation}
\mathbf{F}=\frac{18\pi N^3r_+^{\frac{3}{2}}}{(1-3r_+)\sqrt{N(8-N_f)}}=\frac{2\sqrt{3}\pi}{5}\frac{N^2Ml}{\sqrt{N(8-N_f)}}\sqrt{\frac{Ml}{N+Ml}}\, .
\end{equation}
For $M\sim N$, we find $\mathbf{F}\sim N^{\frac{5}{2}}$ which is typical for the free energy of the five-dimensional SCFT dual to the $AdS_6$ vacuum.
\\
\indent We end this section by considering the singularity at $r=0$. As previously mentioned, the six-dimensional metric when $r\rightarrow 0$ is singular with the behavior given in \eqref{SO(3)d_rto0_SO(2)xSO(2)_AdS4xdisk_metric}. We first note that the metric in this limit takes the same form as that considered in \cite{D4_4D_Orbifolds} upon changing the coordinates as  
\begin{equation}
z=\frac{2\mc{C}}{m}\theta\qquad \textrm{and}\qquad r=m\mc{C}\sqrt{q_1}R
\end{equation}
with
\begin{equation}
q_1=\frac{B^{\frac{3}{2}}}{2^{\frac{4}{3}}m^2}\, .
\end{equation}
As shown in \cite{D4_4D_Orbifolds}, the six-dimensional metric with this asymptotic behavior describes topological disks with enhanced symmetry. This is one of the reasons that we identify the solution in the range $(0,r_+)$ as one example of disk solutions with enhanced symmetry, see more detail in appendix \ref{Spindle_map}.
\\
\indent Following \cite{D4_4D_Orbifolds}, we will take the constrained coordinates $\mu_a$ in the ten-dimensional solution to be
\begin{equation}  
\mu_0=\sqrt{1-\mu^2}\cos\vartheta,\qquad \mu_1=\mu,\qquad \mu_2=\sqrt{1-\mu^2}\sin\vartheta\, .
\end{equation}      
As $r\rightarrow 0$, the ten-dimensional metric is given by
\begin{eqnarray}
ds^2_{10}&=&\frac{B^{\frac{3}{4}}}{2^{\frac{2}{3}}m^{\frac{3}{2}}}\cos^{-\frac{1}{3}}\vartheta(1-\mu^2)^{\frac{1}{3}}\left\{\tilde{r}^{\frac{1}{3}}\left(ds^2_{AdS_4}+\mc{C}^2d\theta^2\right)\phantom{\frac{B^{\frac{3}{2}}}{2^{\frac{4}{3}}}}\right.\nonumber \\ 
& &\left. +\frac{2^{\frac{8}{3}}m^6}{g_1^2B^3\tilde{r}}\left[d\tilde{r}^2+\tilde{r}^2(d\vartheta^2+\sin^2\vartheta d\varphi_2^2)+\frac{B^{\frac{3}{2}}}{2^{\frac{4}{3}}m^3}\frac{d\mu^2+\mu^2d\varphi_1^2}{1-\mu^2}\right]\right\}\quad\label{10D_smeared}    
\end{eqnarray}     
with $\tilde{r}=r^3$. As anticipated in \cite{D4_4D_Orbifolds}, this metric describes the near-horizon limit of D4-branes within the D4-D8-system smeared along two directions. In particular, the ten-dimensional metric corresponding to this smeared D4-branes has already been given in \cite{Smeared_D4} as        
\begin{eqnarray}        
ds^2_{10}&=&\left(\frac{3}{2}Q_8\cos\vartheta\right)^{-\frac{1}{3}}\left\{Q_4^{-\frac{1}{2}}\sigma^{\frac{1}{3}}(-dt^2+dx_1^2+dx_2^2+dx_3^2+dx_4^2)\right. \nonumber \\ 
& &\left. +Q_4^{\frac{1}{2}}\sigma^{-1}\left[dx_5^2+dx_6^2+d\sigma^2+\sigma^2(d\vartheta^2+\sin^2\vartheta d\psi^2)\right]\right\}
\end{eqnarray}
which is of the same form as \eqref{10D_smeared}. Therefore, the singularity at $r=0$ in the six-dimensional metric corresponds to D4-branes smeared along two directions in the uplifted ten-dimensional solution of massive type IIA theory. 
\section{Conclusions}\label{Conclus}
We have studied a number of supersymmetric $AdS_4\times \Sigma$ solutions in which $\Sigma$ is a topological disk with a non-trivial $U(1)$ holonomy on the boundary or a half-spindle from matter-coupled $F(4)$ gauged supergravity in six dimensions with $SO(3)\times SO(3)$ gauge group. The resulting solutions preserve eight supercharges and $SO(2)\times SO(2)$ or $SO(2)_{\text{diag}}$ symmetries and represent a new class of supersymmetric solutions from matter-coupled $F(4)$ gauged supergravity in six dimensions. Although the solutions have the same local form as in the case of $\Sigma$ being a spindle studied in \cite{D4_spindle}, we have extensively discussed various possible ranges of the radial coordinate in which regular solutions exist. The results are very similar to supersymmetric $AdS_5\times \Sigma$ solutions found in seven-dimensional $N=2$ gauged supergravity coupled to three vector multiplets studied in \cite{7D_N2_disk}. 
\\
\indent In the case of $SO(2)\times SO(2)$ symmetric solutions, only a single range of the radial coordinate leads to regular solutions. It turns out that, unlike the solution from pure $F(4)$ gauged supergravity studied previously in \cite{Suh_D4}, the solution is asymptotic to $AdS_6$ geometry at which the circle inside $\Sigma$ decompactifies. This leads to infinite holographic free energy as also pointed out in \cite{D4_4D_Orbifolds}. The solution is expected to describe a codimension-2 conformal defect within the five-dimensional $N=2$ SCFT similar to the solution considered in \cite{Gutperle_coD2_defect} and \cite{Conti} in which a holographic description of codimension-2 defects within six-dimensional $N=(1,0)$ and five-dimensional $N=2$ SCFTs has been given.   
\\
\indent For $SO(2)_{\textrm{diag}}$ symmetric solutions, there are a number of possible ranges of the radial coordinate for which regular solutions exist. Some of the solutions are similar to that of pure $F(4)$ gauged supergravity and lead to finite free energy upon uplited to massive type IIA theory. These solutions should correspond to new three-dimensional SCFTs arising from compactifications of $N=2$ SCFT in five dimensions on a half-spindle. On the other hand, there are solutions with $AdS_6$ asymptotic geometry as in the $SO(2)\times SO(2)$ case. These solutions also lead to infinite free energy and could also be interpreted as gravity duals of codimension-2 defects within $N=2$ SCFT in five dimensions. We have also found that some of the solutions lie outside the space of solutions given in \cite{D4_4D_Orbifolds} and represent new solutions of this type, see more detail in appendix \ref{Spindle_map}. 
\\
\indent As in other cases, the $AdS_4\times \Sigma$ solutions found here are just dual to the IR limit of the full solutions describing holographic RG flows across dimensions from five to three dimensions. It could be interesting to construct the full solutions interpolating between this $AdS_4\times \Sigma$ geometry and the $AdS_6$ vacuum dual to the five-dimensional SCFT. The solutions with $AdS_6$ asymptotic geometry also deserve further investigation in the context of conformal defects by using holographic renormalization as performed in \cite{Gutperle_coD2_defect} to determine vacuum expectation values of the stress tensor and currents in the dual SCFT or entanglement entropy of the defects as in \cite{Conti}. These three-dimensional defects would align nicely with line and surface defects as well as Janus interfaces studied in \cite{6D_Janus_RG,F4_defect,6D_Janus}. It is also interesting to find $AdS_4\times\Sigma$ solutions from matter-coupled $F(4)$ gauged supergravity with $ISO(3)$ gauge group. This gauged supergravity has been studied recently in \cite{PK_ISO3_Janus,PK_ISO3_twist,PK_ISO3_BH} and can be uplifted to type IIB theory as shown in \cite{Malek_AdS7_AdS6}. The solutions could similarly give new three-dimensional SCFTs or codimension-2 defects within $N=2$ SCFTs in five dimensions realized in terms of brane configurations in type IIB theory.  
\vspace{0.5cm}\\
{\large{\textbf{Acknowledgement}}} \\
We would like to thank Carlos Nunez for helpful correspondence.
\appendix
\section{Bosonic field equations}\label{App_field_eqs}
In this appendix, we collect all the bosonic field equations obtained from the Lagrangian given in \eqref{Bos_Lag}. These are given by
\begin{eqnarray}\label{H_field_eqs}
0&=&\frac{3}{8}D_\rho\left(e^{4\sigma}H^{\mu\nu\rho}\right)-m^2e^{-2\sigma}\mathcal{N}_{00}B^{\mu\nu}+me^{-2\sigma}\mathcal{N}_{0\Sigma}F^{\Sigma\mu\nu}\nonumber\\
&&-\frac{1}{16}\epsilon^{\mu\nu\rho\sigma\lambda\tau}\left(\eta_{\Lambda\Sigma}F^\Lambda_{\rho\sigma}F^\Sigma_{\lambda\tau}-2mB_{\rho\sigma}
F^0_{\lambda\tau}+m^2B_{\rho\sigma}B_{\lambda\tau}\right),\\
0&=&D_\nu\left(e^{-2\sigma}\mathcal{N}_{\Lambda\Sigma}\widehat{F}^{\Sigma\nu\mu}\right)+2P^{\mu I\alpha}(L^{-1})_{I\Sigma}f_{\Lambda\phantom{\Sigma}\Pi}^{\phantom{\Lambda}\Sigma}{L^\Pi}_\alpha\nonumber\\&&-\frac{1}{8}\epsilon^{\mu\rho\sigma\nu\lambda\tau}H_{\rho\sigma\nu}(mB_{\lambda\tau}\delta_{\Lambda0}+\eta_{\Lambda\Sigma}F^\Sigma_{\lambda\tau}),\label{F_field_eqs}
\\
0&=&R_{\mu\nu}-4\partial_\mu\sigma\partial_\nu\sigma-P_\mu^{I\alpha}P_{\nu I\alpha}-g_{\mu\nu}V-e^{-2\sigma}\mathcal{N}_{\Lambda\Sigma}\left(\widehat{F}^\Lambda_{\mu\rho}{\widehat{F}^{\Sigma\ \rho}}_{\,\ \nu}-\frac{1}{8}g_{\mu\nu}\widehat{F}^\Lambda_{\rho\lambda}\widehat{F}^{\Sigma\rho\lambda}\right)\nonumber\\
&&-\frac{9}{16}e^{4\sigma}\left(H_{\mu\rho\lambda}{H_\nu}^{\rho\lambda}-\frac{1}{6}g_{\mu\nu}H_{\rho\lambda\tau}H^{\rho\lambda\tau}\right),
\end{eqnarray}
\begin{eqnarray}
0&=&D_\mu D^\mu\sigma-\frac{e^{2\sigma}}{2}\left[\frac{1}{9}A^2+B_iB^i+C_{Ii}C^{Ii}+4D_{Ii}D^{Ii}\right]-6me^{-6\sigma}\mathcal{N}_{00}\nonumber\\
&&+4me^{-2\sigma}\left(\frac{A}{3}L_{00}-L_{0i}B^i\right)-\frac{e^{-2\sigma}}{4}\mathcal{N}_{\Lambda\Sigma}\widehat{F}^\Lambda_{\mu\nu}\widehat{F}^{\Sigma\mu\nu}+\frac{3e^{4\sigma}}{16}H_{\mu\nu\rho}H^{\mu\nu\rho},\qquad \\
0&=&\frac{1}{2}D_\mu {P^\mu}_{I0}+e^{2\sigma}(B_iC_{iI}+K_{iIJ}D^{Ji})-2me^{-2\sigma}L_{0i}C_{Ii}\nonumber\\
& &-e^{-2\sigma}L_{0\Lambda}L_{I\Sigma}\widehat{F}^\Lambda_{\mu\nu}\widehat{F}^{\Sigma\mu\nu},\\
0&=&\frac{1}{2}D_\mu {P^\mu}_{Ii}+2me^{-2\sigma}(L_{00}C_{Ii}-2\varepsilon_{ijk}L_{0j}D_{Ik})-e^{-2\sigma}L_{i\Lambda}L_{I\Sigma}\widehat{F}^\Lambda_{\mu\nu}\widehat{F}^{\Sigma\mu\nu}\nonumber\\&&+e^{2\sigma}\left[\frac{A}{3}C_{Ii}+\varepsilon_{ijk}(B_jD_{Ik}+C_{Jj}K_{kIJ})\right]
\end{eqnarray}
with
\begin{equation}
K_{iIJ}=f_{\Lambda\phantom{\Gamma}\Sigma}^{\phantom{\Lambda}\Gamma}{L^\Lambda}_i(L^{-1})_{I\Gamma}{L^\Sigma}_J\, .
\end{equation}

\section{Relations to spindle solutions}\label{Spindle_map}
$AdS_4\times \Sigma$ solutions with $\Sigma$ being a spindle and a topological disk take the same local form but with different global extensions. A general classification of $AdS_4\times \Sigma$ solutions within the $U(1)\times U(1)$ subsector of matter-coupled $F(4)$ gauged supergravity has been given in \cite{D4_4D_Orbifolds}. In this appendix, we show that the solutions given in this paper are related to the solutions describing D4-branes wrapped on a spindle studied in \cite{D4_spindle}. In addition, we will show that the solutions presented here extend the known spindle and disk solutions classified in \cite{D4_4D_Orbifolds}.
\\
\indent We first recall that the $AdS_4\times \Sigma$ solution for $\Sigma$ being a spindle given in \cite{D4_spindle} consists of the metric, gauge fields, and scalars as in this paper. In the notation of \cite{D4_4D_Orbifolds}, the solution can be written as
\begin{eqnarray}
ds^2_6&=&\left(y^2h_1h_2\right)^{1/4}\left(ds^2_{AdS_4}+\frac{y^2}{F}dy^2+\frac{F}{h_1h_2}dz^2\right),\nonumber\\
\mc{A}_i&=&\left(\alpha_i-\frac{y^3}{h_i}\right)dz,\qquad X_i=\frac{\left(y^2h_1h_2\right)^{3/8}}{h_i}\label{D4_spindle_soln}
\end{eqnarray}
in which
\begin{equation}
F(y)=m^2h_1h_2-y^4\qquad\text{ and }\qquad h_i(y)=\frac{2g}{3m}y^3+q_i\label{Fandhi}
\end{equation}
with $\alpha_i$ and $q_i$ for $i=1,2$ being canstants. Following appendix A.1 of \cite{D4_spindle}, these quantities can be expressed in the present notation as
\begin{eqnarray}
X_1&=&e^{\sigma+\phi_2},\qquad X_2=e^{\sigma-\phi_2},\qquad g=g_1,\qquad m\rightarrow\frac{m}{2},\nonumber\\
\mc{A}_1&=&\frac{1}{2}(A^{3}+A^{6}),\qquad \mc{A}_2=\frac{1}{2}(A^{3}-A^{6}).\label{Uplift_relation}
\end{eqnarray}

\subsection{$SO(2)\times SO(2)$ symmetric solution}
\begin{figure}[h!]
\centering
    \includegraphics[width=0.8\linewidth]{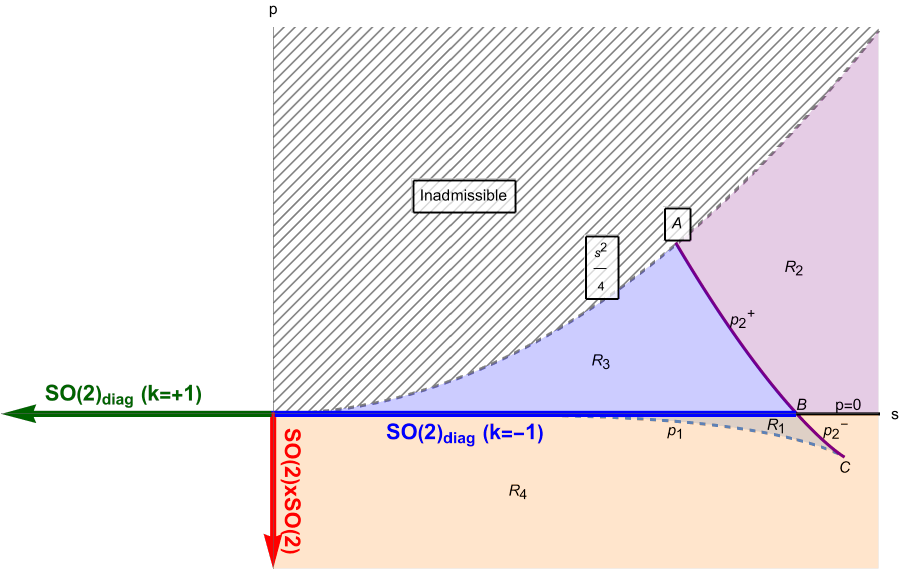}
  \caption{Various domains of validity for $AdS_4\times \Sigma$ solutions in $(s,p)$ space. This plot is the same as that given in \cite{D4_4D_Orbifolds} with four regions $R_1$, $R_2$, $R_3$, and $R_4$ corresponding to different global completions. The regions are separated by different lines with the two non-trivial lines $p=p_1(s)$ and $p=p_2(s)$ given explicitly in \cite{D4_4D_Orbifolds}. A superscript $\pm$ on $p_2$ denotes whether this line lies above or below the $s$ axis. There are three distinguished points arising from the intersection of the boundary lines shown by $A=\left(\frac{8}{27},\frac{16}{729}\right)$, $B=\left(\frac{2}{3\sqrt{3}},0\right)$, and $C=\left(\frac{8\sqrt{2}}{27},-\frac{4}{729}\right)$.}
  \label{Fig9}
\end{figure}

From the $SO(2)\times SO(2)$ symmetric solution given in section \ref{SO(2)xSO(2)_Sol}, we find
\begin{eqnarray}
X_1&=&r^{\frac{1}{8}}\left(G\sqrt{r}+\sqrt{G^2r-1}\right),\\
X_2&=&r^{\frac{1}{8}}\left(G\sqrt{r}-\sqrt{G^2r-1}\right),\\
\mc{A}_1&=&\left[\frac{2\mathcal{C}}{m}\left(Gr+\sqrt{r(G^2r-1)}\right)+\frac{c}{2}-2\left(\frac{q+2\mathcal{C}}{3Gm}\right)\right]dz,\\
\mc{A}_2&=&\left[\frac{2\mathcal{C}}{m}\left(Gr-\sqrt{r(G^2r-1)}\right)-\frac{c}{2}-2\left(\frac{q+2\mathcal{C}}{3Gm}\right)\right]dz.
\end{eqnarray}
Comparing the metric in \eqref{D4_spindle_soln} with \eqref{SO(2)xSO(2)_6D_Met} and \eqref{SO(2)xSO(2)_Sigma_Met} yields the coordinate transformations
\begin{equation}
r=\frac{m^6y^6}{G^2m^6y^6-B^3}\qquad\text{ and }\qquad \theta=\frac{mz}{2\mathcal{C}}
\end{equation}
that map the spindle metric in \eqref{D4_spindle_soln} to the disk solution in \eqref{SO(2)xSO(2)_6D_Met} as follows
\begin{eqnarray}
F(y)&=&\frac{B^2\left[B-r^{\frac{2}{3}}(G^2r-1)^{\frac{1}{3}}\right]}{m^4(G^2r-1)}=m^2h_1h_2-y^4,\\
h_1(y)&=&\frac{B^{\frac{3}{2}}}{m^3\left[G\sqrt{r(G^2r-1)}+G^2r-1\right]}=Gy^3-\frac{B^{\frac{3}{2}}}{m^3},\label{h1_sol11}\\
h_2(y)&=&\frac{B^{\frac{3}{2}}}{m^3\left[G\sqrt{r(G^2r-1)}-G^2r+1\right]}=Gy^3+\frac{B^{\frac{3}{2}}}{m^3}\label{h2_sol11}\, .
\end{eqnarray}
Recalling $W(r)=B-r^{\frac{2}{3}}(G^2r-1)^{\frac{1}{3}}$, we see that $r_1$ given in \eqref{r_Roots} is also a unique real root of $F(y)$ while, at the other endpoint $r=\frac{1}{G^2}$, the function $F(y)$ diverges. Moreover, comparing the functions $h_1$ and $h_2$ in \eqref{h1_sol11} and \eqref{h2_sol11} with \eqref{Fandhi}, we find that the two constants $q_i$ take the form
\begin{equation}
q_1=-\frac{B^{\frac{3}{2}}}{m^3}\qquad\text{ and }\qquad q_2=\frac{B^{\frac{3}{2}}}{m^3}\, .
\end{equation}
\indent For the sake of comparison to the spindle and previous disk solutions, it is convenient to define the following two parameters as in \cite{D4_4D_Orbifolds}
\begin{equation}\label{sandp}
s=m^3(q_1+q_2)\qquad\text{ and }\qquad p=m^6q_1q_2\, .
\end{equation}
In terms of these parameters, we find that $SO(2)\times SO(2)$ symmetric solution given here has $s=0$ and $p=-B^3$. Since $B^3$ is an arbitrary positive real parameter, see \eqref{SO(2)xSO(2)_para_redefine}, this implies that the $SO(2)\times SO(2)$ symmetric solution found in this work is equivalent to the spindle solution in \cite{D4_spindle} with $s=0$ and $p<0$. For clarity, we have added a red line representing solutions with $s=0$ and $p<0$ to the plot of the domains of validity on the $(s,p)$ plane given in \cite{D4_4D_Orbifolds} as shown in figure \ref{Fig9}. As pointed out in \cite{D4_4D_Orbifolds}, the spindle solution of \cite{D4_spindle} exists in the region $R_1$ while the disk solution of \cite{Suh_D4} exists in the regions $R_2$ and $R_3$. On the other hand, the region $R_4$ contains non-compact domain-wall solutions with infinite holographic free energy. Consequently, we conclude that the $SO(2)\times SO(2)$ symmetric solution found here represents a novel case that interpolates between a single root of $F(y)=0$ and a singularity.

\subsection{$SO(2)_{\text{diag}}$ symmetric solution}
For the $SO(2)_{\text{diag}}$ symmetric solution studied in section \ref{SO(2)d_SO(3)xSO(3)gg_Sec}, we have
\begin{eqnarray}
& &X_1=Gr^{\frac{5}{8}},\qquad X_2=\frac{1}{Gr^{\frac{3}{8}}},\nonumber \\
& & \mc{A}_1=\left(-\frac{2(2q+\mathcal{C})}{3Gm}+\frac{2G\mathcal{C}r}{m}\right)dz,\qquad \mc{A}_2=0\, .
\end{eqnarray}
In this case, the metric of the spindle solution given in \eqref{D4_spindle_soln} and that of the disk given in \eqref{SO(2)d_6D_Met} and \eqref{SO(2)d_Sigma_Met} are equivalent after changing the radial and azimuthal coordinates as follows
\begin{equation}
r=\frac{km^3y^3}{kG^2m^3y^3-B^3}\qquad\text{ and }\qquad \theta=\frac{mz}{2\mathcal{C}}\, .
\end{equation}
\\
\indent The various functions appearing in the metric are now given by
\begin{eqnarray}
F(y)&=&\frac{B^2r\left[B-r^{\frac{1}{3}}(k(G^2r-1))^{\frac{2}{3}}\right]}{m^4(G^2r-1)}=m^2h_1h_2-y^4,\\
h_1(y)&=&\frac{B^{\frac{2}{3}}}{kGm^3(G^2r-1)}=Gy^3-\frac{B^{\frac{2}{3}}}{kGm^3},\\
h_2(y)&=&\frac{B^{\frac{2}{3}}Gr}{km^3(G^2r-1)}=Gy^3\, .
\end{eqnarray}
As $W(r)=B-r^{\frac{1}{3}}(k(G^2r-1))^{\frac{2}{3}}$, all three roots determined from $W(r)=0$ in \eqref{r0_def} and \eqref{rpm_def} simultaneously solve $F(y)=0$. Furthermore, it is straightforward to verify that $r=0$ constitutes another root of $F(y)$. As in the $SO(2)\times SO(2)$ case, $r=\frac{1}{G^2}$ leads to a divergence in $F(y)$. In this case, we find $q_2=0$ while the parameter $q_{1}$ is given by
\begin{equation}
q_1=-\frac{B^{\frac{3}{2}}}{kGm^3}\, .
\end{equation}
\indent From \eqref{sandp}, we find that the $SO(2)_{\text{diag}}$ symmetric disk solution given here is equivalent to the spindle solution in \cite{D4_spindle} with $s=-\frac{B^{\frac{3}{2}}}{kG}$ and $p=0$. However, in section \ref{SO(2)d_SO(3)xSO(3)gg_Sec}, we find two different behaviors in the solution depending on whether $k=+1$ or $k=-1$. For $k=+1$, a regular solution exists when $B>0$, $G>0$, and $m>0$. This corresponds to the green line ($s<0$, $p=0$) shown in figure \ref{Fig9} which represents a novel case discovered in this work. On the other hand, for $k=-1$, we find a solution with an $\mathbb{R}^2/\mathbb{Z}_l$ orbifold singularity only when $0<B<\frac{2^{\frac{2}{3}}}{3G^{\frac{2}{3}}}$, $G>0$, and $m>0$. Setting $2g=3m$ in \cite{D4_4D_Orbifolds} which leads to $G=1$ here, we find that this case corresponds to the blue line ($0<s<\frac{2}{3\sqrt{3}}$, $p=0$) in figure \ref{Fig9}. 
\\
\indent As pointed out in \cite{D4_4D_Orbifolds}, there are two types of disk solutions with and without enhanced symmetry. An example of the latter is the solution found in \cite{Suh_D4} and the $SO(2)\times SO(2)$ symmetric solution discussed above. The solution on the blue line in figure \ref{Fig9} closely resembles the disk with enhanced symmetry. However, only solutions interpolating between zero and a single non-vanishing root have been examined in \cite{D4_4D_Orbifolds}. This type of solutions should be identified with $SO(2)_{\text{diag}}$ solution found here in the range $0<r<r_+$ corresponding to the left region of the numerical plots shown in figure \ref{Fig5}. Apart from this, the present work suggests the existence of another type of solutions interpolating between another root and a singularity corresponding to a distinct range $r_-<r<\frac{1}{G^2}$.

\section{Uplift to massive type IIA theory}\label{uplift}
In this appendix, we give a truncation ansatz of massive type IIA theory on a half four-sphere as given in \cite{D4_4D_Orbifolds}, see also \cite{D4_spindle} and \cite{Pope_symmetric_potential} for earlier results. This truncation leads to $U(1)\times U(1)$ invariant sector of $F(4)$ gauged supergravity coupled to vector multiplets. All solutions from $SO(3)\times SO(3)$ gauged supergravity studied in this paper lie within this invariant sector and can be uplifted to ten dimensions by using the aforementioned truncation ansatz, see \cite{D4_spindle} for the uplifted solutions describing D4-branes wrapped on a Riemann surface and a spindle.
\\
\indent In the present convention, the metric ansatz in the string frame is given by
\begin{eqnarray}
ds^2&=&\lambda^2 \mu_0^{-\frac{1}{3}}(X_1X_2)^{-\frac{1}{4}}\Delta^{\frac{1}{2}}\left\{ds^2_6+g_1^{-2}\Delta^{-1}\left[X_0^{-1}d\mu^2_0\right. \right.\nonumber \\
& &\left. \left.+X^{-1}_1(d\mu_1^2+\mu_1^2D\varphi_1^2)+X^{-1}_2(d\mu_2^2+\mu_2^2D\varphi_2^2)\right]\right\}.
\end{eqnarray}
$X_1$ and $X_2$ are related to the dilaton $\sigma$ and $SO(2)\times SO(2)$ singlet scalar $\phi_2$ as in \eqref{Uplift_relation}. On the other hand, $X_0$ is defined by
\begin{equation}
X_0=(X_1X_2)^{-\frac{3}{2}}=e^{-3\sigma}\, .
\end{equation}
The constrained coordinates $\mu_a$, $a=0,1,2$, satisfy
\begin{equation}
\mu_0^2+\mu_1^2+\mu_2^2=1
\end{equation}    
and can be chosen to be
\begin{equation}
\mu_0=\cos\xi,\qquad \mu_1=\sin\xi\sin\eta,\qquad \mu_2=\sin\xi\cos\eta
\end{equation}
with $\eta\in [0,\frac{\pi}{2}]$ and $\xi\in [0,\frac{\pi}{2}]$. The two angular coordinates $\varphi_1$ and $\varphi_2$ have period $2\pi$ with $D\varphi_i=d\varphi_i-g_1\mc{A}_i$, $i=1,2$. The two gauge fields $\mc{A}_{1,2}$ are again related to the two gauge fields $A^3$ and $A^6$ as in \eqref{Uplift_relation}.
\\
\indent The ten-dimensional dilaton is given by
\begin{equation}
e^{\Phi}=\lambda^2\mu_0^{-\frac{5}{6}}\Delta^{\frac{1}{4}}(X_1X_2)^{-\frac{5}{8}}\, .  
\end{equation}
The zero-form Romans mass is 
\begin{equation}
F_{(0)}=\frac{m}{\lambda^3}=\frac{g_1}{3\lambda^3}
\end{equation}
Since the six-dimensional two-form field vanishes in all of the solutions considered here, we will set this two-form field to zero in all subsequent expressions. This results in the following four-form field strength
\begin{eqnarray}
F_{(4)}&=&\sqrt{\frac{g_1}{3m}}\frac{\lambda\mu_0^{-\frac{2}{3}}}{2g_1^3\Delta}\left\{2U\mu_1\mu_2\Delta^{-1}d\mu_1\wedge d\mu_2\wedge D\varphi_1\wedge D\varphi_2\phantom{\frac{X_1}{X_2}} \right.\nonumber \\
& & -2X_1X_2\Delta^{-1}\left[\frac{\mu_1\mu_2\mu_0^2X_0^2}{X_1X_2}\left(\mu_1d\frac{X_1}{X_2}\wedge d\mu_2-\mu_2d\frac{X_2}{X_1}\wedge d\mu_1\right)\right.\nonumber \\
& &\left.\left.-\mu_1^2\mu_2^2\mu_0d\ln\frac{X_1}{X_2}\wedge d\mu_0\right]\wedge D\varphi_1\wedge D\varphi_2\right\}\nonumber \\
& &+\sqrt{\frac{g_1}{3m}}\frac{\lambda\mu_0^{-\frac{2}{3}}}{g_1^2\Delta}\left\{\mc{F}_1\wedge \left[\mu_1\mu_2^2X_2d\mu_1+\mu_2(X_0\mu_0^2+X_2\mu_2^2)d\mu_2\right]\wedge D\varphi_2\right.\nonumber \\
& &\left.+\mc{F}_2\wedge \left[\mu_2\mu_1^2X_1d\mu_2+\mu_1(X_0\mu_0^2+X_1\mu_1^2)d\mu_1\right]\wedge D\varphi_1\right\}.
\end{eqnarray}
All the remaining fields in massive type IIA theory vanish. In all the above expressions, we have included a parameter $\lambda$ such that the uplifted ten-dimensional solutions can be defined globally with all the fluxes properly quantized along the internal compact dimensions, see more detail in \cite{D4_spindle}. In addition, we can also write down the ansatz for the corresponding three-form potential given in \cite{D4_4D_Orbifolds}
\begin{eqnarray}  
C_{(3)}&=&\frac{\lambda}{2g_1^3}\mu_0^{-\frac{2}{3}}\left[\mu_2 d\mu_2-\mu_1d\mu_1\right. \nonumber \\
& &\left. +\frac{2}{\Delta}(X_2\mu_2^2\mu_1d\mu_1-X_1\mu_1^2\mu_2d\mu_2)\right]\wedge D\varphi_1\wedge D\varphi_2\nonumber \\
& &-\frac{3\lambda}{8g_1^3}\mu_0^{\frac{4}{3}}(\mc{F}_1\wedge D\varphi_2+\mc{F}_2\wedge D\varphi_1).
\end{eqnarray}
\indent The $AdS_4\times \Sigma$ geometries are expected to holographically describe three-dimensional SCFTs arising from compactifications of $N=2$ SCFTs in five dimensions. The holographic free energy of the three-dimensional SCFTs can be computed by writing the ten-dimensional metric in string frame as
\begin{equation}
ds^2_{10}=e^{2\mathbf{A}}\left(ds^2_{AdS_4}+ds^2_{M_6}\right).
\end{equation} 
As in \cite{5D_SCFT_on_Riemann} and \cite{D4_spindle}, the free energy is then given by an integral over the internal space $M_6$ of the form
\begin{equation}
\mathbf{F}=\frac{16\pi^3\lambda^4}{(2\pi \ell_s)^8}\int_{M_6}e^{8\mathbf{A}-2\Phi}\textrm{vol}_{M_6}\, .
\end{equation} 
For the six-dimensional metric given in \eqref{6Dmetrix1}, we find the uplifted ten-dimensional metric of the form
\begin{equation}
ds^2_{10}=\cos^{-\frac{1}{3}}\xi e^{-\frac{\sigma}{2}}\Delta^{\frac{1}{2}}f\left[ds^2_{AdS_4}+\frac{h_1}{f}dr^2+\frac{e^{-\sigma}\Delta^{-1}}{g_1^2f}ds^2_{M_2}+ds^2_{M_3}\right]
\end{equation}
in which the metrics on $M_2$ and $M_3$ submanifolds are given respectively by 
\begin{eqnarray}
ds^2_{M_2}&=&\sin^2\xi\left(e^{-\phi_2}\cos^2\eta+e^{\phi_2}\sin^2\eta\right)d\eta^2-\sin2\xi\sin2\eta\sinh\phi_2d\xi d\eta\nonumber \\
& &+\left[e^{4\sigma}\sin^2\xi+\cos^2\xi(e^{-\phi_2}\sin^2\eta+e^{\phi_2}\cos^2\eta)\right]d\xi^2
\end{eqnarray}
and
\begin{eqnarray}
ds^2_{M_3}&=&\frac{h_2}{f}d\theta^2+\frac{e^{-\sigma}\Delta^{-1}\sin^2\xi}{g_1^2f} \left[e^{\phi_2}\cos^2\eta(d\varphi_2-g_1\mc{A}_{2\theta} d\theta)^2\right.\nonumber \\
& &\left.+e^{-\phi_2}\sin^2\eta(d\varphi_1-g_1\mc{A}_{1\theta}d\theta)^2\right]\nonumber \\
&=&\frac{e^{-\sigma}\Delta^{-1}}{g_1^2f}\left[g_1^2h_2e^\sigma \Delta+\sin^2\xi\left(e^{-\phi_2}\mc{A}_{1\theta}^2\sin^2\eta+e^{\phi_2}\mc{A}_{2\theta}^2\cos^2\eta\right)\right]d\theta^2\nonumber \\
& &+\frac{e^{-\sigma}\Delta^{-1}\sin^2\xi}{g_1^2f}\left[e^{-\phi_2}\sin^2\eta d\varphi_1^2+e^{\phi_2}\cos^2\eta d\varphi_2^2\right.\nonumber \\
& &\left.-2g_1\left(e^{-\phi_2}\mc{A}_{1\theta}\sin^2\eta d\theta d\varphi_1+e^{\phi_2}\mc{A}_{2\theta}\cos^2\eta d\theta d\varphi_2\right)\right].
\end{eqnarray}
We can then identify
\begin{equation}
e^{\mathbf{A}}=\cos^{-\frac{1}{6}}\xi e^{-\frac{\sigma}{4}}\Delta^{\frac{1}{4}}f^{\frac{1}{2}}
\end{equation}
and find the volume form on $M_6$ as
\begin{eqnarray}
\textrm{vol}_{M_6}&=&\sqrt{\frac{h_1}{f}}\left(\frac{\Delta^{-1}e^{-\sigma}}{g_1^2f}\right)\sqrt{\textrm{det}g_{M_2}\textrm{det}g_{M_3}}\, dr\wedge d\eta\wedge d\xi\wedge d\theta\wedge d\varphi_1 \wedge d\varphi_2\nonumber \\
&=&\frac{\sqrt{h_1h_2}}{g_1^4f^3}\Delta^{-\frac{3}{2}}e^{-\frac{\sigma}{2}}\sin^3\xi \cos\eta\sin\eta \, dr\wedge d\eta\wedge d\xi \wedge d\theta\wedge d\varphi_1 \wedge d\varphi_2\qquad \,
\end{eqnarray}
in which we have used the expression for the warp factor of the form
\begin{equation}
\Delta=e^{-3\sigma}\left[\cos^2\xi+\sin^2\xi e^{4\sigma-\phi_2}(\cos^2\eta+e^{2\phi_2}\sin^2\eta)\right].
\end{equation}
\indent Using the explicit form of the ten-dimensional dilaton
\begin{equation}
e^{\Phi}=\cos^{-\frac{5}{6}}\xi \Delta^{\frac{1}{4}}e^{-\frac{5\sigma}{4}},
\end{equation}
we eventually obtain the free energy
\begin{equation}
\mathbf{F}=\frac{9\lambda^4}{80\pi^2g_1^4\ell_s^8}\int f\sqrt{h_1h_2}dr\, .
\end{equation}


\end{document}